\documentclass{aa}
\usepackage{amssymb}
\usepackage{graphicx}
\usepackage{longtable}
\usepackage{xcolor}
\usepackage{url} 
\usepackage{amsmath}

\def\Teff{$T_{\rm eff}$}
\def\logg{$\log\,g$}

\def\Vt{V${\rm t}$}
\newcommand{\vsini}{$v$sin$i$}
\newcommand{\kms}{km~s$^{-1}$}

\def\apjl{ApJ}%
\def\apjs{ApJS}%
\def\ao{Appl.~Opt.}%
\def\aap{A\&A}%
\newcommand {\apgt} {\ {\raise-.5ex\hbox{$\buildrel>\over\sim$}}\ }
\newcommand {\aplt} {\ {\raise-.5ex\hbox{$\buildrel<\over\sim$}}\ }

\begin{document} 

   \title{Peculiarities of the chemical enrichment of metal-poor Stars in the Milky Way Galaxy}

%\title[Pecularities Stars ]
%{Peculiarities of the Chemical Enrichment of Metal-Poor Stars in the Galaxy (Milky Way)}
\thanks{Tables A1 and B1 are only available in electronic form at the CDS via the anonymous ftp to cdsarc.u-strasbg.fr (130.79.128.5)
or via http://cdsweb.u-strasbg.fr/cgi-bin/qcat?J/A+A/(vol)/(page)}
   \author{T.~Mishenina\inst{1},
          M.~Pignatari\inst{2,3,4}\thanks{The NuGrid collaboration, http://www.nugridstars.org}*,
          I.Usenko\inst{1},
          C.~Soubiran\inst{5},
          F.-K.~Thielemann\inst{6,7},          
          A.Yu. Kniazev\inst{8},
          S.A.~Korotin\inst{9},
          \and
          T.~Gorbaneva\inst{1},
          }
    
\authorrunning{T.~Mishenina  et al.}

  \institute{Astronomical Observatory, Odesa National University, 1b, Marazlievska str., 65014, Odesa, Ukraine\\ \email{tmishenina@ukr.net}\\
        \and Konkoly Observatory, HUN-REN, Konkoly Thege Miklos ut 15-17, H-1121 Budapest, Hungary\\ \email{mpignatari@gmail.com}\\
        \and MTA Centre of Excellence, Budapest, Konkoly Thege Miklós út 15-17, H-1121, Hungary\\
        \and E.A. Milne Centre for Astrophysics, University of Hull, Hull HU6 7RX, UK\\
        \and Laboratoire d'Astrophysique de Bordeaux, Univ. Bordeaux, CNRS, B18N, all\'ee Geoffroy Saint-Hilaire, 33615, Pessac, France \\
        \and Department of Physics, University of Basel, Klingelbergstrabe 82, CH-4056 Basel, Switzerland \\
         \and GSI Helmholtzzentrum für Schwerionenforschung, Planckstrasse 1, D-64291 Darmstadt, Germany\\
        \and Southern African Large Telescope Foundation, South African Astronomical Observatory, P.O.  box 9, 7935 Observatory,\\ Cape Town, South Africa\\
        \and Crimean Astrophysical Observatory, Nauchny 298409, Crimea}
   \date{Received September 15, 1996; accepted March 16, 1997}

   \date{Received September 15, 1996; accepted March 16, 1997}

\abstract
{The oldest stars in the Milky Way are metal-poor with [Fe/H] $<$ -- 1.0,   displaying peculiar elemental abundances compared to solar values. The relative variations in the chemical compositions among stars is also increasing with decreasing stellar metallicity, allowing for the pure signature of unique nucleosynthesis processes to be revealed. The study of the $r$-process is, for instance, one of the main goals of stellar archaeology and metal-poor stars exhibit an unexpected complexity in the stellar production of the $r$-process elements in the early Galaxy.} 
{In this work, we report the atmospheric parameters, main dynamic properties, and the abundances of four metal-poor stars:  HE 1523-–0901, HD 6268, HD 121135, and HD 195636 (--1.5 $>$ [Fe/H] $>$--3.0).} 
{The abundances were derived from spectra obtained with the HRS echelle spectrograph at the Southern African Large Telescope, using both local and non-local thermodynamic equilibrium (LTE and NLTE) approaches, with the average error between 0.10 and 0.20 dex.} 
{Based on their kinematical properties, we show that HE 1523--0901 and HD 195636 are halo stars with typical high velocities. In particular, HD 121135 displays a peculiar kinematical behaviour, making it unclear whether it is a halo or an accreted star. Furthermore, HD 6268 is possibly a rare prototype of very metal-poor thick disk stars.
The abundances derived for our stars are compared with theoretical stellar models and with other stars with similar metallicity values from the literature.} 
{HD 121135 is Al-poor and Sc-poor, compared to stars observed in the same metallicity range (--1.62 $>$ [Fe/H] $>$--1.12). The most metal-poor stars in our sample, HE 1523 -- 0901, HD 6268, and HD 195636, exhibit anomalies that are better explained by supernova models from fast-rotating stellar progenitors for elements up to the Fe group. Compared to other stars in the same metallicity range, their common biggest anomaly is represented by the low Sc abundances. 
If we consider the elements beyond Zn, HE 1523--0901 can be classified as an r-II star, HD 6268 as an r-I candidate, and HD 195636 and HD 121135 exhibiting a borderline $r$-process enrichment between limited-r and r-I star. Significant relative differences are observed between the $r$-process signatures in these stars.}


   \maketitle

\section{Introduction}

Metal-poor stars ([Fe/H] $<$ -- 1.0) are the first stars ever formed and, thus, they are the oldest in the Galaxy \cite[e.g.][]{beers:05, frebel:15}. These stars show inhomogeneous chemical compositions with broad variations for most of the elements. Therefore, various peculiar abundance signatures have been the main subject of study within stellar and galactic archaeology \citep[e.g.][]{abohalima:18, farouqi:22, hartwig:23}.  
Since low-mass and intermediate-mass stars (M $\lesssim$ 9 M$_{\odot}$) have not had the time to evolve and contribute to the chemical evolution of the Galaxy \citep[e.g.][]{timmes:95, goswami:00, prantzos:18, kobayashi:20}, observations of the light elements up to the Fe group in single unevolved stars have been used to study the production of the elements in the first generations of core-collapse supernovae \citep[CCSNe, e.g.][]{woosley:02, nomoto:13}.  
This progress has enabled us to study the main properties of massive star progenitors evolved and exploded billions of years ago, along with the nature of the supernova engine driving the final collapse of the star and the ejection of the freshly made metals into the interstellar medium. The current generation of stellar models have indeed revealed a number of uncertainties affecting their predictive power \citep[e.g.][]{burrows:21,Curtis.Ebinger.ea:2019,Ghosh.Wolfe.Frohlich:2022}, and the comparison with observations in metal-poor stars highlight these limitations. For instance, stellar models tend to underproduce N, Cl Sc, K, Ti, and V, as compared to observations. Despite the fact that this problem has been recognised for some time, a definitive solution has not been found \cite[e.g.][]{mishenina:17,kobayashi:20, matteucci:21}. 

Beyond Fe, observations of stellar abundances between Ge and Pd are potentially identifying the contribution from a large number of nucleosynthesis components from different stellar sources, such as HD 122563 and HD 88609 \cite[][]{honda:06, hansen:12, roederer:16}. There is a sample of stars carrying the signature of the rapid neutron-capture process or $r$-process, for instance, CS 2892-952, CS31082-001 \cite[][]{cayrel:01, sneden:03, roederer:22, farouqi:22}; these stars have been classified as limited, moderate (r-I), or highly (r-II) $r$-process enhanced stars \citep[][]{christlieb:04}. Some of the so-called carbon-enhanced metal-poor (CEMP) stars are certainly part of a stellar binary system and they exhibit abundance signatures of either the slow neutron-capture process \citep[$s$-process,][and references therein]{kaeppeler:11}, of the intermediate neutron-capture process \citep[$i$-process,][]{cowan:77}, or of a mixture of those over a possible pristine $r$-process enriched composition \citep[e.g.][]{bisterzo:12, dardelet:14, abate:15, choplin:21, choplin:22}.   
 
Stellar archaeology is a major field of study aimed at identifying the stellar sources of the $r$- process active in the early Galaxy. Major stellar nucleosynthesis studies have been focused on neutron star-neutron star (NS-NS) and black hole-neutron star (BH-NS) mergers \cite[e.g.][]{lattimer:74, surman:08, cescutti:15, lippuner:17, thielemann:17, fernandez:20, farouqi:22}, magnetohydrodynamic explosions in massive stars  \cite[MHD CCSNe, e.g.][]{nishimura:06, winteler:12, mosta:18} , 
hypernovae (HNe) and collapsars \cite[e.g.][]{siegel:19}. Some alternative sites of the $r$-process have been proposed in several papers \cite[e.g.][]{travaglio:04, qian:08}.

Recently, there have been many observational efforts aimed at investigating the abundances in 
metal-poor stars 
\citep[e.g. see ][]{ji:24, spite:24}.
In this study, we present a detailed elemental-abundance analysis of four metal-poor stars, HE 1523--0901, HD 6268, HD 121135, and HD 195636, and we discuss their peculiar nucleosynthesis signatures. While previous works in the literature have studied the composition of these stars, inconsistent results have been found for several elements. We are striving to extend the list of elements previously available with new estimated abundances to make better comparisons with nucleosynthesis simulations. In particular, we significantly extended the element abundances available for HE 1523--0901, a well-known r-II star. We show that HD 6268 is possibly a prototype of rare metal-poor thick-disk stars, and that HD 121135 is a peculiar star, with an unclear classification based on its dynamics or on the heavy element abundances. Finally, we broadly extend the available element abundances for HD 195636, a poorly-known metal-poor halo star. 

This work is part of the spectroscopic observational campaign of the different stellar sub-systems in our Galaxy known as the \underline{\bf M}ilky W\underline{\bf A}y \underline{\bf G}alaxy w\underline{\bf I}th SALT spe\underline{\bf C}troscopy ({\bf MAGIC})  and including more than 100 stars \citep{2019AstBu..74..208K}.

\par
The paper is organised as follows. 
The observations and selection of stars are presented in 
\S \ref{sec: stellar target}.
The definition of the main stellar 
parameters and comparison with the results of other authors are described in \S \ref{sec: stellar param}. 
The abundance determinations  and the error analysis are presented in 
\S \ref{sec: abundance determination}. 
The kinematic parameters and membership of galactic structures are presented in \S \ref{sec: kinematics}. 
The impact of intrinsic stellar evolution are analysed in 
 \S \ref{sec: result, stellar}. 
The application of the results  in the theory of nucleosynthesis and the chemical evolution of the Galaxy in stars is reported in \S \ref{sec: element nucleosynthesis}. 
Our conclusions are drawn in 
\S \ref{sec: conclusions}.

\section{Observations: Main characteristics of stars and spectra}
\label{sec: stellar target}

For this study, we selected four peculiar stars which differ in parameters and features of the spectrum with metallicities ranging from [Fe/H] = -1.5 to -3.0. The star HE 1523--0901 is one of the oldest stars in the Galaxy with enhanced $r$-process element abundances \cite[][]{frebel:07, frebel:13}; the star HD 6268 shares similar physical parameters with HE 1523--0901, but shows milder enrichments beyond iron \cite[][]{roederer:14}. We then studied HD 121135, whose metallicity was previously estimated to be about [Fe/H] = -1.5 \cite[][]{pilachowski:96}. 
Finally, HD 195636 has a high projected rotational velocity \vsini~ \citep[][]{preston:97}, with a current value \vsini~ = 20 \kms which is classified as HB star in \cite{behr:03}.

The sample of stars discussed in this work are part of the Gaia DR3 catalogue \citep{gaiadr3}, with parallaxes ranging from 0.3 mas to 1.7 mas, with a good astrometric quality (parallax over error from 16.6 to 98.7 and renormalised unit weight error, RUWE, lower than 1.05). Table~\ref{main} presents the coordinates with B, V, and K magnitudes adopted from the database SIMBAD, and parallaxes (P) from Gaia DR3.

Spectral observations were made during 2017--2021 years at the Southern African Large Telescope \citep[SALT;][] {2006SPIE.6267E..0ZB,2006MNRAS.372..151O} using the fibre-fed \'echelle spectrograph HRS \citep[][]{2008SPIE.7014E..0KB,2010SPIE.7735E..4FB,2012SPIE.8446E..0AB,2014SPIE.9147E..6TC}.
The HRS is a thermostabilised, fibre-fed, dual-beam \'echelle spectrograph covering the wavelengths 3700--5500~\AA\ and 5500--8900~\AA~ in the blue and red arms respectively. The HRS is equipped with four pairs of large-diameter optical fibres (object and sky fibres) and can be used in low- (LR), medium- (MR), high-resolution (HR), and high-stability (HS) modes. All spectral observations of the target objects were made in MR mode (R=$\lambda / \delta\lambda$=36 500--39 000), using object and sky fibres of 2.23 arcsec diameter. Both the blue and red arm CCDs were used with 1$\times$1 binning.
As the HRS is an \'echelle spectrograph housed in a vacuum tank in a temperature controlled enclosure, all standard calibrations are performed once a week, which is sufficient to achieve an accuracy of 300~m/s in MR mode. The HRS primary data reduction was performed automatically using the general purpose SALT pipeline  \citep{2010SPIE.7737E..25C}. Further \'echelle data reduction was performed using the standard HRS data pipeline \citep{2016MNRAS.459.3068K,2019AstBu..74..208K}. Each observed HRS spectrum was additionally corrected for bad columns and pixels, and for the spectral sensitivity curve obtained on the date closest to the observation. Spectrophotometric standard stars for the HRS are observed once a month as part of the HRS calibration plan. The rotational velocity projection \vsini~ was measured by fitting the observed spectrum with models from \cite{coelho:14}. The target stars, Julian dates, exposure times of the observations performed, signal-to-noise ratios S/N, \vsini~ and radial velocities (RV)  obtained from the SALT spectra are given in Table~\ref{target}.

\begin{figure}
\begin{tabular}{c}
\includegraphics[width=8cm]{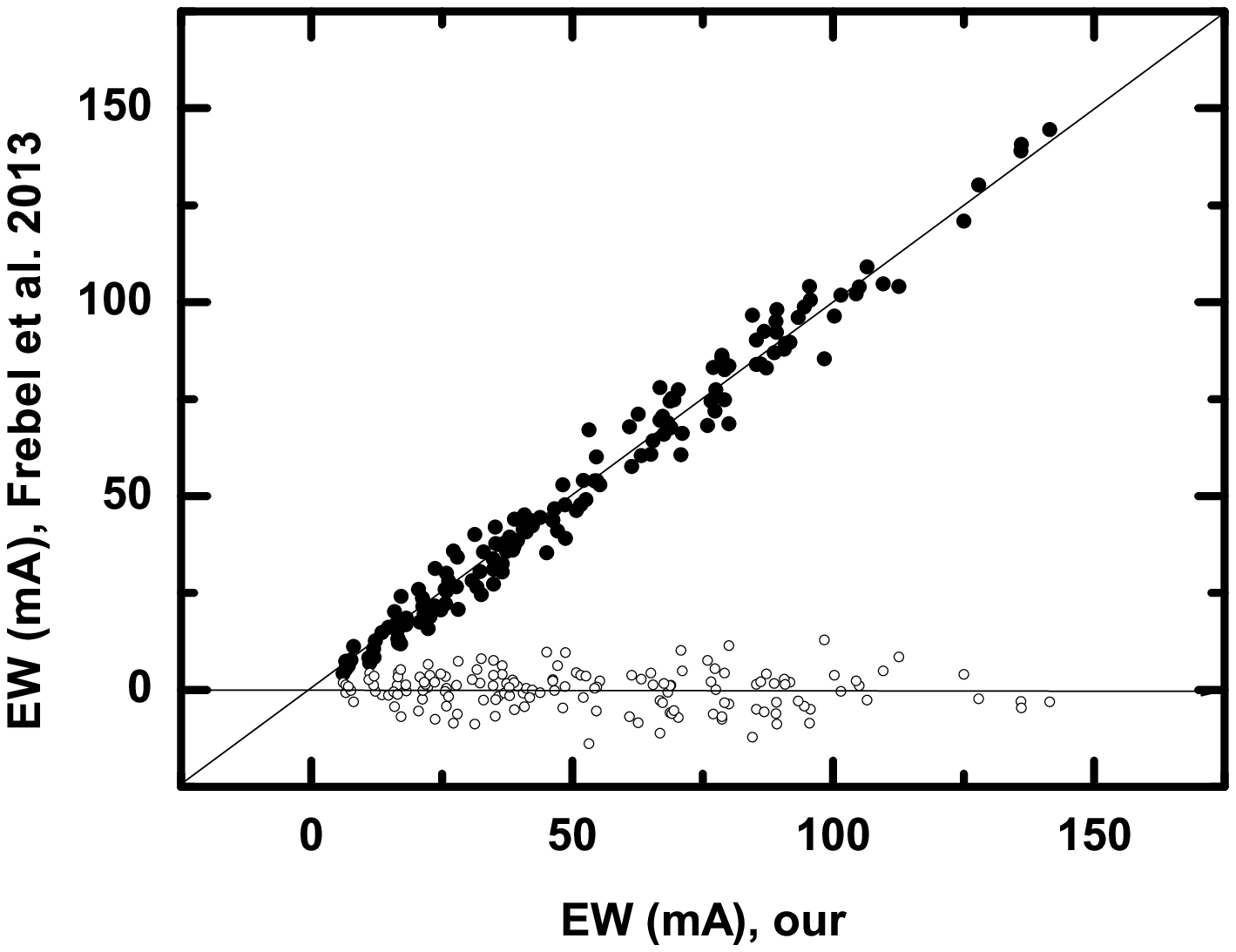}\\
\end{tabular}
\caption{Comparison between EWs measured in the present study and those reported by \protect\cite{frebel:13}. The EW differences between them are also reported at the bottom along the horizontal line.
}
\label{ew_comp1}
\end{figure}

Apart from spectra obtained using SALT, we used spectra from the UVES/VLT archive, based on the data obtained from the ESO Science Archive Facility\footnote{DOI(s): https://doi.org/10.18727/archive/50.} to perform control analysis in the blue region for HE 1523-0901, HD 6268, and HD 121135 on the basis of the spectrum with highest signal-to-noise ratio (S/N). The target stars, the observation dates, resolution, R, exposure times of the conducted observations, region of wavelengths $\lambda$ $\lambda$, and S/N values are presented in Table \ref{target2}.
Figure \ref{ew_comp1} illustrates the comparison between equivalent widths measured in the present study and those reported by \citet{frebel:13} for HE 1523 -- 0901. The mean difference and standard error of the mean are -0.015 $\pm4.8$ m\AA~ (for 150 iron lines).
\par
Spectral processing, which included the individual spectrum normalisation to the local continuum, identification of spectral lines of different chemical elements, measurements of the line depth and equivalent widths (EWs), was performed for each star using the DECH30 software package developed by G. A. Galazutdinov\footnote{\url{http://gazinur.com/DECH-software.html}}.

\begin{table*}
\caption{Coordinates and magnitudes of the studied stars, adopted from SIMBAD, and their parallaxes from Gaia DR3.}
\label{main}
\begin{tabular}{lrlcrcc}
\hline
Star         &  $\alpha$&   $\delta$ &       B   &    V&  K & Parallax         \\
             &  (2000)&   (2000)  &   (m)    &   (m) & (m) &  (mas)    \\
\hline                                          
HE 1523--0901 &  15 26 01& --09 11 39 & 12.37& 11.50 &8.352  &0.3278 [0.0197]\\ 
HD 6268      &  01 03 18& --27 52 49 &9.725 & 9.046 &5.714   &1.4720 [0.0250]] \\  
HD 121135    &  13 53 33& +02 41 41  &10.14 & 9.37  &7.182   &1.2564 [0.0171] \\  
HD 195636    &  20 32 49& --09 51 52 &10.13 & 9.57  &7.638   &1.6942 [0.0172] \\  
\hline
\end{tabular}
\end{table*}

\begin{table*}
\caption{Targets and their observational log (SALT). }
\label{target}
\begin{tabular}{lrcrccc}
\hline
Star           & $V$ &     HJD    & Exp  & S/N & \vsini  & RV\\
                 & (m) &    2450000+  & (s)  &   &  (\kms) &  (\kms) \\
\hline     
HE 1523--0901  & 11.500   & 9592.4103 & 2148 & 161 & 8.08 $\pm0.07$ &--163.455 $\pm0.052$     \\                  
HD 6268         &  9.046   & 8094.4032 &  111 & 107 & 6.45 $\pm0.03$  &  39.082 $\pm0.012$   \\
HD 121135       &  9.370   & 8540.6213 &  337 & 112 & 12.00 $\pm0.001$ & 124.002 $\pm0.024$  \\
HD 195636       &  9.570   & 9450.3022 & 1100 & 193 & 22.64 $\pm0.06$ & --256.797 $\pm0.025$ \\
\hline
\end{tabular}
\end{table*}

\begin{table*}
\caption{Targets and their observational log (ESO-VLT-U2).}
\label{target2}
\begin{tabular}{lrcrccc}
\hline
Star           & $V$ &   Observational  & R &$\lambda$ $\lambda$  & Exp  & S/N  \\
                 & (m) &   data           &   &  (\AA)        & (s)  &     \\
\hline     
 HE\ 1523--0901   & 11.500   & 2006-04-22T07:31:05.178 & 68 040  &3732 --  4999 &3900.0 & 132.9   \\                  
HD\,6268         &  9.046   & 2001-11-07T23:45:01.767 & 58 640  &3304 -- 4607 & 20.0   &  26.2  \\
HD\,6268         &  9.046   & 2016-10-19T04:15:45.547 &107 200  &4143  -- 5159 &1600.0 & 351.4   \\
HD\,121135       &  9.370   & 2021-02-09T06:56:35.946 &140 000  &3772 -- 7899 & 2939.0 & 345.2 \\
\hline
\end{tabular}

\end{table*}

\section{Atmospheric parameters}
\label{sec: stellar param}

\subsection{The effective temperature: \Teff}

\begin{figure*}
\begin{tabular}{cc}
\includegraphics[width=8.0cm]{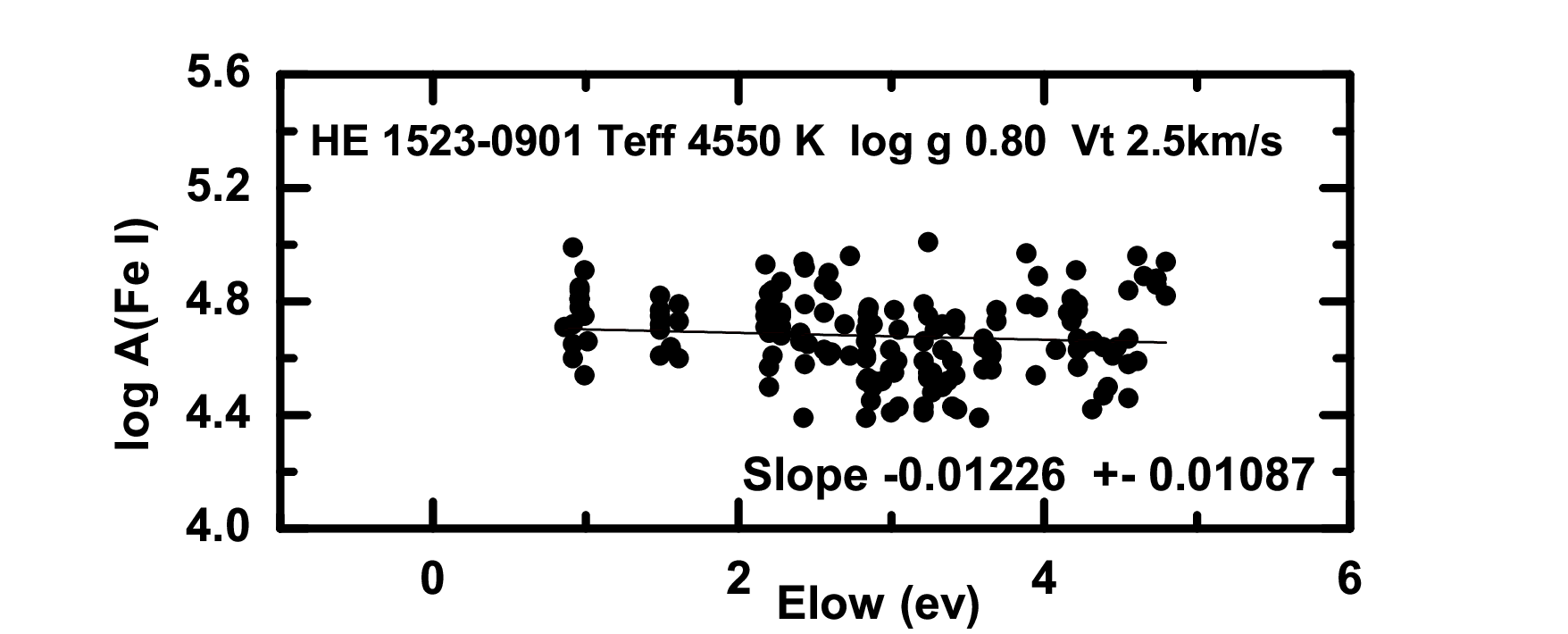} &\includegraphics[width=8.0cm]{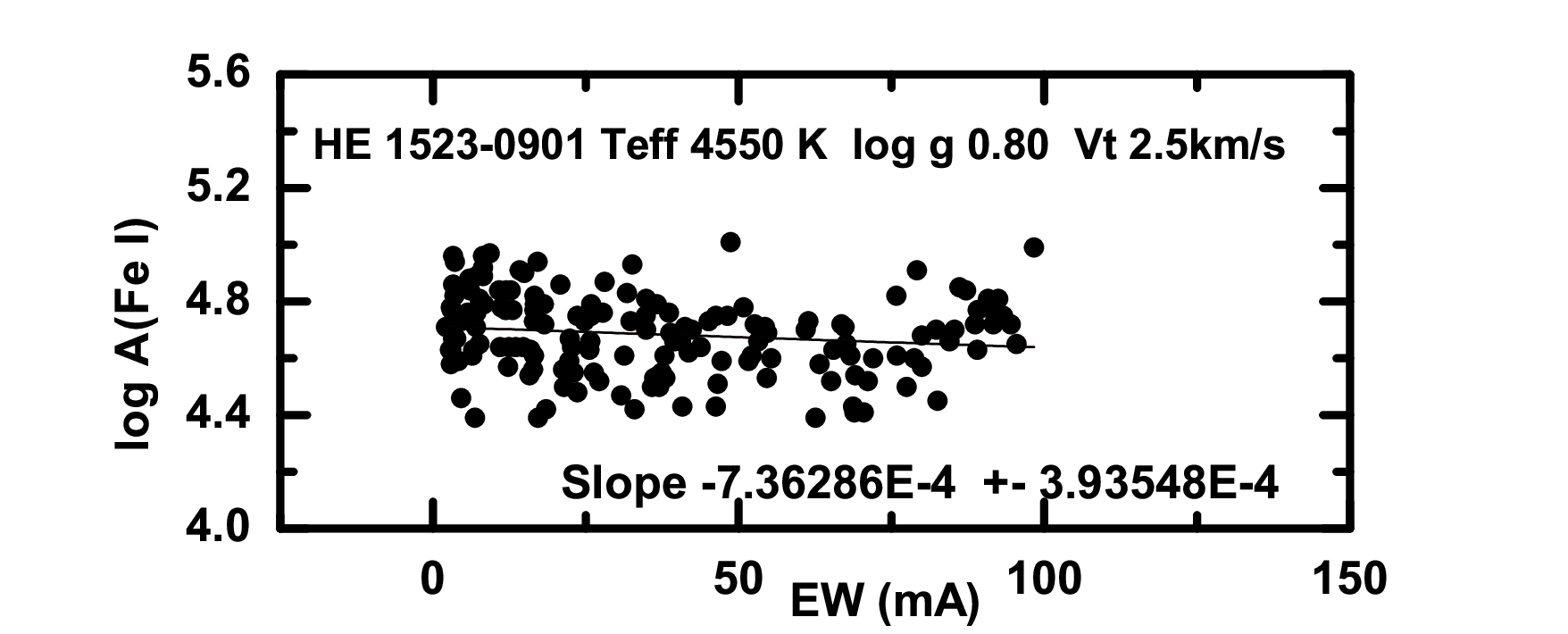}\\
\includegraphics[width=8.0cm]{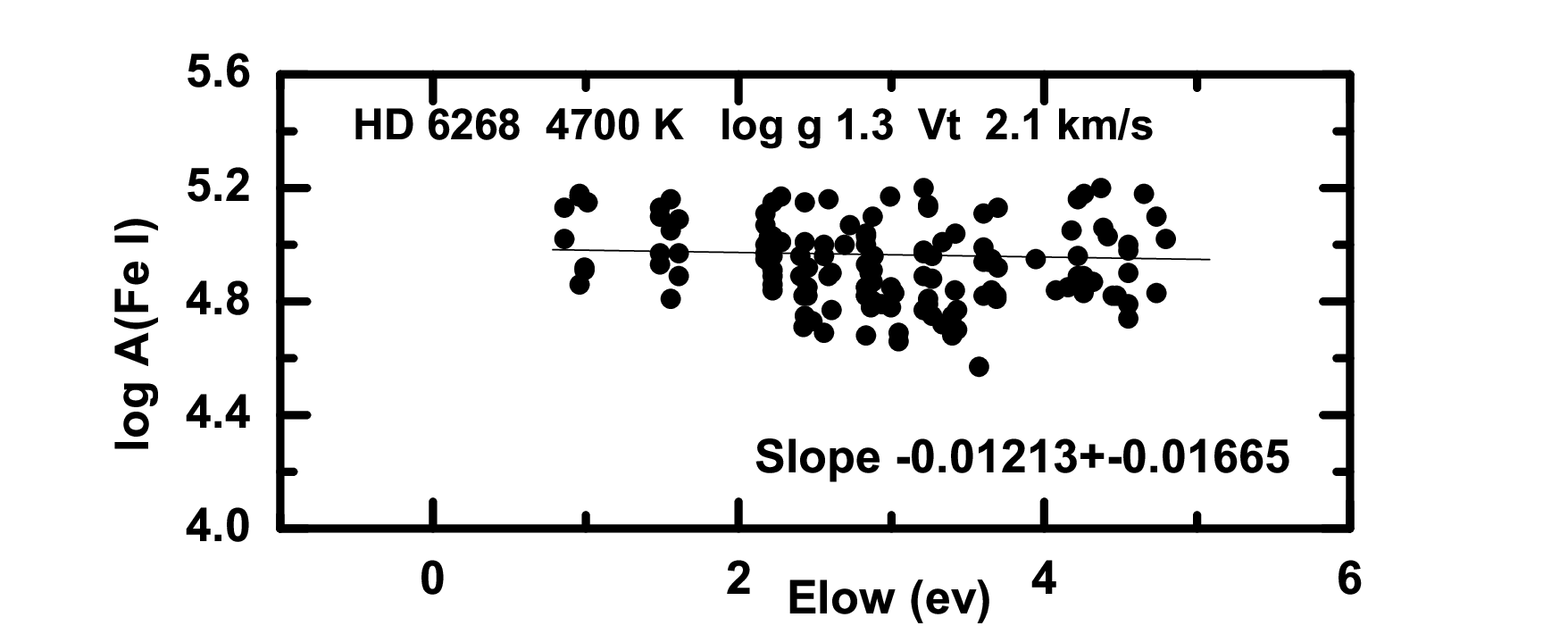}   &\includegraphics[width=8.0cm]{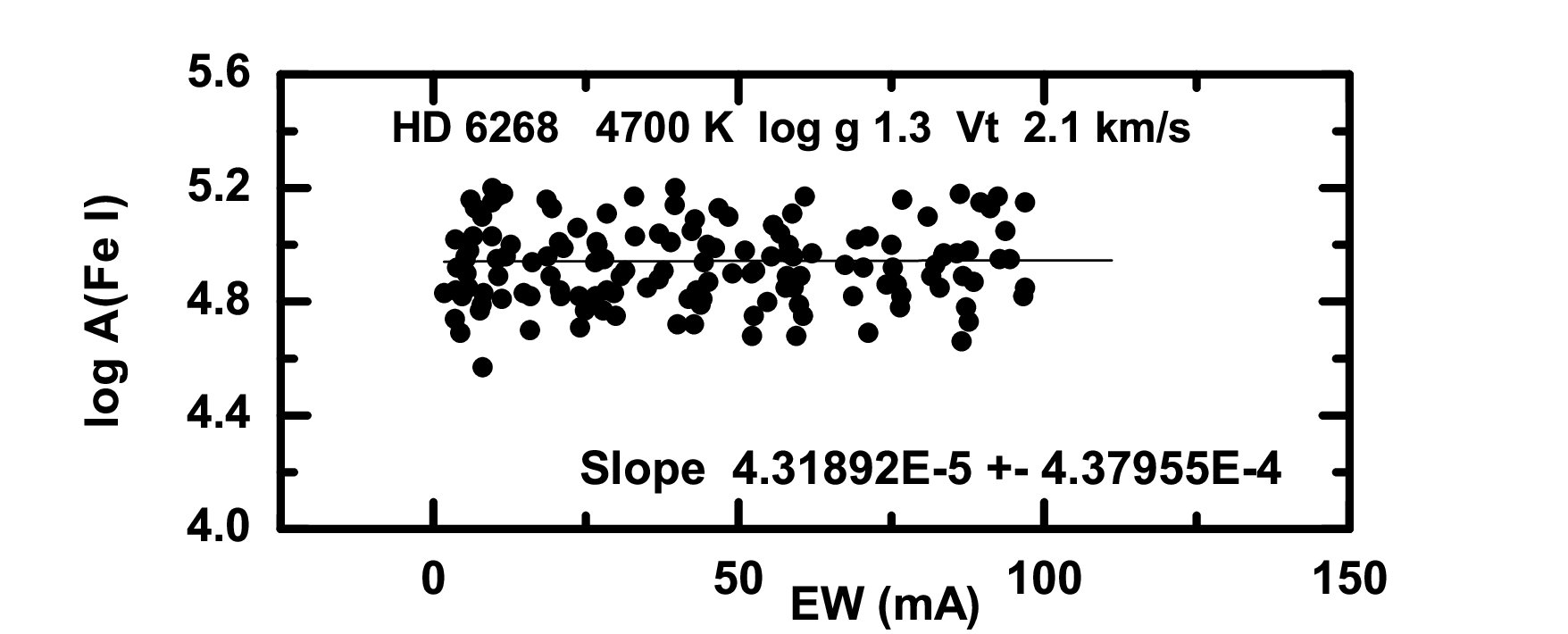}\\
\includegraphics[width=8.0cm]{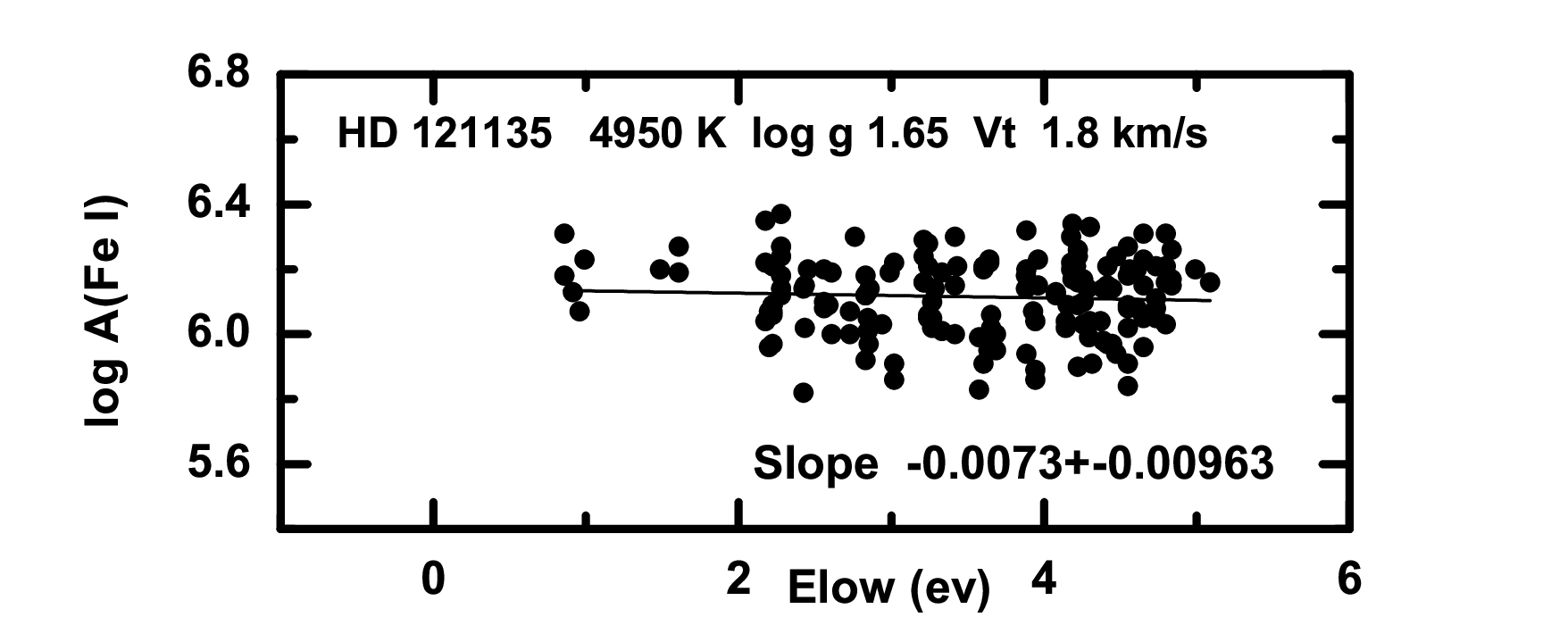} &\includegraphics[width=8.0cm]{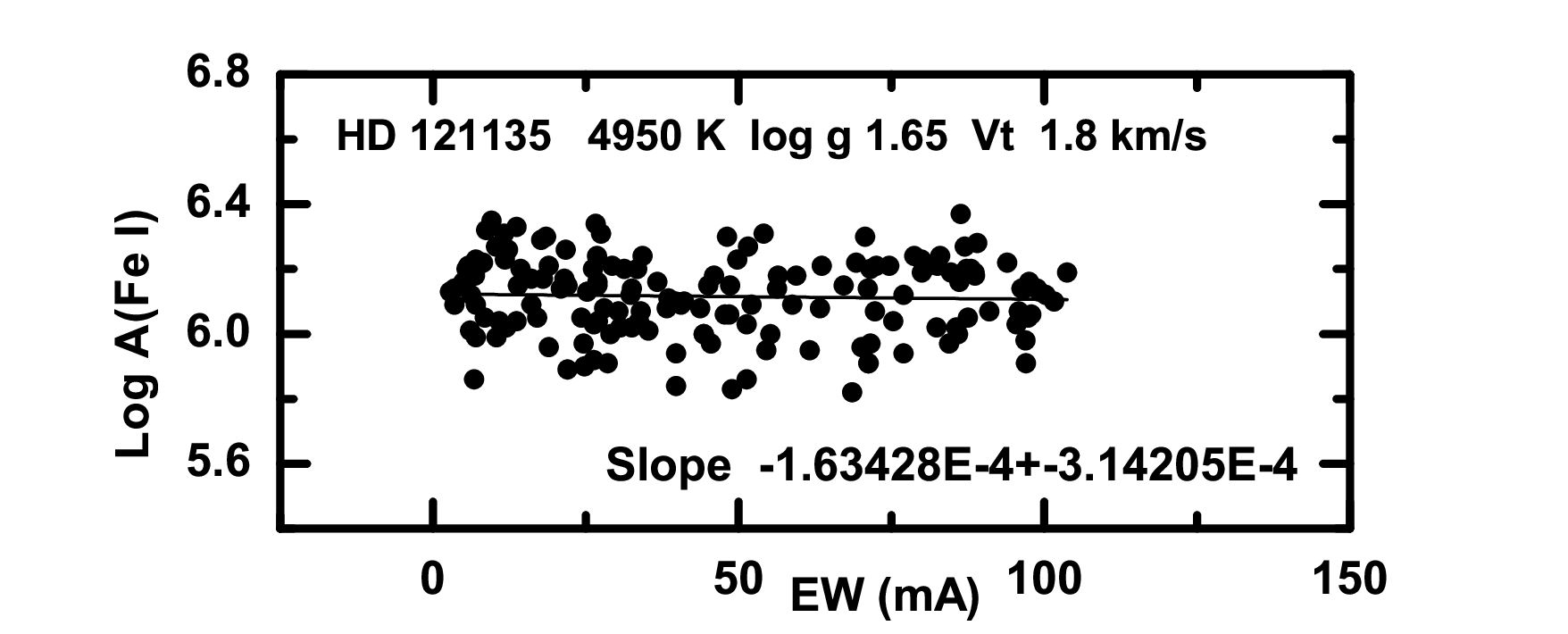}\\
\includegraphics[width=8.0cm]{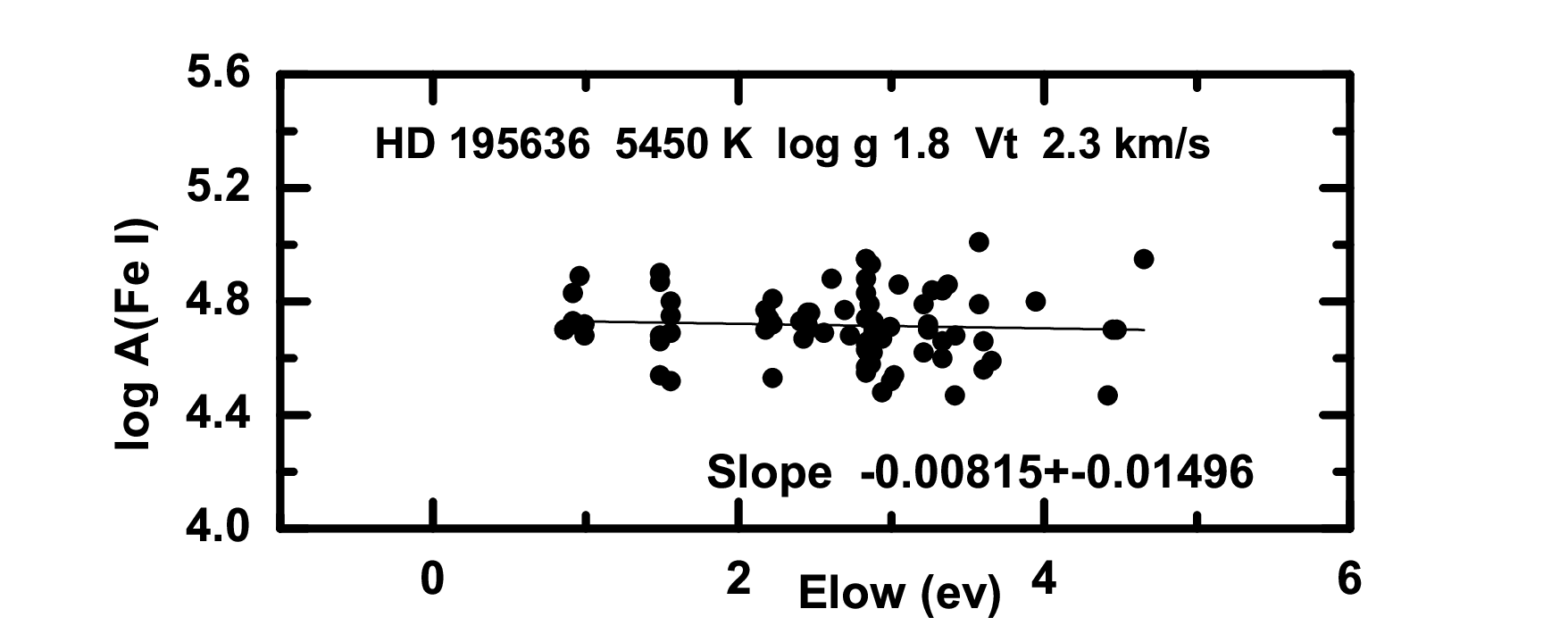} &\includegraphics[width=8.0cm]{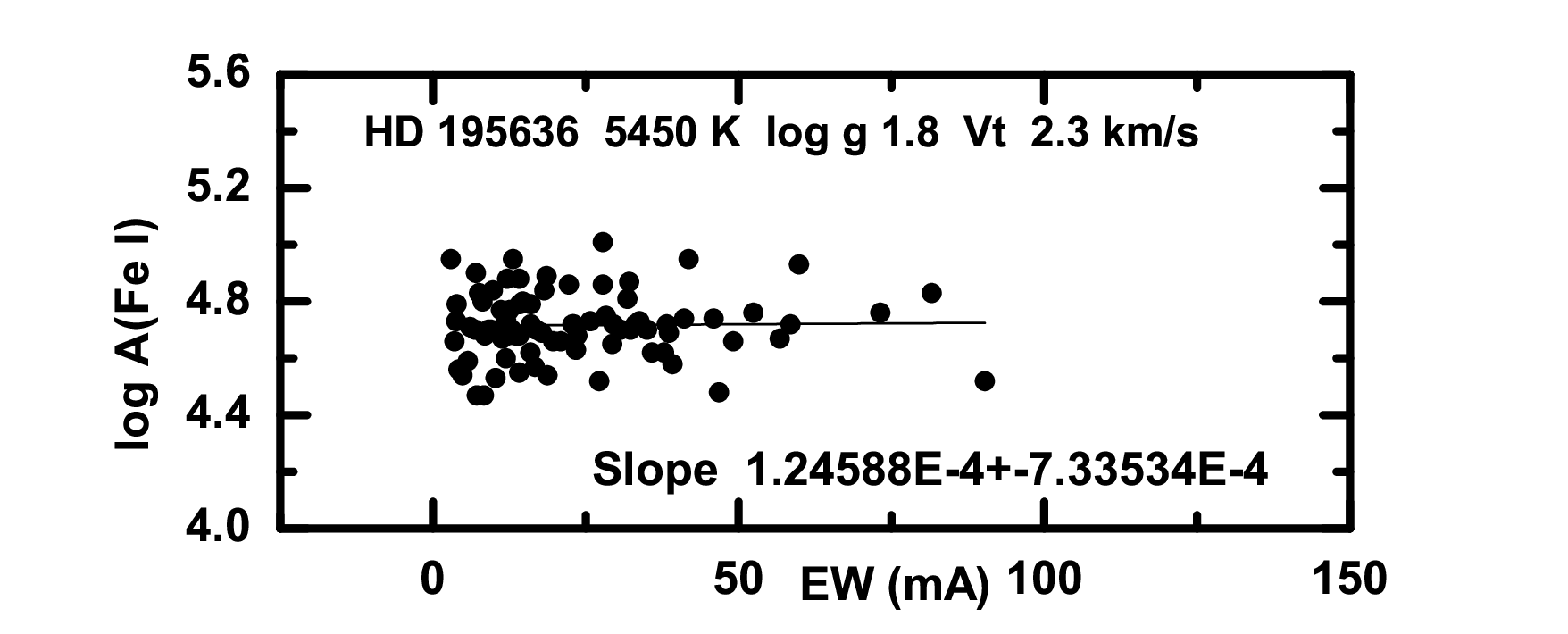}\\
\end{tabular}
\caption{Dependence of Fe~{\sc i} element abundance on $E_{low}$ and EW.}
\label{fe_elow1}
\end{figure*}

The most common methods for determining the effective temperature, \Teff,\ involve the stellar flux measurements, photometric calibrations, and spectroscopic techniques. 
In this work, we adopted photometric B-V and V-K colour indices (from the SIMBAD database) and calibrations for giants (dependent on [Fe/H]) from \citet[][]{alonso:99a}, given in Table \ref{param}.
The use of various sources of stellar magnitudes and respective color indices, as well as taking into account interstellar reddening E(B-V), affects the determination of the \Teff\ and other parameters based on photometric calibrations. We illustrate this for HE 1523-0901 and HD 6268 in Table \ref{param}, by displaying results from different sources of photometric data and using different reddening values from the literature (see details later in the text). 
The obtained \Teff\ values significantly differ from each other and also from the data reported by other authors (see Table \ref{compar}), with variations up to 300 K for HD 195636. As shown in the work of \cite{mucciarelli:20}, the use of photometric methods is more preferable compared to the spectral method for determining \Teff\ values for giant stars. For photometric temperature estimates, we used SIMBAD data,  and we obtained the differences in \Teff\ values. Therefore the next step was also using the spectroscopic method for determining the effective temperature. For its validation, we used a comparison of the obtained temperatures with those  determined both photometrically and in the literature. 
The accuracy of this method depends on the oscillator strengths used in calculations, and on the assumptions made with LTE and NLTE approximations.  
The \Teff\ were determined by requiring that the abundances derived from Fe~{\sc i} lines are showing no trend with the excitation potential of the lower level $E_{low}$ of the transition line for a given temperature (see, Fig. \ref{fe_elow1}, left panel). The iron abundance derived from Fe~{\sc i} lines in metal-poor stars tends to deviate from LTE, as reported, for instance, by \citet[][]{mash:11, sitn:15, berg:12}. Deviations from LTE decrease with increasing excitation potential $E_{low}$ \citep{berg:12}. Aiming to reduce such an effect, we examined the lines with the excitation potential ($E_{low}$) higher than 1 eV, most of the neutral iron lines used in our calculations have $E_{low}$ of more than 2 eV. 
In addition, the expected effect of deviations from LTE on the lines of Fe~{\sc i} differs significantly in different studies. This is due to the complexity of the multilevel model of the iron atom, which requires a large amount of atomic data that are known with high uncertainty. This is also one of the arguments in favor of the application of LTE analysis.  
Moreover, to eliminate uncertainty in the EW measurements (the Gaussian approximation was applied), we relied on the lines with EW $<$ 100 m\AA.
The achievable precision of determining \Teff\ by this method (the accuracy of establishing independence of Fe~{\sc i} lines on $E_{low}$ ) is about $\pm$50 K.

\begin{table*}
\caption{Stellar parameters of the stars of this study.
 }
\label{param}
\begin{tabular}{lcccccccccc}
\hline
 Star  &     V     &B-V    & V-K       &        [Fe/H]& \Teff\ spec &\Teff\ (B-V) &\Teff\ (V-K)  & log g$_{P}$ & log g$_{IE}$ & \Vt\\ 
 &      (m)     & (m)   &(m)    &       &(K)    &       (K)&     (K)&                   &        &       (\kms) \\ 
\hline 
HE 1523 --      0901&   11.50&  0.87&   3.15&           --2.82& 4550&   4695&   4621    &       1.36&   0.80 &2.5     \\ 
      (1)       &       11.13&1.07      &              &                --3.14 &4644   &       &               &1.21   &       &        \\ 
        HD 6268   & 9.05 &      0.68 &  3.33&           --2.54& 4700&   4947&   4412    &       1.82&   1.30 & 2.1    \\ 
                  (2)    & 8.08 &       0.84 &  --&     --2.69&         4728&   &       --      &       1.43&    &        \\ 
        HD 121135&      9.37 &  0.77 &  2.19&           --1.40&4950&    4909&   4971    &       1.90&   1.65&1.8         \\ 
        HD 195636&      9.57 &  0.56 &  1.93&           --2.78& 5450&   5206&   5392    &       2.54&   1.80&2.3         \\
\hline                                                                            
\end{tabular}  
\\
Notes. The photometric data and [Fe/H] marked by (1) and (2) were taken from \citet{placco:18}  and \citet{roederer:14}, respectively. 
\end{table*}

\begin{table*}                                                     
\caption{Comparison of the atmospheric parameters derived in this work with the literature.}
\label{compar}
\begin{tabular}{lccccl}
\hline                                                                  
Star&\Teff &\logg       & \Vt & [Fe/H]& sources  \\  
    & (K) &     &  (\kms)&  &   \\     
\hline                                                                  
\hline  
HE 1523-0901 &  4630  &    1.00 &  2.60 & -2.95 &        \cite{frebel:07}\\
             &  4612    &    0.80       &  2.70 & -2.85 &         \cite{frebel:13} \\
             &  4502    &    0.78 &     - & -3.05 &       \cite{beers:17} \\                                      
&4499   &       0.76 &       2.41       &     -2.83  &          \cite{hansen:18}  \\
&4642 &        0.64 &        -  &           -3.14 &         \cite{placco:18} \\
&4530  &       0.83&         2.70 &         -2.85 &           \cite{sakari:18} \\
&4742 &        1.29 &       1.83 &         -2.69 &          \cite{reggiani:22}  \\
mean   &        4594 $\pm89$&   0.87 $\pm0.21$&  &      --2.91 $\pm0.15$&           \\ 
\hline  
HD 6268 & 4818 &   0.84         &       1.50 &         -2.40    &      \cite{gratton:89}   \\
&4670  &        0.75 &          2.73 &        -2.58 &         \cite{mcwilliam:95}  \\
&4700 &     1.60 &              1.30  &        -2.36 &       \cite{pilachowski:96}  \\
&4800 &         0.84&   2.50  &        -2.56&        \cite{francois:96}     \\
&4700 &     1.60 &        1.60  &      -2.36 &       \cite{burris:00} \\
&4800 &      0.80 &   -   &         -2.5 &       \cite{medeiros:06}  \\ 
&4740 &     1.20  &  -   &         -2.32&            \cite{cenarro:07} \\
&4696 &         1.20&           -      &       -2.74    &     \cite{yong:13} \\
&4570 &     0.70 &      1.85 &        -2.89     &     \cite{roederer:14}  \\
&4696 &     1.20 &              2.75 &        -2.74 &     \cite{placco:14} \\
&4726 &     1.14 &              2.05 &       -2.63 &       \cite{wu:15} \\
mean   &        4720 $\pm70$&   1.08 $\pm0.32$&  &      --2.55 $\pm0.19$&           \\ 
\hline 
HD 121135 &4925          &    1.50      &     2.10 &    -1.57    &       \cite{pilachowski:96}  \\
&4925    &      1.50    &     2.10 &       -1.57        &         \cite{burris:00}   \\
&4910    &     1.90  &       - &          -1.83  &        \cite{carney:03}  \\
&4934     &     1.91    &  1.60  &      -1.54 -1.37     &        \cite{simmerer:04}    \\
&4934   &           &        &                  &    \cite{alonso:99a}\\
&4927   &          &        &                  & \cite{ramirez:05}\\
&5105   &    1.50      &         &   -1.57            & \cite{gonzalez:09}\\
mean   &        4960 $\pm81$&   1.70 $\pm0.23$&  &      --1.63 $\pm0.14$&           \\ 
\hline 
HD 195636 & 5500 &   3.40  &    2.00    &  -2.79    &  \cite{gratton:88}   \\
& 5370     &    2.40     &   -     &      -2.77 &    \cite{carney:03}  \\
& 5399      &   1.93   &    1.80   &    -2.74  &   \cite{behr:03}   \\
& 5242    &     3.40   &         1.50  &    -2.86   &  \cite{zhang:05}     \\
&5550   &   3.40       &         &   -2.79         & \cite{gonzalez:09}\\
mean   &        5412 $\pm120$&  2.91 $\pm0.7$&  &       --2.79 $\pm0.04$&           \\
\hline                                                                                    
\end{tabular}
\end{table*}

\subsection{Surface gravities: \logg}
\label{surf_grav}

The surface gravity, \logg 
, is determined from the ionisation equilibrium (IE), which implies similar elemental abundances derived from the two ionization stages, and using stellar parallaxes with the standard formulae:\\

\begin{equation}
 \begin{aligned}
log \left( \frac{g}{g_{\odot}} \right) = 
log \left( \frac{M}{M_{\odot}} \right) + 4 log \left( \frac{T_{eff }}{T_{eff \odot}} \right) + \\
+ 0.4(M_{bol} - M_{bol \odot}) %, 
\end{aligned}
\end{equation}

\begin{equation}
 \begin{aligned}
M_{bol} = V + BC + 5logP +5 %, 
\end{aligned}
\end{equation}

where M is the stellar mass, M$_{bol}$, is the absolute bolometric magnitude, V is the 'visual'  magnitude, 
BC is a bolometric correction \cite[][]{flower:96}, and P is the parallax from Gaia DR3. The values used as solar references are \logg$_{\odot}$\ = 4.44, \Teff$_{\odot}$\ = 5770 K and M$_{bol}$$_{\odot}$ = 4.75.
When determining \logg\ from the parallax (\logg$_{P}$), the primary error is introduced by the accuracy of the parallax itself and by factoring the reddening in. When using a spectroscopic method to determine \logg\ (\logg$_{IE}$), it is essential to use oscillator strengths and take deviations from LTE into account. As shown in \citet{berg:12}, a small NLTE correction is needed to fulfil the ionisation equilibrium at solar metallicities, while in very metal-poor stars these effects can reach up to +0.2 dex or more on Fe~{\sc i} lines. Fe~{\sc ii} lines are not sensitive to the departures from LTE.
The coincidence of the iron content in two stages of ionization up to 0.05 dex gives an accuracy of the \logg$_{IE}$ determination of the order of $\pm0.1$ dex or less.  

\subsection{Turbulent velocity \Vt\ and metallicity [Fe/H]}

The microturbulent velocity, \Vt, is given by imposing that the Fe abundance, log A(Fe), is independent of the EW of the respective neutral iron (Fe~{\sc i}) line (figure \ref{fe_elow1}, right panel). In general, parameters are derived using a method of successive approximations that enable to reach more accurate estimates of effective temperature (\Teff), surface gravity (\logg$_{IE}$) and microturbulent velocity (\Vt).
The obtained accuracy for the turbulent velocity \Vt\ is about $\pm0.1$ \kms.
The adopted [Fe/H] value is calculated using the Fe abundance derived from the Fe~{\sc i} lines.

\subsection{Description of the model selection and comparison of the results with the literature}

\textbf {HE 1523 - 0901}. This is one of the oldest stars in the Galaxy according to \citet{frebel:07}. Adopting the SIMBAD data (B, V, and K), as well as calibrations for B-V and V-K indices from \cite{alonso:99a}, we obtain \Teff\ = 4695 K and 4301 K, respectively. The correction for the reddening, color-excess E(B-V) = 0.138, is applied according to \citet{schlegel:98}. Based on the parallax of 0.3278 [0.0197] mas (Gaia DR3) and bolometric corrections (BC) \cite[][]{flower:96}, using the formula given above we obtain the surface gravity, \logg\ = 1.36 (V= 11.50, SIMBAD). The parallax value of 0.2772 [0.0434] mas (Gaia DR2) with the same data results in \logg\ = 1.22.
 To proceed with the parameter determination based on spectroscopic techniques, the initial calculation of the Fe abundance is done using the model parameters \Teff\ = 4500 K and \logg\ = 0.75. \Teff\ derived under the condition of the independence of the Fe abundance derived from neutral Fe (Fe~{\sc i}) lines on $E_{low}$ was equal to 4550 K. 
 Setting condition of the Ionisation Equilibrium (IE) for Fe resulted in the surface \logg\ =0.80. 
 The final parameter values adopted for further calculations of elemental abundances are given in Table \ref{param}.
 
Various methods and techniques were employed to determine atmospheric parameters in different studies. 
For instance, in \citet{frebel:07} the BVRI CCD photometry and JHK 2MASS data were used for \Teff, \Vt~ was derived from 
calculating the Fe abundance over several lines with different EWs, 
and \logg\ was derived from IE. \citet{frebel:13} introduced \Teff\ corrections, in order to bring spectroscopic temperature in agreement with the photometrically derived ones. 
\citet{beers:17} applied the SEGUE Stellar Parameter Pipeline (SSPP) to medium-resolution (R = 2000). 
\citet{hansen:18} adopted the \Teff\ corrections reported by \citet{frebel:13}. Surface gravities \logg\ were derived through IE by ensuring agreement between abundances derived from Fe~{\sc i} and Fe~{\sc ii} lines.
Microturbulent velocities \Vt\ were determined by removing any trend inline abundances with reduced EWs for both Fe~{\sc i} and Fe~{\sc ii} lines. 
In \cite{placco:18}, the result was obtained in a medium-resolution (R = 2000) follow-up spectroscopic campaign on low-metal targets selected based on stellar parameters published in Data Releases 4 and 5 of the RAdial Velocity Experiment (RAVE). 
 Using the magnitude V = 11.131, colour index B-V = 1.071 and the reddening correction E(B-V) = 0.138 adopted from \citet{schlegel:98},  together with the calibrations from \citet{alonso:99a}, the derived \Teff\ is 4644 K. The SSPP, as in \citet{beers:17}, was applied to estimate the parameters of the stellar atmosphere and carbon and $\alpha$-elements abundances. 

In \citet{sakari:18}, the atmospheric parameters were determined spectroscopically from the Fe~{\sc i} lines, taking into account (3D) NLTE corrections and using differential abundances with respect to a set of standards. The values reported in the above study are very close to those obtained in the present study (see table \ref{compar}). 
In \citet{reggiani:22}, stellar parameters were obtained using a hybrid isochrone/spectroscopy approach. High-quality multiwavelength photometry was employed to determine \Teff\, and Gaia EDR3 parallaxes were used to calculate \logg\ via isochrones. 
The \Vt\ was estimated through the empirical relation from \citet{kirby:09}. Such a method produced results different from the other authors. The mean values of the parameters from different studies are: \Teff\ = 4594 $\pm89$ K; \logg\ = 0.87$\pm21$; [Fe/H] = --2.91 $\pm0.15$ (see Table \ref{compar}), which is in very good agreement with our determinations. 

\textbf{HD 6268}. The star has been investigated in several studies (see Table \ref{compar}). From adopting the SIMBAD data B, V, and K, and using the B-V and V-K index calibrations from \cite{alonso:99a}, we obtain \Teff\ = 6153 K (from B-V = 0.679) and 4411 K (from V-K = 3.332), respectively. 
A correction for the reddening E(B-V) = 0.017 was taken from \citet{roederer:14}. The use of the SIMBAD data V and P =  0.3278 [0.0197] mas (Gaia DR3) yielded \logg\ =1.82. The SIMBAD photometric data for this star differ from those reported by \citet{mcwilliam:95} (V = 8.16, B-V = 0.813, d(B$-$V) = 0.22) and \citet{roederer:14} (V = 8.08, (B-V)$_{0}$ = 0.84, the reddening correction E(B-V) = 0.017, (V-K)$_{0}$ = 2.32). The use of parameters reported in \citet{roederer:14} and calibrations from \cite{alonso:99a} resulted in \Teff\ = 4728 K. Then, 
we employed a spectroscopic method of successive approximations using \Teff\ = 4709 K and \logg\ = 1.43 as initial parameters. 
The final values from our calculations are given in Table \ref{param}: \Teff\ = 4700 K, \logg\ = 1.30 and \Vt\ = 2.1 \kms. 
\par
For HD 6268, \citet{mcwilliam:95} used standard colour vs. \Teff\ relations for solar neighbourhood giants 
to calculate the temperatures. The \Vt\ and \logg\ were determined using abundances derived from multiple iron lines in spectra of individual stars.
To determine \Teff, \cite{pilachowski:96} used photometric index calibrations; \logg\ resulted from the average of three values obtained using three methods: 1) photometry; 2) the C-M (colour - magnitude) diagram of the globular cluster; 3) the empirical correlation between surface gravity and \Teff (using the reddening E(B-V) = 0.03). \cite{francois:96} took the average of parameter values from literature. \cite{burris:00} adopted stellar parameters from \cite{pilachowski:96}. In \citet{roederer:14}, the \Teff\ were derived by requiring that abundances derived from Fe~{\sc i} lines should show no trend with the excitation potential of the lower level of the transition. \Vt\ was determined under the condition that abundances derived from Fe~{\sc i} lines should show no trend with line strengths, and the \logg\ were obtained from theoretical isochrones in the YY grid \cite[][]{demarque:04}. 
A comparison with the mean parameter values obtained by different authors (Table \ref{compar}) with our data shows a good agreement  within determination errors: $<$\Teff$>$ = 4720 $\pm70$ K, $<$\logg$>$ = 1.08 $\pm0.32$, $<$[Fe/H]$>$ = --2.55 $\pm0.19$. 

\textbf{HD 121135}. 
Adopting the SIMBAD data (B, V, and K) and calibrations for B-V and V-K indices from \cite{alonso:99a}, we obtained \Teff\ = 4909 K and 4970 K, respectively. By using the parallax of 1.2564 [0.0171] mas (Gaia DR3) and the specified formula, we obtained \logg\ = 1.90. 
Using \Teff\ = 4900 K and \logg\ = 1.90 as initial parameters and applying the method of successive approximations, 
we  obtained the final parameters shown in Table \ref{param}: \Teff\ = 4950 K, \logg\ = 1.65 and \Vt\ = 1.8 \kms.  

\cite{carney:03}, adopting the colour index b-y = 0.530 and a reddening correction E(b-y) = 0.008 from \cite{anthony:98}, obtained \Teff\ = 4895 K.
\cite{burris:00} adopted stellar parameters from \cite{pilachowski:96}, \Teff\ = 4925 K and \logg\ = 1.50. \cite{simmerer:04} used the InfraRed Flux Method (IRFM) determination from \cite{alonso:99b} and the calibration by \cite{alonso:99a}, obtaining the final value \Teff\ = 4934 K.  In the case of \logg\, by using the formula based on luminosity (see Section \ref{surf_grav}) as well as the IE condition, we obtained the same value \logg\ = 1.91.
 
The IRFM  was employed to determine \Teff\ for the target star in a series of studies: \Teff\ = 4934 K \cite[][]{alonso:99b}; \Teff\ = 4927 K \cite[][]{ramirez:05}; \Teff\ = \textbf{5105} K and \logg\ = 1.50 \cite[][]{gonzalez:09}. The mean parameters for this star obtained in the literature (Table \ref{compar}) are \Teff\ = 4960 $\pm81$ K; \logg\ = 1.70 $\pm23$; [Fe/H] = --1.63 $\pm14$. They are in good agreement with our determinations. 
\par
\textbf{HD 195636}. The star is classified as High Proper Motion Star (SIMBAD) with a high rotation velocity, \vsini~ = 25 \kms \cite[][]{preston:97}, and as an HB star by  \cite{behr:03} with \vsini~ =  20 \kms. Adopting the SIMBAD data (B, V, and K) 
and E(B-V) 0.06 \cite[][]{bond:80}, as well as calibration for giants from \cite{alonso:99a}, we found  \Teff\ = 5206 and 5392 K, respectively. By employing calibrations for dwarf stars from \cite{alonso:96}, we obtained \Teff\ = 5575 K. The use of the parallax 1.6942 [0.0172] mas (Gaia DR3) and the formula we obtain \logg\ = 2.54. Then, by applying a spectroscopic approach based on successive approximations and using \Teff\ = 5500 K and \logg\ = 3.0 as initial parameters, 
we obtained the results given in Table \ref{param}: \Teff\ = 5450 K, \logg\ = 2.50 and \Vt\ = 2.3 \kms.    

In the work by \citet{carney:03}, adopting b-y = 0.467 an E(b-y) = 0.044 \cite[][]{anthony:98} resulted in \Teff\ = 5370 K and \logg\ = 2.4; the authors also determined the rotational velocity \vsini~ = 20.6 \kms, in agreement with the results of \cite{behr:03} for the same star.
Determinations through the IRFM method yielded \Teff\ = 5550 K and \logg\ = 3.40 \cite[][]{gonzalez:09}. In \citet{gratton:88}, the adopted \Teff\ were the mean values of temperatures deduced from V-R and V-K colours, while the adopted gravities were the mean values of \logg\ derived from the IE for Fe, Ti, Cr lines and from the wings of strong Fe~{\sc i} lines. In \citet{zhang:05}, \Teff\ was determined from colour indices and [Fe/H] using the calibration from \citet{alonso:96}; the surface gravity was determined via Hipparcos parallaxes. 

As can be seen from Table \ref{compar}, the scatter in the definitions of temperature (up to $\pm120$ K) and gravity (up to $\pm0.7$ dex) in the literature for HD 195636 is greater than for the other stars analyzed in this work. 
In particular, the large \logg\ spread is due to different derivation methods implemented. Notice that while our definitions of \Teff\ (5450 K) and [Fe/H] ([Fe/H] = --2.78) are in agreement with the mean values, 
(\Teff\ = 5412 $\pm120$ K and [Fe/H] = --2.79 $\pm0.04$), 
our \logg\ (\logg\ = 1.8 dex) is different from mean value ($<$ \logg\ $>$  = 2.91 $\pm0.70$). 

\par 
Summing up, it can be concluded that our parameter determinations are 
overall in good agreement with the results obtained earlier. 
The respective errors are $\delta$\Teff = $\pm100~K$, 
$\delta$\logg = $\pm0.2$, and $\delta$\Vt = $\pm0.1$ \kms.
Note that, however, when we calculate the elemental abundances we considered 2$\sigma$ error for \logg\, 
taking into account both the greater scatter of literature data, and the fact that we did not take into account possible influence of deviations from LTE on the determination of this parameter.

\section{Determination of the elemental abundances}
\label{sec: abundance determination}

Elemental abundances are derived with the LTE and NLTE approximations using the atmosphere models by \cite{castelli:04}. Each stellar model was obtained by interpolating the grid in accordance with the required combination of stellar parameters \Teff\ and \logg. To create the list of chemical element lines, we adopted data from several studies \cite[][etc]{frebel:13, roederer:14}, and also used the synthetic spectrum computations based on the VALD3 database \citep{kupka:99, pakhomov:19}. 

The Kurucz WIDTH9 code was used to determine LTE abundances based on the equivalent widths (EWs) of Ca, Ti, Cr, Fe, and Ni lines. We did not use strong lines with EWs $>$ 100 m\AA, due to noticeable effects of damping. The line list, their EW and the obtained abundances are
provided in machine readable format at the Centre de Donn\'ees astronomiques de Strasbourg (CDS) in Table A1.

The spectral synthesis code STARSP by \cite{tsymbal:96} was employed to calculate the line profiles for other investigated elements (C, N, O,  Na, Mg, Al, Si, K, Sc, Mn, Co, Cu, Zn, Sr, Y, Zr, Mo, Ru, Ba, La, Ce, Pr, Nd, Sm, Eu, Gd, Tb, Dy, Ho, Er, Tm, Hf, Os, Ir, and Th). 
The C and N abundances were determined by molecular lines at 4300 \AA\ (CH) and at 3883 \AA\ (CN). 
To measure the O abundance, both the 6300 \AA\ line and the IR triplet at 7770 \AA\ were used. The Na lines at 5889.951,  5895.924 and also 5682.633, 5688.205 \AA\ (for HD121135), the Mg lines at 4571.096, 5172.684, 5183.604, 5528.405, 5711.088 \AA, the Al line at 3961.5 \AA, the Si line at 4102.93 \AA, the K line at 7698.974 \AA\ were applied to obtain the abundances.
The determination of the Sc, Co, Mn, Ba, Eu and Pr abundances took into account the line hyperfine structure (HFS). 

We used lines in the region of 4030 \AA\ for Mn and those at 4129.7 and 6645.1 \AA\ for Eu, along with the HFS data from \citet{prochaska:00, ivans:06}, respectively. The HFS data for Pr II lines and log gf were taken from the study by \citet{sneden:09}. For Sc, Co and Ba we adopted the HFS parameters from the latest version of the VALD3 database, 
six Sc lines at 4246.81, 4314.08, 4320.73, 4324.99, 4400.38 and 5526.81 \AA\ and three Co lines at 4092.38,
4118.7, 4121.3 \AA\ were used. 

For Ba, we explored five lines at 4554.029, 4934, 5853.7, 6141.7, and 6496.9 \AA. The lines at 5853.7 and 6141.7 \AA\ had a practically insignificant effect on the HFS, while the lines at 4554.029 and 6496.9 \AA\ required factoring the HFS in \cite{kor:15}. The damping constants for Ba lines we adopted  from \cite{kor:15}. All oscillator strengths were scaled by the solar isotopic ratios. 
The Cu abundance was only determined for the star HD 121135 using the line 5105.53 \AA. The Zn abundances are based on three lines, 4680.134, 4722.153, 4810.528 \AA, for all stars.
The abundance of Ho and Tb were determined by taking into account the HFS, using the Ho lines at 3796.75, 3810.7 \AA,\ and the
Tb lines at 3874.17, 4002.57, and 4005.47 \AA\ . To compute the Th abundance, the line list in the region of 4019 \AA\ 
from \cite{mishenina:22} were used. The list of lines of Y, Zr, Mo, Ru, La, Ce, Pr, Nd, Sm, Gd, Tb, Dy, Ho, Er, Tm, Hf, Os, and Ir (adopted from the latest version VALD3) and the derived abundance by applying the synthetic spectrum method are provided in machine-readable format at the Centre de Donn\'ees astronomiques de Strasbourg (CDS) in Table B1.
Within the range of metallicities of the target stars, spectral lines of a number of chemical elements may be subject to deviations from LTE. The abundances of O, Na, Mg, Al, K, Cu, Sr, and Ba were also computed under the NLTE approximation. 
?

\begin{figure}
\begin{tabular}{c}
\includegraphics[width=8cm]{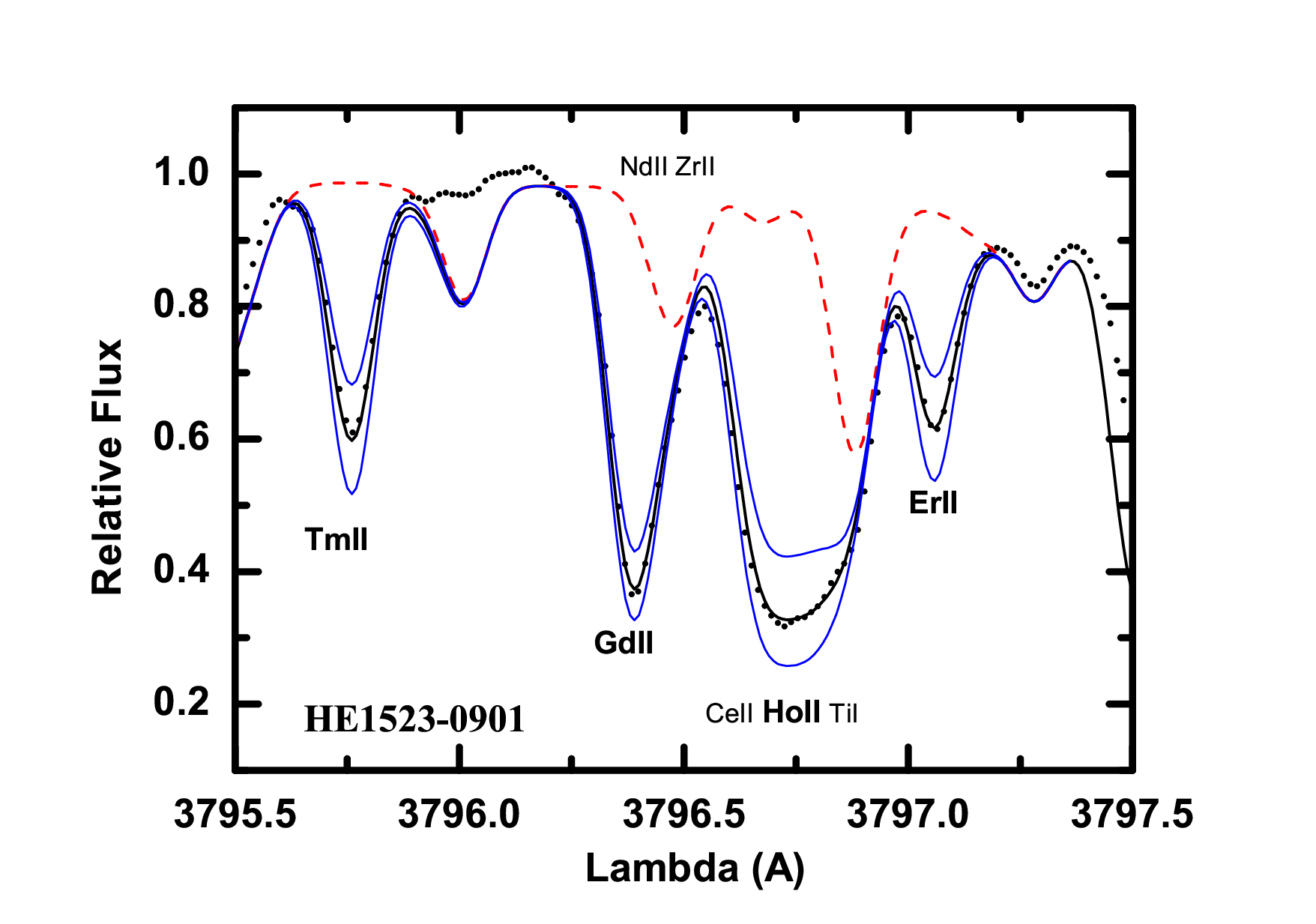}\\
\end{tabular}
\caption{Observed (dots) and calculated (solid and dashed lines) spectra
in the region 
of Ho II line for HE 1523-0901. The impact of the Tm, Gd, Er and Ho abundance
variation of 0.20 dex and without any Tm, Gd, Er and Ho contributions (dashed
line) is shown.}
\label{ho}
\end{figure}

\begin{figure}
\begin{tabular}{c}
\includegraphics[width=8cm]{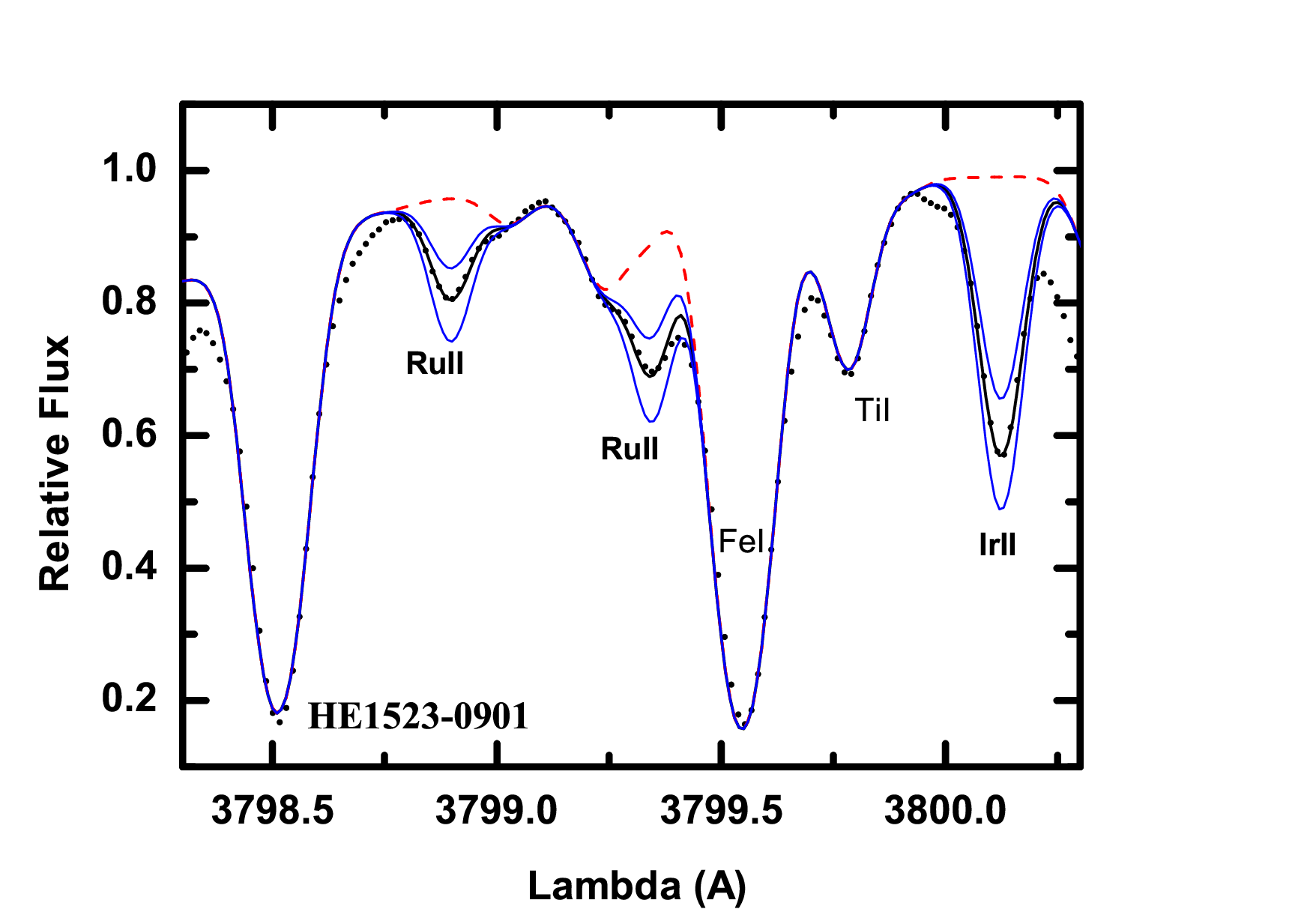}\\
\end{tabular}
\caption{Observed (dots) and calculated (solid and dashed lines) spectra
in the region 
of Ru I and Ir I lines for HE 1523-0901. The impact of the Ru and Ir abundance
variation of 0.20 dex and without any Ru and Ir contributions (dashed line)
is shown.}
\label{ru}
\end{figure}

\begin{figure}
\begin{tabular}{c}
\includegraphics[width=8cm]{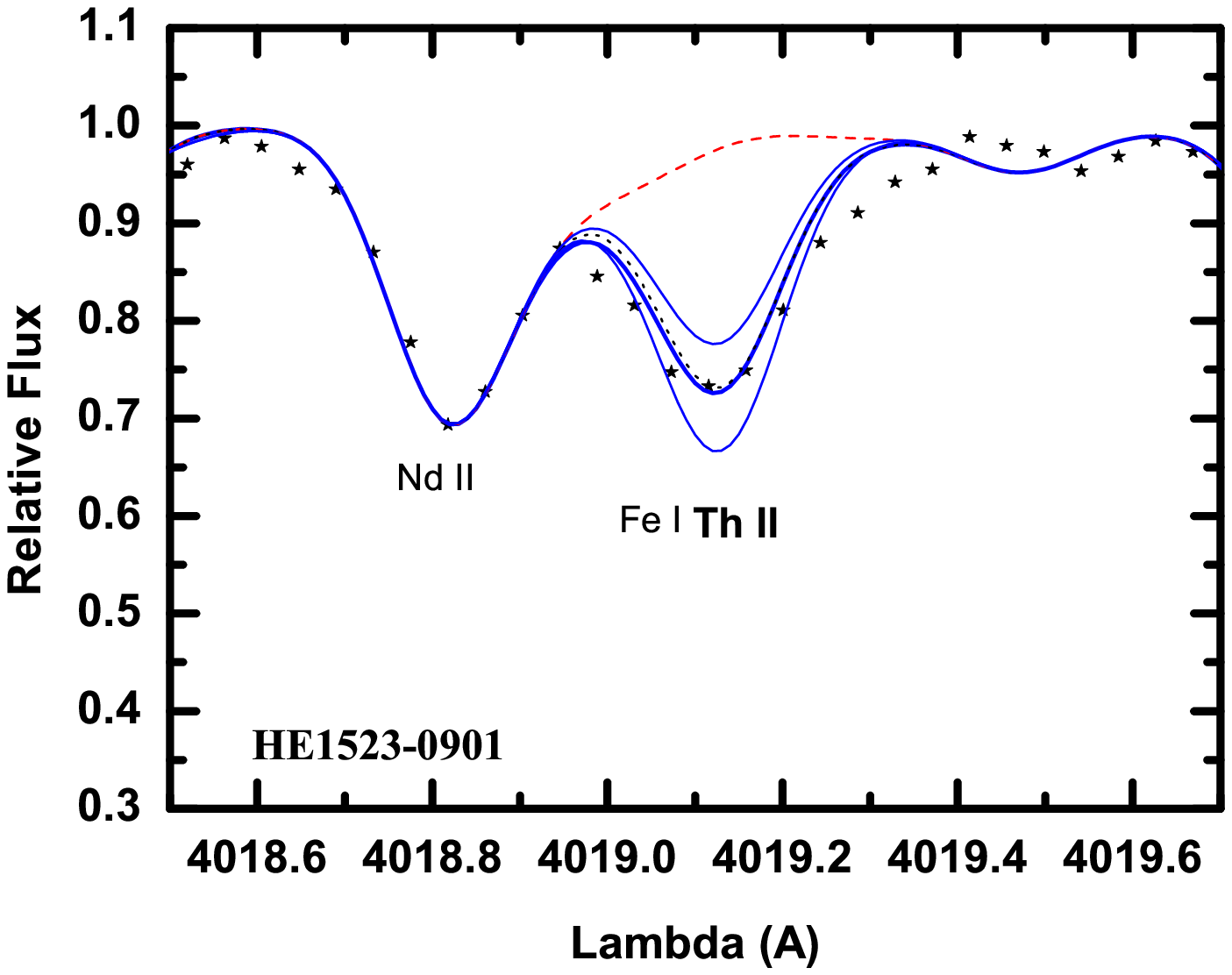}\\
\end{tabular}
\caption{Observed (asterisks) and calculated (solid and dashed lines) spectra
in the region 
of Th II line for HE 1523-0901. The impact of the Th abundance variation of 0.20
dex and without any Th contribution (dashed line) is shown.}
\label{th}
\end{figure}

To perform the NLTE calculations, we employed a version of the NLTE code MULTI \cite[][]{carl:86} modified by S. Korotin \cite[][]{kor:99}, based on the data and atomic models for O \cite[][]{mish:00, kor:14}, Na \cite[][]{kor:99,  dobr:14}, Mg \cite[][]{mish:04, cern:17}, Al \cite[][]{andr:08, caffau:19}, K \cite{andr:10, kor:20}, Cu \cite[][]{andr:18}, Sr \cite[][]{andr:11}, and Ba \cite[][]{andr:09}.
As reported in \cite{berg:08}, the NLTE effect is more pronounced for the Mn abundance derived from the resonance lines at 4030 \AA. We performed an approximate estimation of the NLTE abundance effects using the data from \cite{berg:08}. The mean value of the abundance deviation (correction) for the stars with [Fe/H] of about -2.5 (and giants as well) reaches 0.4-0.5 dex while for those with [Fe/H] of about -1.5 it is $\lesssim$ 0.2-0.3 dex. We factored these corrections, extracted from \cite{berg:08}, in the resulting Mn abundance. The NLTE Cu abundance was determined using the lines 5105.537 and 5782.127 \AA\ only for star HD 121135.
The NLTE correction for Eu lines were considered \cite[e.g.][]{mash:00}. Those corrections did not exceed 0.1 dex for the range of the target star parameters, well within the current observational errors. The spectrum synthesis fitting of some lines to the observed profiles is shown in Figs. \ref{ho}, \ref{ru}, and \ref{th}.

The element abundances log A(El), where log A(H) = 12) and those relative
to iron [El/Fe], have solar abundances taken from \cite{asplund:09}). These are presented in Table \ref{abund}.

\begin{table*}
\begin{center}
\caption{Elemental abundances of our target stars.}
\label{abund}
{\fontsize{5}{8} \selectfont 
\begin{tabular}{lcccccccccccccccc}
\hline
 &      HE 1523-0901    & &       &             &       HD 6268 &                       &               &               & HD 121135&              &        & &            HD 195636&      &               &                       \\
\hline
 
  El&   log A   &       $\sigma$        &       n       &       [El/Fe] &       log A       &       $\sigma$        &       n       &       [El/Fe] &log A& $\sigma$        &       n & [El/Fe]       &       log A   &       $\sigma$        &       n       &       [El/Fe]                 \\
   &        &  $\pm$    &   &         &       &  $\pm$    &   &         &     &  $\pm$    &   &         &       &  $\pm$    &   &      \\
\hline
C       &       4.99    &       --      &       --      &       -0.62   &       5.05    &       --      &       --      &       -0.84   &       6.55    &       --      &       --      &       -0.48   &       4.89    &       --      &       --      &       -0.76   \\
N       &       5.79    &       --      &       --      &       0.78    &       6.20    &       --      &       --      &       0.91    &       --      &       --      &       --      &       --      &       --      &       --      &       --      &       --      \\
O       &       6.65    &       --      &       1       &       0.78    &       7.05    &       --      &       1       &       0.90    &       8.28    &       --      &       1       &       0.99    &       7.14    &       --      &       1       &       1.23    \\
NLTE    &       6.60    &       --      &       1       &       0.73    &       7.15    &       0.15    &       2       &       1.00    &       7.90    &       0.04    &       2       &       0.61    &       7.04    &       --      &       1       &       1.13    \\
Na      &       3.60    &       --      &       2       &       0.18    &       4.15    &       --      &       2       &       0.45    &       5.20    &       --      &       2       &       0.36    &       4.05    &       --      &       2       &       0.59    \\
NLTE    &       3.10    &       --      &       2       &       -0.32   &       3.43    &       --      &       2       &       -0.27   &       4.75    &       0.10    &       4       &       -0.09   &       3.50    &       --      &       2       &       0.04    \\
Mg      &       5.14    &       0.05    &       5       &       0.36    &       5.38    &       0.07    &       5       &       0.32    &       6.27    &       0.05    &       5       &       0.07    &       5.40    &       0.06    &       5       &       0.58    \\
NLTE    &       5.14    &       0.12    &       5       &       0.36    &       5.44    &       0.12    &       5       &       0.38    &       6.53    &       0.12    &       5       &       0.33    &       5.46    &       0.12    &       5       &       0.64    \\
Al      &       3.00    &       --      &       1       &       -0.63   &       3.05    &       --      &       1       &       -0.86   &       4.10    &       --      &       1       &       -0.95   &       3.01    &       --      &       1       &       -0.66   \\
NLTE    &       3.16    &       --      &       2       &       -0.47   &       3.27    &       --      &       1       &       -0.64   &       4.77    &       --      &       2       &       -0.28   &       3.69    &       --      &       1       &       0.02    \\
Si      &       5.18    &       --      &       --      &       0.49    &       5.18    &               &       --      &       0.21    &       6.00    &       --      &       --      &       0.44    &       5.20    &       --      &       1       &       0.47    \\
K       &       2.68    &       --      &       1       &       0.47    &       2.94    &       --      &       1       &       0.45    &       4.05    &       --      &       1       &       0.42    &       2.75    &       --      &       1       &       0.50    \\
NLTE    &       2.51    &       --      &       1       &       0.30    &       2.80    &       --      &       1       &       0.31    &       3.71    &       --      &       1       &       0.08    &       2.81    &       --      &       1       &       0.56    \\
Ca      &       3.78    &       0.12    &       20      &       0.26    &       4.10    &       0.11    &       20      &       0.30    &       5.24    &       0.12    &       20      &       0.30    &       4.10    &       0.13    &       12      &       0.54    \\
Sc      &       0.10    &       0.09    &       5       &       -0.23   &       0.35    &       --      &       3       &       -0.26   &       1.52    &       --      &       3       &       -0.23   &       0.37    &       0.11    &       4       &       0.00    \\
Ti I    &       2.35    &       0.12    &       25      &       0.22    &       2.70    &       0.12    &       29      &       0.29    &       3.78    &       0.15    &       36      &       0.23    &       2.68    &       0.13    &       5       &       0.49    \\
Ti II   &       2.34    &       0.11    &       24      &       0.21    &       2.74    &       0.10    &       23      &       0.33    &       3.82    &       0.16    &       13      &       0.25    &       2.56    &       0.12    &       23      &       0.40    \\
Cr      &       2.67    &       0.14    &       5       &       -0.15   &       2.96    &       0.08    &       5       &       -0.14   &       4.13    &       0.15    &       12      &       -0.11   &       2.71    &       0.06    &       2       &       -0.15   \\
Mn      &       1.60    &       --      &       4       &       -1.01   &       1.80    &       --      &       4       &       -1.09   &       3.20    &       --      &       4       &       -0.83   &       1.80    &       --      &       --      &       -0.85   \\
NLTE    &       2.10    &       --      &       --      &       -0.51   &       2.30    &       --      &       --      &       -0.59   &       3.70    &       --      &       --      &       -0.33   &       2.30    &       --      &       --      &       -0.35   \\
Fe I    &       4.68    &       0.15    &       174     &       --      &       4.96    &       0.15    &       158     &       --      &       6.10    &       0.12    &       151     &       --      &       4.72    &       0.12    &       91      &       --      \\
Fe II   &       4.67    &       0.12    &       31      &       --      &       4.96    &       0.10    &       26      &       --      &       6.11    &       0.15    &       33      &       --      &       4.73    &       0.10    &       21      &       --      \\
Co      &       2.00    &       0.03    &       3       &       -0.17   &       2.15    &       --      &       3       &       -0.30   &       3.30    &       0.08    &       3       &       -0.29   &       2.18    &       0.01    &       2       &       -0.03   \\
Ni      &       3.55    &       0.16    &       22      &       0.15    &       3.74    &       0.09    &       15      &       0.06    &       4.87    &       0.12    &       47      &       0.05    &       3.69    &       0.11    &       4       &       0.25    \\
Cu      &       --      &       --      &       --      &       --      &       --      &       --      &       --      &       --      &       2.11    &       --      &       1       &       -0.68   &       --      &       --      &       --      &       --      \\
NLTE    &       --      &       --      &       --      &       --      &       1.30    &       0.15    &       2       &       -0.35   &       2.60    &       0.10    &       2       &       -0.19   &       --      &       --      &       --      &       --      \\
Zn      &       1.88    &       0.04    &       2       &       0.14    &       2.15    &       --      &       3       &       0.13    &       3.17    &       0.05    &       3       &       0.01    &       2.10    &       --      &       2       &       0.32    \\
Sr      &       0.45    &       --      &       2       &       0.40    &       0.48    &       --      &       2       &       0.15    &       1.43    &       0.03    &       2       &       -0.04   &       -0.12   &       --      &       2       &       -0.21   \\
NLTE    &       0.39    &       0.20    &       2       &       0.34    &       0.44    &       0.12    &       3       &       0.11    &       1.46    &       --      &       1       &       -0.01   &       -0.10   &       0.12    &       2       &       -0.19   \\
Y       &       -0.28   &       0.08    &       15      &       0.34    &       -0.54   &       0.07    &       13      &       -0.21   &       0.47    &       0.07    &       18      &       -0.34   &       --      &       --      &       --      &       --      \\
Zr      &       0.34    &       0.13    &       12      &       0.58    &       0.29    &       0.07    &       9       &       0.25    &       1.35    &       0.05    &       8       &       0.17    &       --      &       --      &       --      &       --      \\
Mo      &       -0.22   &       --      &       1       &       0.72    &       --      &       --      &       --      &       --      &       0.75    &       --      &       1       &       0.27    &       --      &       --      &       --      &       --      \\
Ru      &       0.10    &       0.09    &       3       &       1.17    &       --      &       --      &       --      &       --      &       0.40    &       --      &       2       &       0.05    &       --      &       --      &       --      &       --      \\
Ba      &       0.19    &       0.05    &       4       &       0.83    &       0       &       0.16    &       4       &       0.36    &       0.84    &       0.22    &       4       &       0.06    &       -1.26   &       0.10    &       4       &       -0.66   \\
NLTE    &       -0.11   &       0.10    &       4       &       0.53    &       -0.27   &       0.10    &       4       &       0.09    &       0.62    &       0.10    &       4       &       -0.16   &       -1.03   &       0.10    &       4       &       -0.43   \\
La      &       -0.52   &       0.13    &       19      &       1.20    &       -1.22   &       0.08    &       10      &       0.22    &       -0.44   &       0.08    &       14      &       -0.14   &       --      &       --      &       --      &       --      \\
Ce      &       -0.35   &       0.07    &       35      &       0.89    &       -0.75   &       0.06    &       20      &       0.21    &       -0.12   &       0.08    &       17      &       -0.30   &       --      &       --      &       --      &       --      \\
Pr      &       -0.58   &       0.13    &       12      &       1.52    &       -1.36   &       0.04    &       4       &       0.46    &       -0.89   &       0.18    &       7       &       -0.21   &       --      &       --      &       --      &       --      \\
Nd      &       -0.11   &       0.08    &       53      &       1.29    &       -0.77   &       0.08    &       25      &       0.35    &       -0.13   &       0.09    &       24      &       -0.15   &       --      &       --      &       --      &       --      \\
Sm      &       -0.43   &       0.08    &       35      &       1.43    &       -1.08   &       0.13    &       21      &       0.50    &       -0.50   &       0.11    &       24      &       -0.06   &       --      &       --      &       --      &       --      \\
Eu      &       -0.63   &       0.09    &       6       &       1.67    &       -1.55   &       --      &       4       &       0.47    &       -0.85   &       0.12    &       6       &       0.03    &       -2.00   &       --      &       2       &       0.26    \\
Gd      &       -0.33   &       0.11    &       21      &       1.42    &       -0.95   &       0.05    &       6       &       0.52    &       -0.39   &       0.05    &       7       &       -0.06   &       --      &       --      &       --      &       --      \\
Tb      &       -1.05   &       0.08    &       3       &       1.47    &       --      &       --      &       --      &       --      &       --      &       --      &       --      &       --      &       --      &       --      &       --      &       --      \\
Dy      &       -0.16   &       0.10    &       12      &       1.56    &       -0.83   &       0.10    &       8       &       0.61    &       -0.26   &       0.10    &       8       &       0.04    &       --      &       --      &       --      &       --      \\
Ho      &       -0.75   &       --      &       2       &       1.59    &       -1.75   &       --      &       1       &       0.31    &       -1.08   &       0.11    &       2       &       -0.16   &       --      &       --      &       --      &       --      \\
Er      &       -0.46   &       0.09    &       4       &       1.44    &       -1.10   &       0.07    &       4       &       0.52    &       -0.49   &       0.03    &       4       &       -0.01   &       --      &       --      &       --      &       --      \\
Tm      &       -1.23   &       0.21    &       2       &       1.33    &       -2.00   &       --      &       1       &       0.44    &       -1.40   &       --      &       1       &       -0.10   &       --      &       --      &       --      &       --      \\
Hf      &       -0.80   &       0.07    &       2       &       1.17    &       --      &       --      &       --      &       --      &       -0.75   &       --      &       1       &       -0.20   &       --      &       --      &       --      &       --      \\
Os      &       0.40    &       0.13    &       3       &       1.82    &       -0.70   &       --      &       1       &       0.44    &       0.38    &       --      &       1       &       0.38    &       --      &       --      &       --      &       --      \\
Ir      &       0.15    &       --      &       1       &       1.59    &       -0.65   &       --      &       1       &       0.51    &       0.05    &       --      &       1       &       0.07    &       --      &       --      &       --      &       --      \\
 Th     &       -1.30   &       --      &       1       &       1.50    &       --      &       --      &       --      &       --      &       -1.22   &       --      &       1       &       0.16    &       --      &       --      &       --      &       --      \\
\hline
\end{tabular} 
}
\end{center}       
\end{table*}

\textbf{Errors in the abundance determinations}.
As a representative example of 
errors in the elemental abundance due to the uncertainties in the atmospheric parameters, we derived the elemental abundances of the star HE 1523--0901 for different sets of stellar parameters (\Teff = 4550 K$\pm100$K, \logg = 0.80$\pm0.4$, \Vt = 2.5$\pm0.1$ \kms) and metallicity [Fe/H] = --2.82. 
Abundance variations associated with the modified parameters 
are given in Table \ref{errors}. We also took into account an 0.03 dex error from the fitting of the calculated or observed spectral lines.
In general, the errors are mostly given by uncertainties in \Teff\ and \logg, when neutral or ionised atomic lines, respectively, were used to determine abundances. The typical abundance error due to uncertainties in stellar parameters and spectral measurements varies within 0.05–0.18 dex. the Root Mean Square (rms) value of the iron determination from Fe~{\sc i} and  Fe~{\sc ii} lines ranges from 0.09 to 0.13 dex, respectively.
In Table \ref{comp_abund} we compare the stellar abundances from this work with other results from the literature for HE 1523-0901, HD 6268 and HD 195636. \begin{table}
\caption{Abundance errors due to atmospheric parameter uncertainties as examples for HE 1523-0901 
with the stellar parameters \Teff = 4550 K, \logg = 0.80, \Vt = 2.5 \kms and [Fe/H] = --2.82. }
\label{errors}
\begin{tabular}{llcccc}
\hline
 AN & El  & $\Delta$ \Teff+  & $\Delta$ \logg+ & $\Delta$ \Vt+ & tot+ \\
\hline
06  &C    &--0.1 &0.15  &--0.01 &0.18    \\
07  &N    & 0.19 & --0.10& --0.01& 0.21  \\
08&OI   & 0.07 &0.12 &--0.01   & 0.14    \\
11      &NaI    &0.05   &--0.03 &--0.01 &0.07                            \\ 
12      &MgI    &0.13   &--0.13 &--0.03 &0.19                            \\ 
13      &AlI    &0.13   &--0.04   & --0.01 &0.14                                  \\ 14  &SiI    &0.04   &--0.01 &--0.01 &0.05      \\
19      &KI     &0.05   &--0.03 &--0.01 &0.07                     \\
20      &CaI    &0.08   &--0.04 &--0.01 &0.09                     \\ 
23.01   &ScII   &0.04   &0.12  & --0.01 &0.13      \\
22      &TiI    &0.13   &--0.04 &--0.01 &0.14                     \\ 
22.01   &TiII   &0.02   &0.12   &--0.01 &0.13                     \\  
23      &VI     &0.13   &--0.05  &--0.01        &0.14     \\  
24      &CrI    &0.09   &--0.03 &--0.01 &0.10                              \\ 
25      &MnI    &0.08   &--0.03  &-0.01 &0.09             \\ 
26      &FeI    &0.09   &--0.03 &--0.01 &0.10                    \\ 
26.01   &FeII   &0.01   &0.13   &0.01   &0.13                     \\  
27      &CoI    &0.13   &--0.04 &--0.01 &0.14                            \\  28  &NiI    &0.09   &--0.02  &--0.01        &0.10                            \\ 
29      &CuI    &0.08   &--0.03         &0.0    &0.09   \\
30      &ZnI    &0.04   &0.06   &--0.01 &0.08                     \\ 
38      &SrII   &0.05   &0.12  &--0.02  &0.14                    \\ 
39      &YII    &0.05   &0.12  &--0.01  &0.13             \\  
40      &ZrII   &0.04   &0.12   &--0.01  &0.13                   \\ 
44      &RuI    &0.14   &--0.03 &--0.01  &0.15                   \\  
56      &BaII   &0.12   &0.07   &--0.05 &0.15                    \\ 
57      &LaII   &0.08   &0.12   &--0.01 &0.15                     \\ 
58  &CeII &0.07 &0.12   &--0.01 &0.14   \\
59      &PrII   &0.07   &0.12  &--0.01  &0.14             \\ 
60      &NdII   &0.07   &0.12  &--0.01  &0.14                   \\ 
62      &SmII   &0.084  &0.12  &--0.01  &0.15                     \\  
63      &EuII   &0.05   &0.12  &--0.03  &0.14            \\  
64  &GdII &0.05  & 0.12  &--0.01  & 0.13                        \\
65  &TbII &0.07  & 0.14  &--0.01  & 0.16                        \\
66  &DyII &0.07  & 0.13  &--0.01  & 0.15                        \\
67  &HoII &0.06  & 0.13  &--0.01  & 0.15                        \\
68  &ErI &0.15 & --0.03  &--0.01  & 0.16                        \\
69  &TmII &0.09  & 0.13  &--0.01  & 0.16                        \\
72  &HfI &0.17  &--0.03  &--0.01  & 0.18                        \\
76  &OsI &0.15  & 0.02  &--0.01  & 0.15                 \\
77  &IrI &0.14  & 0.08  &--0.01  & 0.16                 \\
90  &ThII &0.06  & 0.14  &--0.01  & 0.15                        \\
\hline                             
\end{tabular}
\end{table}

\begin{table*}
\begin{center}
%\begin{small} 
\caption{[El/Fe] ratios of our stars are compared with the literature for HE 1523-0901, HD 6268, and HD 195636. 
}
\label{comp_abund}

{\fontsize{5}{8} \selectfont  
\begin{tabular}{lcccccccccccccccc}
\hline
 &      HE 1523-0901 &          &       &               &               &                       &               &       HD 6268    &&              &        &HD 195636&    &               &               &                       \\
\hline
 El &   1       &       2       &3      &       4       &       5       &       6       &       7       &       8       &9&     10      &       11 & 12 &  13      &       14      &       15      &       16                      \\
\hline
 
C       &       --      &       -0.82   &       0.39[b] &       --      &       --      &       -0.62   &       --      &       -0.68   &       -0.91   &       -0.84   &       --      &       --      &       --      &       --      &       -0.76   &       --      \\
N       &       --      &       --      &       --      &       --      &       --      &       0.78    &       --      &       --      &       0.80    &       0.91    &       --      &       --      &       --      &       --      &       --      &       --      \\
O       &       --      &       --      &       --      &       --      &       --      &       0.78    &       0.73    &       --      &       0.66    &       0.90    &       1.00    &       --      &       --      &       --      &       1.23    &       1.13    \\
Na      &       0.135   &       --      &       --      &       -0.035  &       3.342   &       0.18    &       -0.32   &       0.22    &       --      &       0.45    &       -0.27   &       0.57    &       -0.14   &       -0.38   &       0.59    &       0.04    \\
Mg      &       0.33    &       --      &       0.47    &       0.156   &       --      &       0.36    &       0.36    &       0.44    &       0.56    &       0.32    &       0.38    &       0.28    &       0.82    &       --      &       0.58    &       0.64    \\
Al      &       --      &       --      &       --      &       -1.155  &       2.98    &       -0.63   &       -0.47   &       0.48    &       -0.36   &       -0.86   &       -0.64   &       -0.90   &       --      &       --      &       -0.66   &       0.02    \\
S       &       --      &       --      &       --      &       --      &       --      &       --      &       --      &       0.50    &       --      &       0.32    &       --      &       0.32    &       --      &       --      &       --      &       --      \\
Si      &       0.42    &       --      &       --      &       0.032   &       --      &       0.49    &       --      &       --      &       --      &       0.21    &       --      &       --      &       --      &       --      &       0.47    &       --      \\
K       &       --      &       --      &       0.39    &       --      &       --      &       0.47    &       0.30    &       --      &       --      &       0.45    &       0.31    &       --      &       1.35    &       1.04    &       0.50    &       0.56    \\
Ca      &       0.29    &       --      &       0.26    &       0.237   &       --      &       0.26    &       --      &       0.44    &       0.45    &       0.30    &       --      &       0.32    &       0.63    &       --      &       0.54    &       --      \\
Sc      &       -0.03   &       --      &       0.27    &       -0.243  &       --      &       -0.33   &       --      &       0.07    &       -0.07   &       -0.26   &       --      &       0.45    &       0.69    &       --      &       0.0     &       --      \\
Ti I    &       0.16    &       --      &       0.25    &       0.314   &       --      &       0.22    &       --      &       0.29    &       0.12    &       0.29    &       --      &       0.76    &       --      &       --      &       0.49    &       --      \\
Ti II   &       --      &       --      &       0.32    &       0.273   &       --      &       --      &       --      &       0.29    &       0.28    &       0.33    &       --      &       --      &       --      &       --      &       0.40    &       --      \\
V       &       --      &       --      &       -0.12   &       --      &       --      &       --      &       --      &       -0.09   &       --      &       --      &       --      &       --      &       --      &       --      &       --      &       --      \\
Cr      &       -0.14   &       --      &       0.01    &       -0.333  &       --      &       -0.15   &       --      &       -0.08   &       -0.25   &       -0.14   &       --      &       --      &       --      &       --      &       -0.15   &       --      \\
Mn      &       --      &       --      &       -0.32   &       --      &       --      &       -1.01   &       -0.51   &       -0.22   &       -0.49   &       -1.09   &       -0.59   &       0.10    &       --      &       --      &       -0.85   &       -0.35   \\
Co      &       0.34    &       --      &       0.41    &       --      &       --      &       -0.17   &       --      &       0.02    &       -0.26   &       -0.30   &       --      &       --      &       --      &       --      &       -0.03   &       --      \\
Ni      &       0.03    &       --      &       0.09    &       0.225   &       --      &       0.15    &       --      &       0.04    &       0.08    &       0.06    &       --      &       --      &       --      &       --      &       0.25    &       --      \\
Cu      &       --      &       --      &       --      &       --      &       --      &       --      &       --      &       --      &       -0.56   &       --      &       -0.35   &       --      &       --      &       --      &       --      &       --      \\
Zn      &       0.21    &       --      &       0.35    &       --      &       --      &       0.14    &       --      &       --      &       0.32    &       0.13    &       --      &       --      &       --      &       --      &       0.32    &       --      \\
Sr      &       --      &       0.90    &       0.57    &       0.395   &       --      &       0.40    &       0.34    &       -0.14   &       0.15    &       0.15    &       0.11    &       0.39    &       --      &       --      &       -0.21   &       -0.19   \\
Y       &       --      &       --      &       0.50    &       0.335   &       --      &       0.34    &       --      &       --      &       -0.41   &       -0.21   &       --      &       --      &       --      &       --      &       --      &       --      \\
Zr      &       --      &       --      &       0.95    &       --      &       --      &       0.58    &       --      &       0.17    &       0.07    &       0.25    &       --      &       --      &       --      &       --      &       --      &       --      \\
Mo      &       --      &       --      &       --      &       --      &       --      &       0.72    &       --      &       --      &       0.08    &       --      &       --      &       --      &       --      &       --      &       --      &       --      \\
Ru      &       --      &       --      &       --      &       --      &       --      &       1.02    &       --      &       --      &       --      &       --      &       --      &       --      &       --      &       --      &       --      &       --      \\
Cd      &       --      &       --      &       --      &       --      &       --      &       --      &       --      &       --      &       --      &       0.36    &       0.09    &       --      &       --      &       --      &       --      &       --      \\
Ba      &       --      &       0.69    &       1.27    &       0.969   &       --      &       0.83    &       0.53    &       0.13    &       0.05    &       --      &       --      &       --      &       -0.14   &       --      &       -0.66   &       -0.43   \\
La      &       --      &       --      &       1.44    &       1.20    &       --      &       1.20    &       --      &       -0.03   &       0.09    &       0.22    &       --      &       --      &       --      &       --      &       --      &       --      \\
Ce      &       --      &       --      &       1.26    &       --      &       --      &       0.89    &       --      &       0.34    &       -0.02   &       0.21    &       --      &       --      &       --      &       --      &       --      &       --      \\
Pr      &       --      &       --      &       1.64    &       --      &       --      &       1.52    &       --      &       --      &       0.30    &       0.46    &       --      &       --      &       --      &       --      &       --      &       --      \\
Nd      &       --      &       --      &       1.56    &       --      &       --      &       1.29    &       --      &       0.57    &       0.17    &       0.35    &       --      &       --      &       --      &       --      &       --      &       --      \\
Sm      &       --      &       --      &       1.74    &       --      &       --      &       1.43    &       --      &       --      &       0.25    &       0.50    &       --      &       --      &       --      &       --      &       --      &       --      \\
Eu      &       --      &       1.70    &       1.82    &       1.90    &       --      &       1.67    &       --      &       0.68    &       0.31    &       0.47    &       --      &       --      &       --      &       --      &       0.26    &       --      \\
Gd      &       --      &       --      &       --      &       --      &       --      &       1.42    &       --      &       0.85    &       0.40    &       0.52    &       --      &       --      &       --      &       --      &       --      &       --      \\
Tb      &       --      &       --      &       --      &       --      &       --      &       1.47    &       --      &       --      &       --      &       --      &       --      &       --      &       --      &       --      &       --      &       --      \\
Dy      &       --      &       --      &       2.47    &       --      &       --      &       1.56    &       --      &       1.12    &       0.49    &       0.61    &       --      &       --      &       --      &       --      &       --      &       --      \\
Ho      &       --      &       --      &       --      &       --      &       --      &       1.59    &       --      &       --      &       --      &       0.31    &       --      &       --      &       --      &       --      &       --      &       --      \\
Er      &       --      &       --      &       --      &       --      &       --      &       1.44    &       --      &       --      &       0.32    &       0.52    &       --      &       --      &       --      &       --      &       --      &       --      \\
Tm      &       --      &       --      &       --      &       --      &       --      &       1.33    &       --      &       --      &       0.28    &       0.44    &       --      &       --      &       --      &       --      &       --      &       --      \\
Yb      &       --      &       --      &       --      &       --      &       --      &       --      &       --      &       --      &       0.41    &       --      &       --      &       --      &       --      &       --      &       --      &       --      -- \\
Hf      &       --      &       --      &       --      &       --      &       --      &       1.17    &       --      &       --      &       0.41    &       --      &       --      &       --      &       --      &       --      &       --      &       --      \\
Os      &       1.73[a] &       --      &       1.97    &       --      &       --      &       1.82    &       --      &       --      &       --      &       0.44    &       --      &       --      &       --      &       --      &       --      &       --      \\
Ir      &       1.81[a] &       --      &       --      &       --      &       --      &       1.59    &       --      &       --      &       --      &       0.51    &       --      &       --      &       --      &       --      &       --      &       --      \\
Th      &       1.73[a] &       --      &       --      &       --      &       --      &       1.50    &       --      &       --      &       0.41    &       --      &       --      &       --      &       --      &       --      &       --      &       --      \\
U       &       1.43[a] &       --      &       --      &       --      &       --      &       1.41    &       --      &       --      &       --      &       --      &       --      &       --      &       --      &       --      &       --      &       --      \\
\hline
\end{tabular} 
}
\\
Notes.The first column (1) gives the values calculated by  \cite{frebel:07,  frebel:13},  (2)  -- \cite{hansen:18}, (3) -- \cite{sakari:18}, (4, 5)  -- \cite{reggiani:22} (LTE and NLTE, respectively), (8) -- \cite{mcwilliam:95},  (9)  -- \cite{roederer:14},  
(12)  --  \cite{gratton:88}, 
(13, 14)  --  \cite{zhang:05} (LTE and NLTE, respectively),  (6 and 7; 10 and 11; 15 and 16) --  our determinations (LTE and NLTE, respectively).
The values labeled with [a] are by \cite{frebel:07} (1)  
and with [b] we refer to \cite{sakari:18} (3) 
but recalculated for changes due to stellar evolution.
\end{center} 
\end{table*}

\textbf{HE 1523--0901}. Table \ref{comp_abund} shows the values of the abundances of elements determined in the following papers-- \cite{frebel:07,  frebel:13} (column 1), \cite{hansen:18} (column 2), \cite{sakari:18} (column 3), \cite{reggiani:22} (LTE and NLTE, respectively, columns 4 and 5), and our determinations (LTE and NLTE, respectively, columns 6 and 7). As can be seen from the table, for all the stars Co abundance is varying up to 0.5 dex considering our numbers and the literature. 
If we do not consider Co, the mean abundance differences 
between our results and \cite{frebel:07,  frebel:13} is smaller than rms deviations, with $<\Delta[El/Fe]>$ = -0.04 $\pm 0.13$.
Such a mean difference is derived by using the abundances of 13 elements available in both studies.
If we focus the comparison with \cite{hansen:18}, we obtain $<\Delta[El/Fe]>$ = -0.015  $\pm 0.33$ with 4 common elements: C, Sr, Ba and Eu. The variation larger than average is due to the different Sr abundance, which also differs from \cite{sakari:18} and \cite{reggiani:22}. 
In comparison with \cite{sakari:18}, we also observe a discrepancy in the C abundance. This is because in \cite{sakari:18} the reported C value is already corrected for the effect of stellar evolution (see details in Section \ref{sec: result, stellar}). 
For Co, Mn and Dy there are differences in LTE values due to the strong influence of NLTE on the resonance lines used for this study. If we exclude the elements mentioned above, $<\Delta[El/Fe]>$ = -0.18 $\pm 0.17$ (19 common elements). Our [Eu/Fe] = 1.67 is in agreement with those of \cite{frebel:07} ([Eu/Fe]= 1.70), and within 1 $\sigma$ we are consistent with other works. Finally, we obtain $<\Delta[El/Fe]>$ = 0.08 $\pm0.21$ (14 common elements) compared to \cite{reggiani:22}.

\textbf{HD 6268}. In Table \ref{comp_abund}, our results (10, 11) are compared with \cite{mcwilliam:95} (8) and \cite{roederer:14} (9). If we exclude the Mn values where the \cite{mcwilliam:95} abundance was calculated using the EW from stellar spectra with LTE assumption,
we obtain  $<\Delta[El/Fe]>$ = -0.06 $\pm$ 0.22 for 19 common elements. For \cite{roederer:14} we have $<\Delta[El/Fe]>$  = 0.11 $\pm$ 0.15 (24). On the other hand, the [Eu/Fe] ratio shows a much larger variation between different works: [Eu/Fe] = 0.65 for \cite{mcwilliam:95}, 0.31 for \cite{roederer:14} and 0.47 according to our measurement. 

\textbf{HD 121135}. The comparison for this star is not presented in the Table \ref{comp_abund}, since data are only available in the previous literature for a limited number of elements.
\cite{pilachowski:96} obtained [Na/Fe] = -0.10, [Mg/Fe] = 0.30 and [Ca/Fe] = 0.3, while we derived [Na/Fe] = -0.12 (including NLTE corrections), [Mg/Fe] = 0.3 and [Ca/Fe] = 0.27.
We also find a good agreement with \cite{simmerer:04}: they obtained [C/Fe] = -0.45 as compared to our value of [C/Fe] =-0.46, and log A(La/Eu) = 0.37 compared to 0.41. 
\cite{burris:00} obtained [Ba/Fe] = 0.20 without NLTE corrections. We obtain LTE [Ba/Fe] = 0.08, and [Ba/Fe] = -0.14 dex taking into account the NLTE correction. 

\textbf{HD 195636}. We highlight here two papers with abundance determinations for several chemical elements: \cite{gratton:88} (column 12) and  \cite{zhang:05} (columns 13 and 14). Relevant differences are found between the abundances reported in these studies and also compared to our results. 
For example, to determine the Na abundance the resonant D lines were used in all studies, which required NLTE corrections for the considered metallicities to be introduced. \cite{gratton:88} obtained a value of [Na/Fe] = 0.57 (LTE); in the study by \cite{zhang:05} [Na/Fe] = --0.14 (LTE) and --0.38 (NLTE); in the present study, [Na/Fe] = 0.59 (LTE) and 0.04 (NLTE). For Mg, the NLTE corrections are much smaller compared to Na. The [Mg/Fe] varies between 0.28 (LTE, \cite{gratton:88}), 0.82 (LTE, \cite{zhang:05}) and 0.58 (LTE) and 0.64 (NLTE) in present study. As for the neutron-capture elements, the Sr abundance was the only one determined in \cite{gratton:88}. Their value is [Sr/Fe] = +0.39, as compared to our values --0.21 (LTE) and --0.19 (NLTE), respectively. The Ba abundance was the only one reported by \cite{zhang:05}, [Ba/Fe] = --0.14 (LTE), while we have --0.66 (LTE) and --0.43 (NLTE) respectively. These discrepancies are probably due to the different methods and techniques employed in the different studies. 

\section{Kinematics and population membership}
\label{sec: kinematics}

\begin{figure}
\begin{tabular}{c}
\includegraphics[width=8cm]{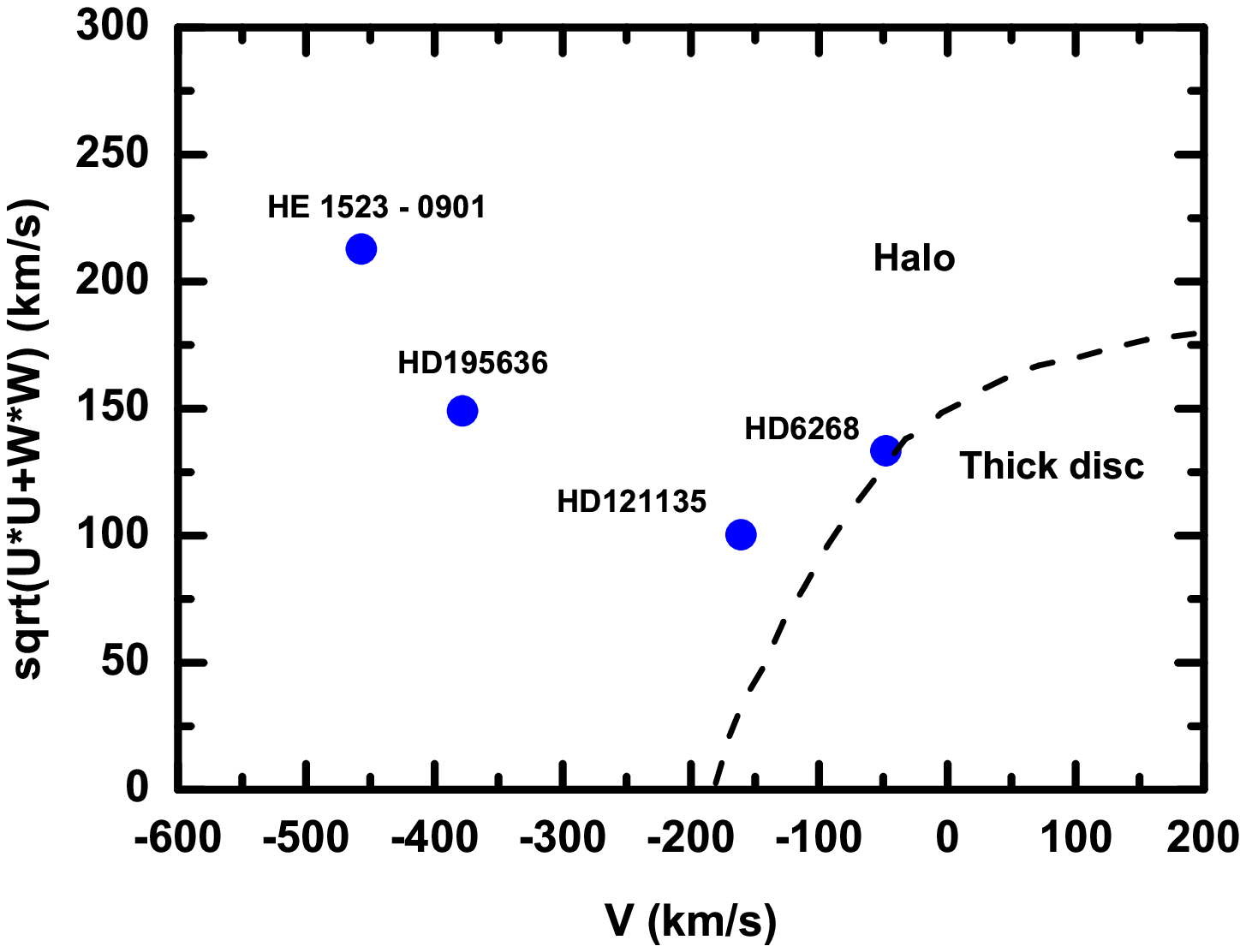}\\
\end{tabular}
\caption{Toomre diagram  of studied stars. The dashed curve corresponds to a total velocity of 180 km s$^{-1}$ marking the kinematical separation between the thick disc and the halo.} \label{uvw}
\end{figure}

 In this section we examine the stellar velocities of our sample to define whether they belong to different Galactic populations. We adopt the photogeometric estimates of distance from \cite{bailer:21} (listed in Table \ref{velocity}), since we expect them to have higher accuracy and precision than the inverse of the parallax. We obtain distances between $\sim$500 pc and $\sim$2.8 kpc from the Sun. 
Combined with coordinates and proper motions from Gaia DR3 and radial velocities determined by us from the SALT spectra (see Table \ref{target}), we compute the three components of the velocities (U, V, W) with respect to the Local Standard of Rest (LSR), with U towards the Galactic centre, V in the direction of Galactic rotation, and W towards the North Galactic Pole. 
We adopt the solar motion with respect to the LSR by \cite{robin:22} from Gaia DR3: (U, V, W)$_\odot$ = 10.79$\pm$0.56, 11.06$\pm$0.94,7.66$\pm$0.43 \kms. 
We obtain a total velocity larger than 100 \kms\ for all our stars, which would indicate that they are part of the halo or the thick disc. However the Toomre diagram in Fig. \ref{uvw} shows a diversity in the kinematics of the targets. 
HE 1523--0901 and HD 195636 have large total velocities, about 500 and 400 \kms\ respectively, making them likely members of the Galactic halo. They also lag well behind the rotation of the LSR, which is also compatible with their belonging to an accreted population, such as the Thamnos streams \citep{koppelman:19}. Their low [Fe/H] (-2.82 and -2.78 dex respectively) also supports their membership to the halo or an accreted population \citep{dimatteo:20}. 
The kinematical separation between the halo and the thick disc is around a total space velocity of 180 \kms\ \citep[e.g.][]{buder:19}, making HD121135 a possible thick disc star, at the kinematical transition with the halo. Its rotation velocity of -161 \kms\ is also consistent with the thick disk. However its large vertical velocity of 100 \kms\ would be very unusual, owing to the velocity ellispoid of the thick disc which is centered on W$\simeq$0 with a dispersion of $\sim$20 \kms\ (22 \kms\, according to the recent work by \cite{vieira:22}. 
Therefore HD 121135 is more likely a halo star or an accreted star.  
Furthermore, its metallicity ([Fe/H] = -1.40) is lower than typical thick disc stars \citep{li:17}, although it would be still compatible with the metal-weak thick disc described by \cite{beers:14}. 
HD 6268 has the lowest space velocities of the four stars discussed here, with a rotation velocity and a vertical velocity typical of the disc. Its large velocity of 124 \kms\ in the radial direction is still consistent with the thick disc given the dispersion of 49 \kms\ determined for that population in that direction by \cite{vieira:22}. However, its metallicity [Fe/H]=-2.54 would make it a very special member of the thick disc. According to \cite{beers:14}, the metal-weak thick disc does not extend below [Fe/H]=-1.8 dex but more recent studies found very metal-poor stars with typical disc kinematics \citep[e.g.][]{dimatteo:20,sestito:20}. The origin of such stars is not yet clear and different scenarios have being proposed. Among others, they could be the relics of merged satellites or of a pristine prograde disc in the Milky Way \citep{bellazzini:23}.
HD 6268 could therefore be a prototype of a very metal-poor star remaining close to the galactic plane with a non circular orbit. Its detailed chemical composition may shed light on its origin.

 \begin{table*}
\caption{Distance of the targets from \protect \cite{bailer:21} and their space velocities with respect to the LSR.} 
\label{velocity}
\begin{tabular}{lrrrrr}         
\hline                 
star          & distance &      U &     V &  W &        Total velocity \\                         
              & pc     & km s$^{-1}$&km s$^{-1}$&km s$^{-1}$&km s$^{-1}$ \\
            \hline                
  HE1523-0901 & 2692 & -153 & -457 & -148 & 505 \\
  HD6268      & 666  &  124 &  -48 &  -49 & 142 \\
  HD121135    & 778  &   -8 & -161 &  100 & 190 \\
  HD195636    & 578  &   -1 & -378 &  149 & 406 \\
\hline 
\end{tabular}  
\end{table*}

\begin{figure}
\begin{tabular}{c}
\includegraphics[width=8cm]{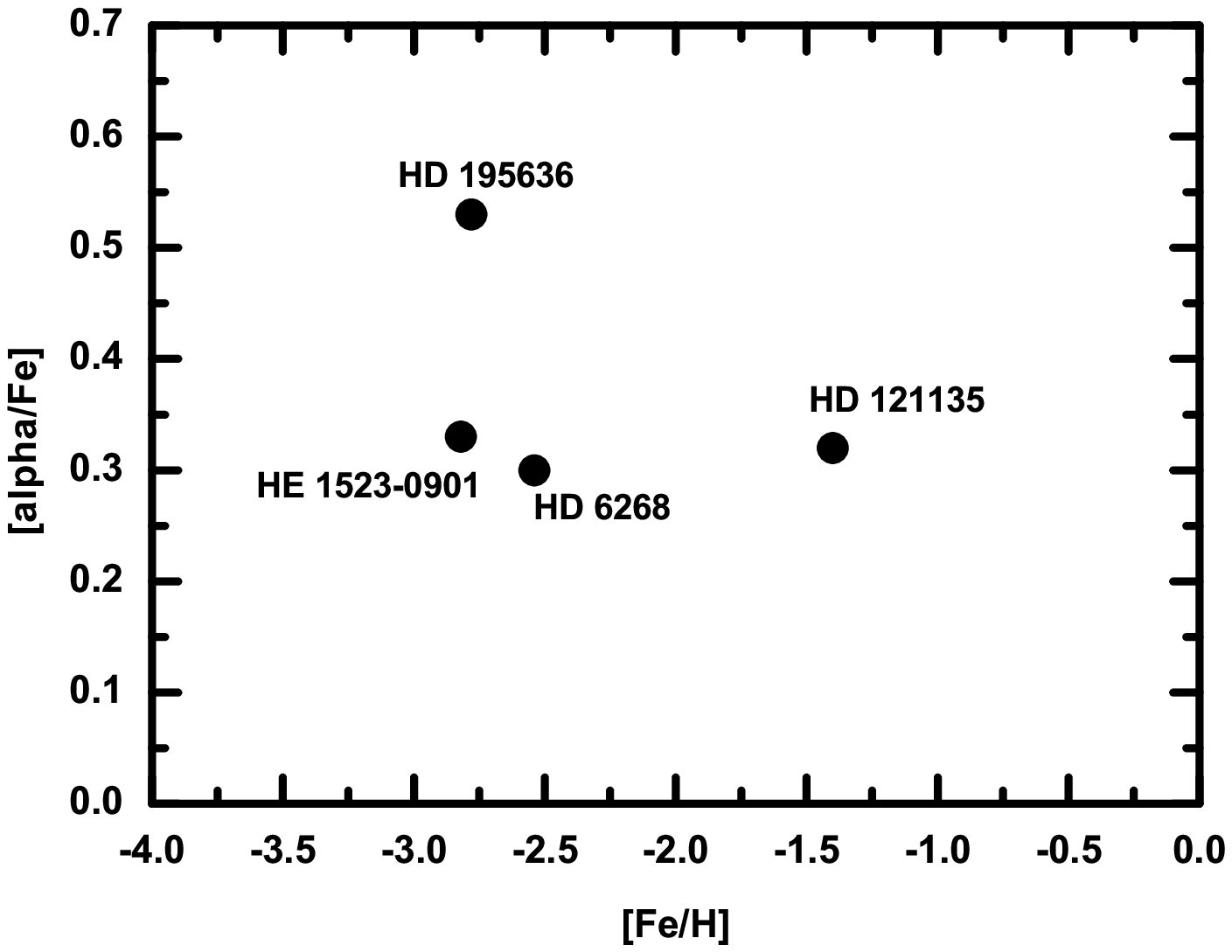}\\
\includegraphics[width=8cm]{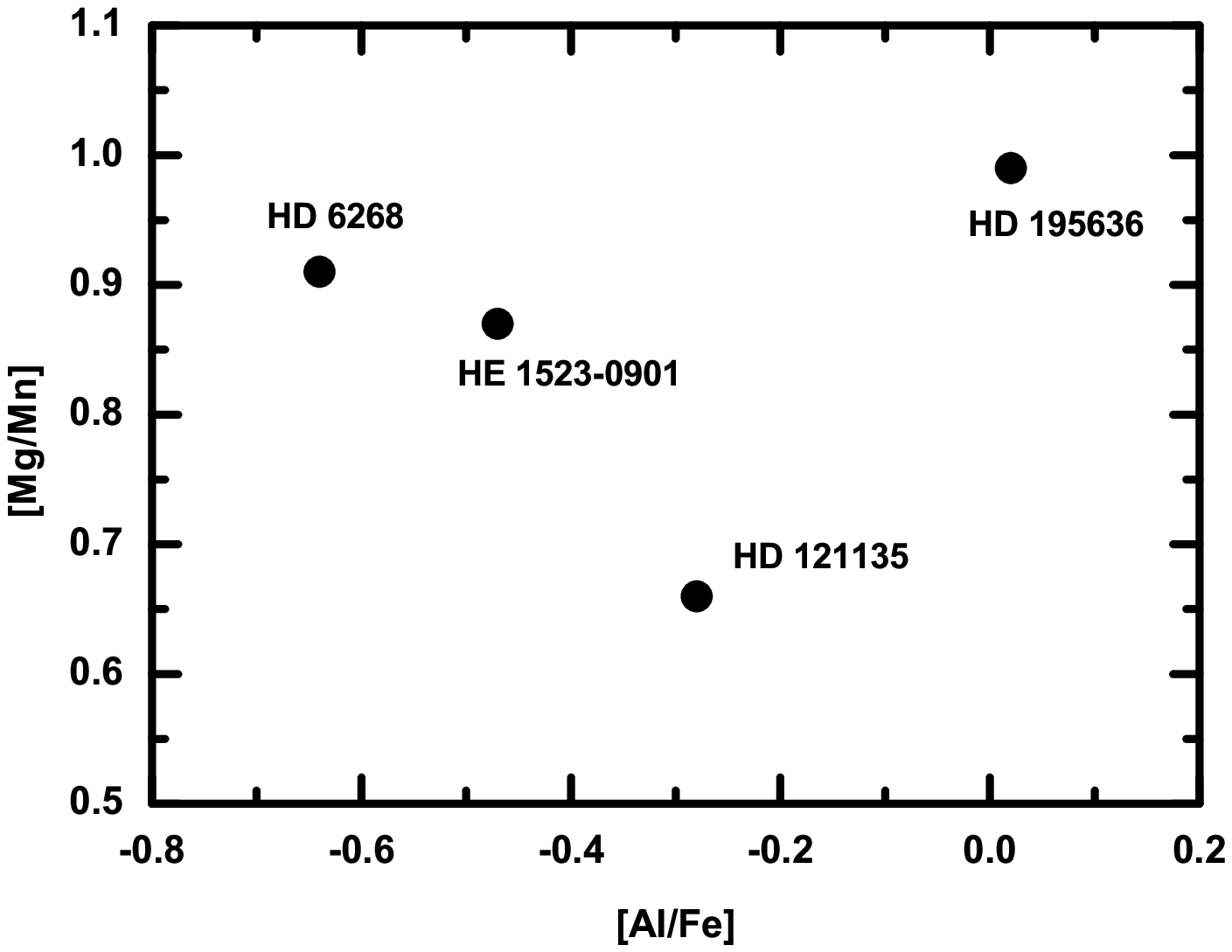}
\end{tabular}
\caption{Abundance distributions of the stars presented in this work for [$\alpha$/Fe] with respect to [Fe/H] (top panel), and [Mg/Mn] vs [Al/Fe] (bottom panel).}
\label{alpha}
\end{figure}

 \begin{table*}
\caption{Marking of stars to belong to galactic substructures and to different r-process enrichment classes. }
\label{belong}
\begin{tabular}{lllcc}
\hline
Star  &                              [$\alpha$/Fe] -- [Fe/H]  &    [Mg/Mn]-[Al/Fe] &   kinematics   &          nucleosynthesis \\
\hline
HE 1523-–0901  &          halo    &      ? &         halo or accreted  &                                 r-II  \\
HD 6268    &      halo &     accreted? &  thick    disc &               r-I or r-II (actinide?)\\
HD 121135   &         halo &    accreted &      halo or  accreted&              limited-r or r-I  \\
HD 195636  &                halo or accreted  &    ? &              halo or accreted &                       limited-r or r-I (?)\\
\hline
\end{tabular} 
\end{table*}

Based on their kinematic properties and their low metallicities, we have seen that HE 1523--0901 and HD195636 can be classified as halo stars, possibly accreted, owing to their very large retrograde motion, while the identification for both HD 121135 and HD6268 is less clear. 

If the stellar metallicity is not too low, stellar abundances are also used as additional diagnostics to define stellar origin. For instance, \cite{das:20} used the [Mg/Mn]-[Al/Fe] chemical abundance plane to identify nearby stars included in the APOGEE survey and that have been accreted on the Milky Way: they identify a group (or blob) of stars with high--[Mg/Mn], low--[Al/Fe] component as a sample of accreted stars peaked at [Fe/H] $\sim$ --1.6. \cite{belokurov:22}, considered instead [Al/Fe] as a unique indicator to distinguish between stars formed in the canonical halo and those that have been accreted: metal-poor stars with [Al/Fe]$<$-0.075 are allocated to the accreted population, while the in-situ stars have higher Al-to-Fe ratios.

For HD 121135, the composition and metallicity ([Fe/H]=-1.40) would well fit as accreted star ([Mg/Mn] = 0.66 dex and [Al/Fe] = -0.28 dex), based on both the criteria discussed above.

Table \ref{belong} summarizes the possible membership of each star to the different populations according to chemical and kinematical criteria, with  the various $r$-process contributions identified for them (see discussion in the following sections). 

\section{Considering whether all the stellar abundances are pristine: Impact of the intrinsic stellar evolution }
\label{sec: result, stellar}

\begin{figure}
\begin{tabular}{c}
\includegraphics[width=8cm]{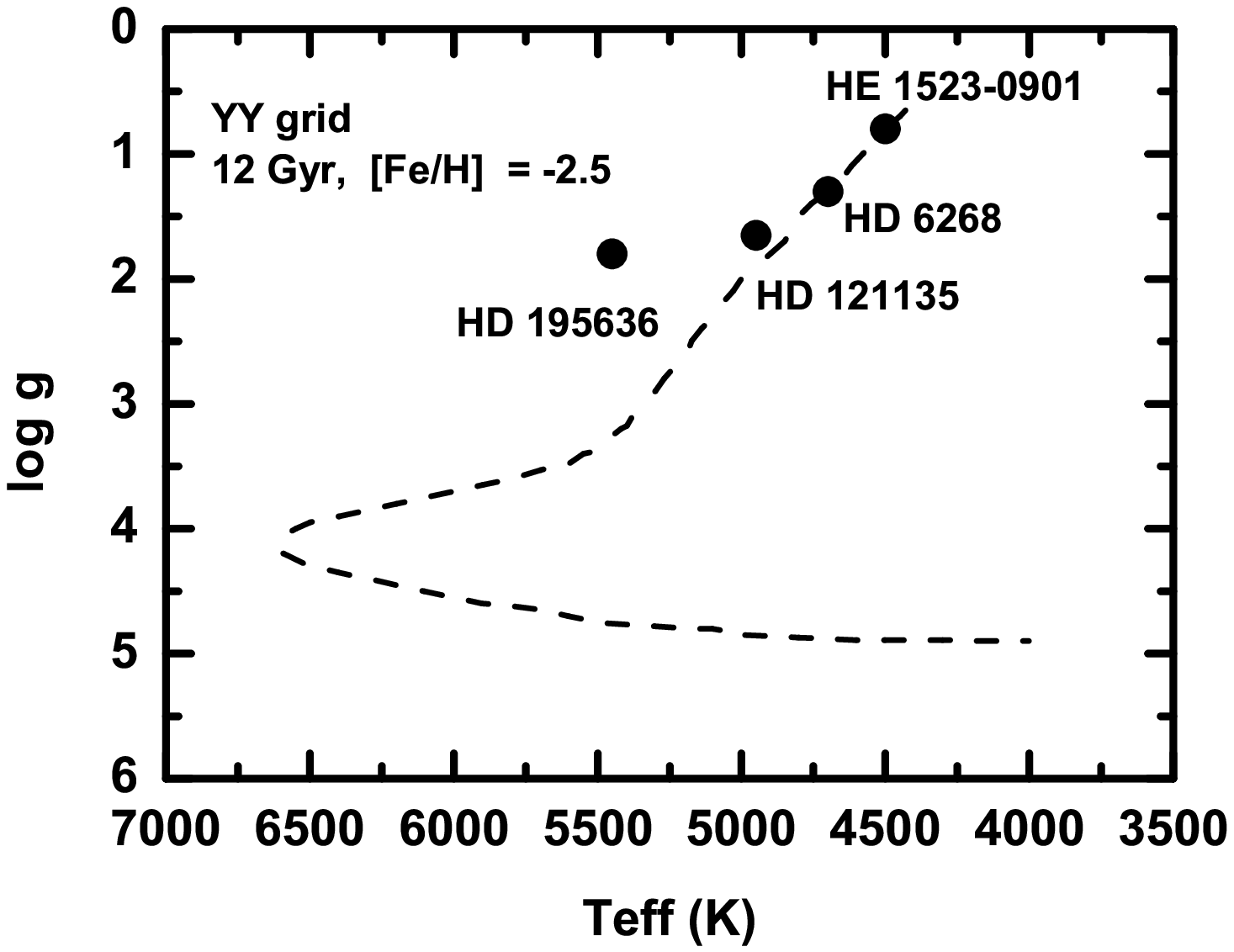}\\
\end{tabular}
\caption{Diagram of \logg\ versus~ \Teff\ from \cite{demarque:04} and the position of the four stars studied in this work. }
\label{treck}
\end{figure}

Depending on the evolutionary stage of the observed stars, some of the pristine abundances, inherited from the interstellar medium where the stars have formed, could have been modified by internal nucleosynthesis and mixed all the way to the stellar surface. In this case, the affected elements such as C, N, and O are not indicative anymore of the initial stellar composition, but they instead carry the signature of internal stellar evolution processes \citep[e.g.][]{charbonnel:94, gratton:00, spite:06}. In order to infer the evolutionary stage of the observed star, the first key diagnostic is given by the surface temperatures and densities. 
A template stellar isochrone with \logg\ with respect to~ \Teff\ from \cite{demarque:04} is shown in figure~\ref{treck}, in comparison to the positions of the studied stars: three stars are on the red giant branch (RGB) phase; whereas HD 195636 appears to have passed the RGB and is now on the  HB or even at the onset of the asymptotic giant branch (AGB) phase. We note, for example, that the YY tracks \cite{demarque:04}  are in good agreement with the BaSTI tracks, as reported in \cite{pietrinferni:13}.
 
According to earlier studies by \cite{gratton:00}, light element abundances in lower-RGB stars (i.e. stars brighter than the first dredge-up luminosity and fainter than that of the RGB bump) seem to be consistent with predictions from stellar evolutionary models. 
\cite{spite:06} performed a LTE analysis of several extremely metal-poor (EMP) giants in order to investigate their CNO abundance. The C-N anti-correlation obtained in this study is in agreement with the expectation from theoretical models that surface abundances are affected by the CNO-processed material from the inner regions. More recently, \cite{khan:18} found that standard stellar models may underestimate the extra-mixing efficiency at the base of the convective envelope, which seems to increase with decreasing metallicity. Because of its apparent universality, \cite{denis:03} even proposed to call such a non-convective mixing process the 'canonical extra mixing'. The physics mechanisms that are responsible for such extra-mixing processes are a matter of debate. 
For instance, 
thermohaline mixing was proposed as a possible explanation for the anomalous abundances in the envelopes of red giants 
\citep[e.g.][]{charbonnel:07}. 
\cite{stancliffe:09} showed that by adopting the prescription introduced by \cite{charbonnel:07} it is possible to reproduce the abundance trends observed across a wide range of metallicities for both carbon-rich and carbon-normal stars in the upper RGB phase, while \cite{angelou:11} shows that this solution would work also for globular cluster stars. More modern parametrizations of thermohaline mixing in stellar models seem to be more compatible also with the Li abundances observed in these stars \citep[][]{henkel:18}, but a definitive theoretical solution is not yet defined for this problem \citep[e.g.][]{tayar:22, fraser:22}, and it is unclear if more mixing mechanisms are needed \citep[e.g.][]{mcmormick:23, aguileragomez:23}.   

\begin{figure}
\begin{tabular}{c}
\includegraphics[width=8cm]{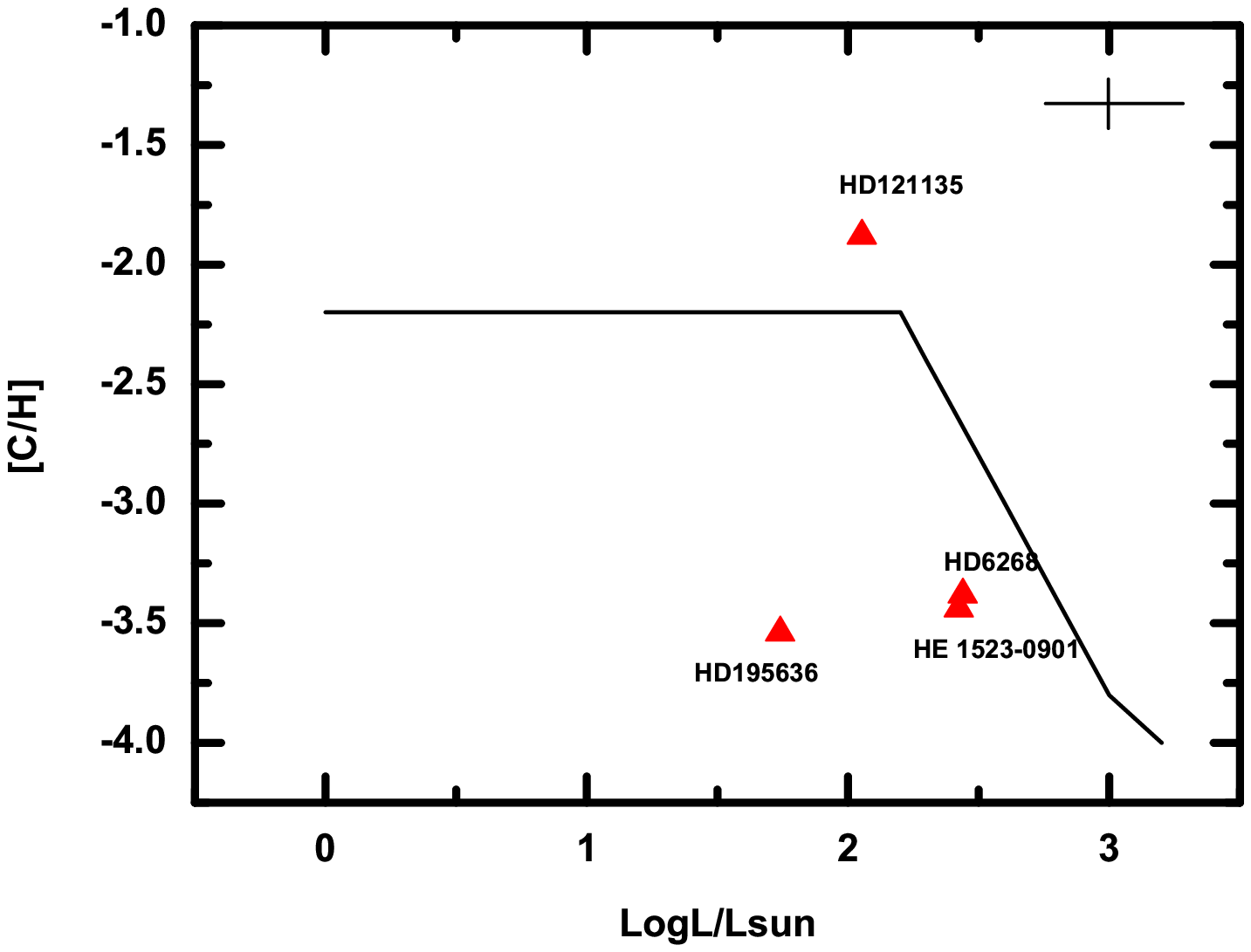}\\
\end{tabular}
\caption{C abundances ([C/H] = -3.44, -3.38, -1.88, and -3.54 for HE 1523--0901, HD 6268, HD 121135, and HD 195636 respectively) are shown with respect to the derived luminosities log$_{10}$(L/L$_{\odot}$) (see the text for details).    
The black line takes into account the effect of to canonical extra mixing for non C-enhanced VMP stars \citep[][see their Fig. 2]{denis:08}.}
\label{chl}
\end{figure}

\cite{denis:08} examined changes in C and N abundances in Carbon Enhanced Metal Poor stars and non Carbon-enhanced Very Metal Poor (VMP) stars (which is the case of interest here) that could be associated with the canonical extra mixing in RGB stars. The C abundances measured in our stars are given in Table \ref{abund}. In Fig. \ref{chl}, we compare them with the predictions by \cite{denis:08}. 
 For our investigated stars, we calculated the value of absolute luminosity given as log(L/L$_{\odot}$), and based on the classical formula: 
log(L/L$_{\odot}$) = -2log(P) -0.4V +0.4A(v),
 where V is the magnitude and P is the parallax taken from the SIMBAD database (Gaia DR3, \cite{gaia:20}). To take into account the interstellar absorption, A(v), color excess data from the following works were used, for HE1523 -0901: E(B-V) = 0.138 \citep[][]{schlegel:98}; HD 6268: E(B-V) = 0.017 \citep[][]{roederer:14}; HD195636:  E(B-V) =  0.06 \citep[][]{bond:80}. Finally, for HD121135 the E(B-V) is not taken into account.
 
 The calculations yielded the following values, also used in Figure \ref{chl}:   
 log$_{10}$(L/L$_{\odot}$) = 2.42, 2.44, 2.05, and 1.74 for HE 1523-0901, HD 6268, HD 121135, and HD 195636, respectively. 
Based on their C abundances and luminosities, HD 121135, HD 6268 and HE 1523-0901 are consistent with the predictions by \cite{denis:08}. In particular, the initial C abundance of HD 6268 and HE 1523-0901 seems to have been strongly depleted to produce N. On the other hand, HD 195636 seems also to be C depleted, but it does not follow the trend of the other stars. Therefore, 
we can derive that HD 195636 has already passed the RGB stage and has moved to the HB stage (see Fig. \ref{treck}).

In order to assess the frequencies of carbon-enhanced stars, \cite{placco:14} took into account the expected depletion of the surface carbon abundance, occurring due to CN processing on the upper RGB, and recovered the initial carbon abundance of those stars using the STARS stellar evolution code \cite[][]{eggleton:71, staneldr:09}. 
If we consider the mean corrections due to stellar mixing by \cite{placco:14}, the pristine [C/Fe] ratios for the four target stars would be +0.28, -0.32, -0.11 and -0.69 for HE 1523-0901, HD 6268, HD 121135, and HD 195636, respectively\footnote{Notice that by using the corrections given by \cite{henkel:18}, we would instead obtain [C/Fe] = -0.19 -- -0.05 (HE1523-0901), -0.51 -- -0.37 (HD6268), -0.3 -- -0.2 (HD121135) and  -0.76 (HD 195636, with no corrections to be applied). These different corrections do not change our discussion presented in this section.}. 
Even taking into account mixing corrections, HD 195636 seems to be significantly depleted in carbon.

We notice that very faint C molecular bands in the spectrum of this star may also be detected, yielding the same low C values ([C/Fe] = -0.76).  
HD 195636 was also studied by \cite{takeda:13} together with other 46 stars. 
Based on NLTE analysis of C I lines at 1.068-1.069$\mu$, 
for HD 195636 they reported [C/Fe] = -0.47, which is 0.3 higher than our value. For HD 6268 they obtained [C/Fe] = -0.61 instead of [C/Fe]=-0.84 in this study and -0.91 in the study by \cite{roederer:14}. Finally, for HD 121135 \cite{takeda:13} measured [C/Fe] = -0.05, which is 0.4 dex higher than this study. As pointed out by \cite{takeda:13}, their C abundances were appreciably higher than those from CH lines, especially for very metal-poor giants of low gravity. Such a difference was due to the NLTE correction taken into account for the C I lines, which progressively increases with decreasing metallicity. 
In our calculations we did not take into account departures from LTE for C abundances. Therefore, it is not surprising to obtain such a large variations between the [C/Fe] values derived by different authors. 
It should also be noted that, due to the formation of molecules in higher layers of the atmosphere compared to atomic lines, dwarfs and giants show C abundance discrepancies up to -0.5 to -0.7 dex in 3D hydrodynamics models with respect to synthetic spectra from standard 1D model atmospheres \citep[][]{collet:07, behara:10, steffen:18}. This introduces an additional significant error in the determination of C abundance. So, based on all of these considerations, we did not take into account the carbon content for the nucleosynthesis analysis discussed in the next sections.

\section{Nucleosynthesis signatures}
\label{sec: element nucleosynthesis} 

\begin{figure}
\begin{tabular}{c}
\includegraphics[width=8cm]{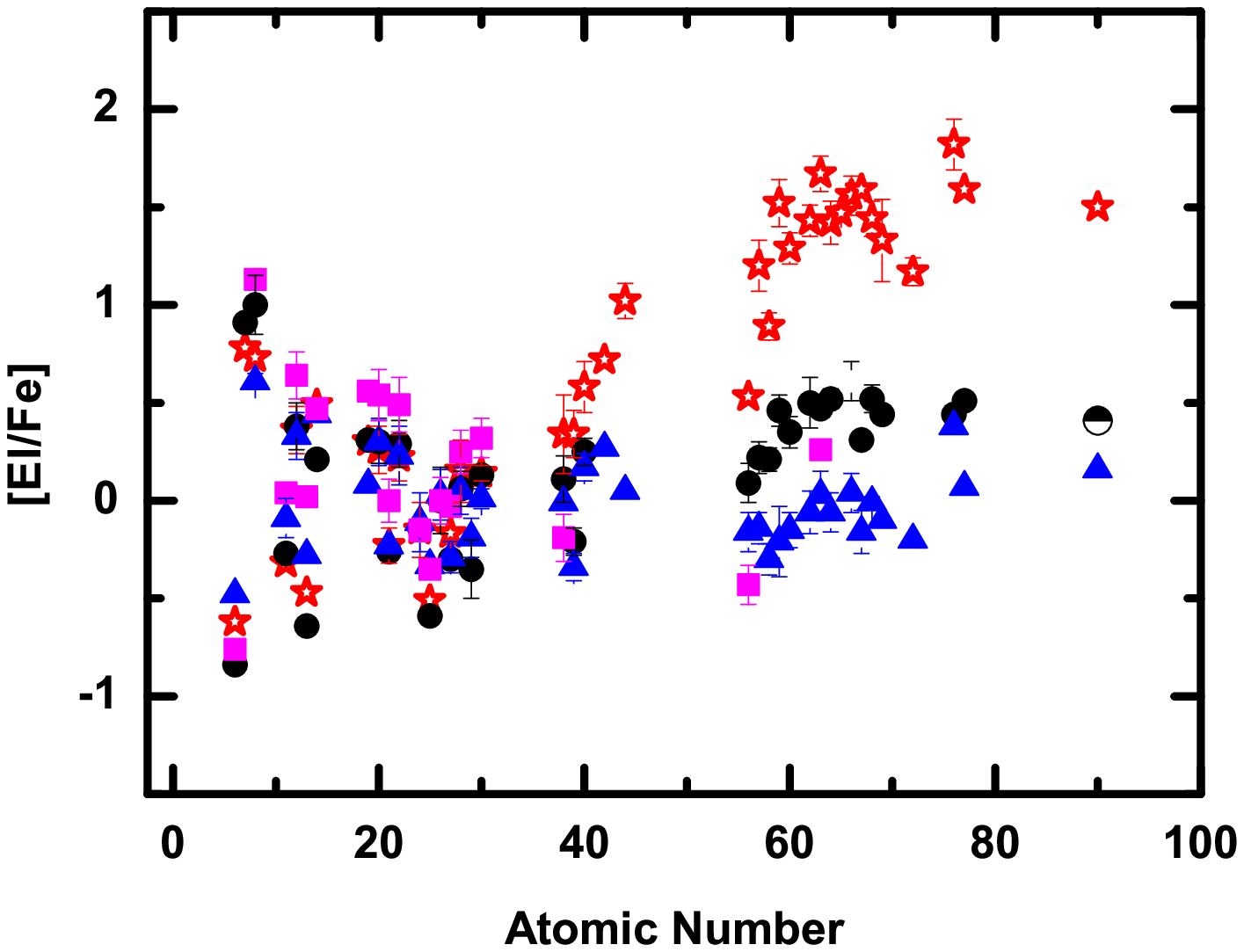}\\
\end{tabular}
\caption{Abundance distribution of [El/Fe] with respect to atomic number for HE 1523--0901 (red empty asterisks), HD 6268 (black full circles), HD 121135 (blue full triangles) and HD 195636 (purple full squares). The Th abundance for HD 6268 is taken from \protect\cite{roederer:14} (black semicircle). }
\label{elfeat}
\end{figure}

The elemental abundance patterns derived for our four peculiar stars are shown in figure~\ref{elfeat}. NLTE corrections are considered for O, Na, Mg, Al, K, Mn, Cu, Ba, Eu (see the details in Section \ref{sec: abundance determination}). 
Various works on the influence of NLTE on iron determinations present different estimates as mentioned in the Introduction. If we rely on the work of \cite{berg:12}, which considered the influence on stars similar in parameters to those studied here (for example, HD122563 with \Teff = 4665 $\pm$ 80 K, \logg = 1.64 $\pm$ 0.16, [Fe/H] = -2.61), the authors found that the iron abundance determination for the Fe II lines are practically unaffected by deviations from LTE, and for the lines of neutral iron the deviations do not exceed 0.15 dex. 
These uncertainties are close to the errors in determining the abundance of elements, but the NLTE correction is positive. If we take this correction into account when comparing with nucleosynthesis calculations, we would roughly obtain a systematic shift of 0.10 - 0.15 dex in figure~\ref{elfeat}, and the overall distribution of elements is still within the reported error bars.
Here we discuss the observed nucleosynthesis signatures in comparison to theoretical stellar predictions and to other stars with similar metallicities. 

One of the major advantages of galactic archaeology studies is that the element content of old metal-poor stars 
typically show the abundance pattern they inherited at their birth. The ongoing nuclear burning and production of new elements taking place in the interior does not show up in the surface abundances which are observed. Thus, such stars are the witnesses of all nucleosynthesis events which took place prior to their formation \citep[e.g.][]{Wheeler.Sneden.Truran:1989,sneden:08,frebel:10,Matteucci:2012}. Exceptions from this general behavior can be found in evolved stars that experienced already dredge-up of processed material from the stellar interiors, and - as the stars presented in this paper are giants - this should be considered here as well \citep[e.g.][]{busso:07,denis:08}. 
With the CNO-cycle a significant amount of C and some O is converted to N. All of our four stars show such a depleted C-abundance. i.e. [C/Fe] =  -0.62 for HE 1523--0901, -0.84 for HD 6268, -0.48 for HD 121135, and -0.76 for HD 195636. We can report enhanced N of 0.78 for HE 1523-0901 and 0.91 for HD 6268 (see Section \ref{sec: result, stellar}). However, most of the elements beside C and N (and possibly F and Na) will not be significantly modified \citep[e.g.][]{palmerini:11}. Therefore, these stars can still be used to study their pristine abundance signatures for most of the observable elements.  
The star HD 121135 has [Fe/H] = $-$1.40, which is the highest among the stars considered in this work. With such a metallicity, the pristine composition of the star is expected to be mostly a product of GCE, where a few generations of CCSNe contributed to build the chemical abundances of the interstellar medium to some degree of homogeneous signature \citep[e.g.][]{prantzos:18, kobayashi:20,  matteucci:21}. On the other hand, the other three peculiar stars considered in this work have [Fe/H] $<$ $-$2.5. For these much lower metallicities, only a few CCSNe had the time to contribute (and even less so rare stellar sources), and therefore inhomogeneous abundances are expected (and observed) in metal-poor stars \citep[e.g.][]{argast:00, gibson:03}. While simple GCE simulations do not have much predictive power to study these conditions, inhomogeneous GCE models and multi-dimensional cosmological chemodynamical simulations can be used for comparison with the observations \citep[e.g.][]{brook:12, mishenina:17, kobayashi:23}. The abundances of the most metal-poor stars or of the most anomalous metal-poor stars are also used for stellar archaeology studies, where observations are directly compared with the abundance yields of theoretical stellar models \citep[e.g.][]{sneden:03, nomoto:06, aoki:07,  frebel:10, keller:14, Yong.Kobayashi.ea:2021, farouqi:22, Placco.Almeida.ea:2023}. This approach becomes a powerful tool to study stellar simulations and nucleosynthesis once we assume that the abundances observed today were mostly shaped by one (or very few) stellar sources or nucleosynthesis processes before the formation of the star. Within this context, below we discuss the abundances of HE 1523--0901, HD 6268, HD 12113, and HD 195636.

\subsection{Elements up to Zn (Z=30) }

\begin{figure}
\begin{tabular}{c}
\includegraphics[width=8cm]{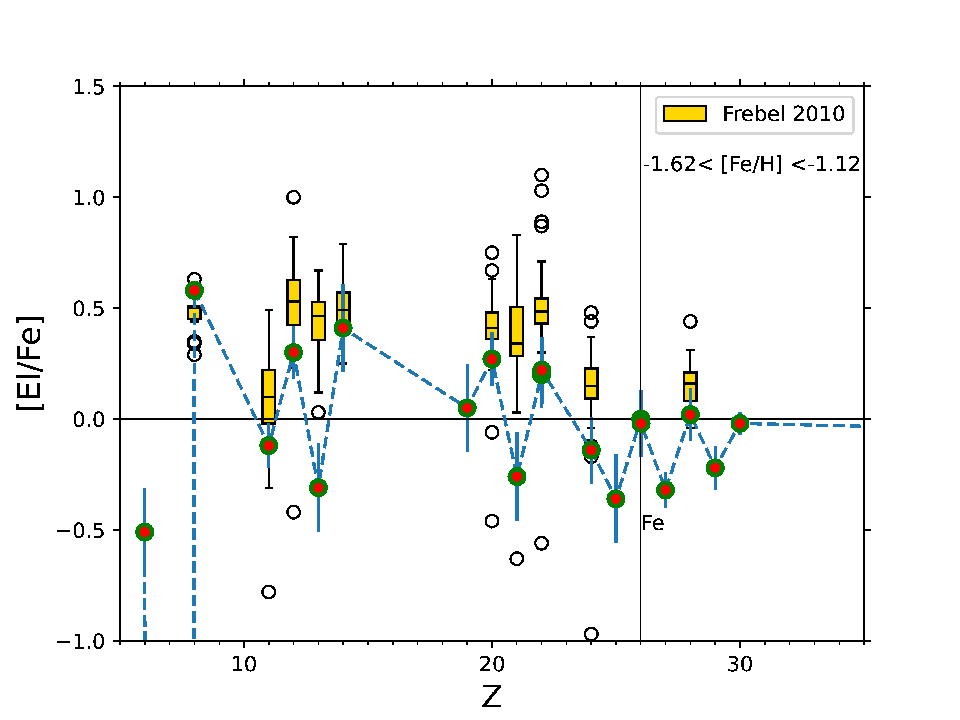}\\
\end{tabular}
\caption{Observed abundances of HD 121135 are shown in comparison with the abundance distribution of stars with analogous metallicity (-1.62 $\gtrsim$ [Fe/H] $\gtrsim$ -1.12) by \protect\cite{frebel:10}. The rectangle boxes cover the observed values between 25$\%$ and 75$\%$ of all the data points, for elements with more than 15 observed stars. The median values of the models are also shown in the boxes. The error bars cover 1.5 times the range of the box plot. Any points outside the error bars (outliers) are shown as open circles.}
\label{fig: hd121135_vs_obs}
\end{figure}

The four stars considered in this work have [Fe/H]$<$-1.3. In this metallicity range the GCE of elements up to the iron group in the Milky Way is still mostly dominated by CCSNe. Indeed, Type Ia supernovae 
did not have time to give a significant contribution to the overall galactic chemical inventory \citep[e.g.][]{matteucci:86, Wheeler.Sneden.Truran:1989,timmes:95}. 

\textbf{HD 121135}. As mentioned earlier, we expect that HD 121135 abundances are representative of the GCE at the time of its formation, with the exception of C and N.  In Fig. \ref{fig: hd121135_vs_obs}, the HD 121135 composition is shown compared to the abundance distribution of stars within the same metallicity range \citep[][]{frebel:10}. For most of the elements, the HD 121135 abundances seem to be consistent with \cite{frebel:10} data. We do not consider C in the comparison, since this element is available only for few stars and we have seen that during the HD 121135 evolution the initial C has been modified. On the other hand, the observed abundance of the odd elements Al (Z=13) and Sc (Z=21) are significantly lower compared to the typical Galactic disk stars in the same metallicity range. Such variation of these elements compared to nearby even elements (e.g. Mg and Ca respectively) would be expected for stars at much lower metallicities than HD 121135, belonging to the galactic halo \citep[e.g.][]{francois:20, lombardo:22}.

\begin{figure*}
\begin{tabular}{cc}
\includegraphics[width=8.0cm]{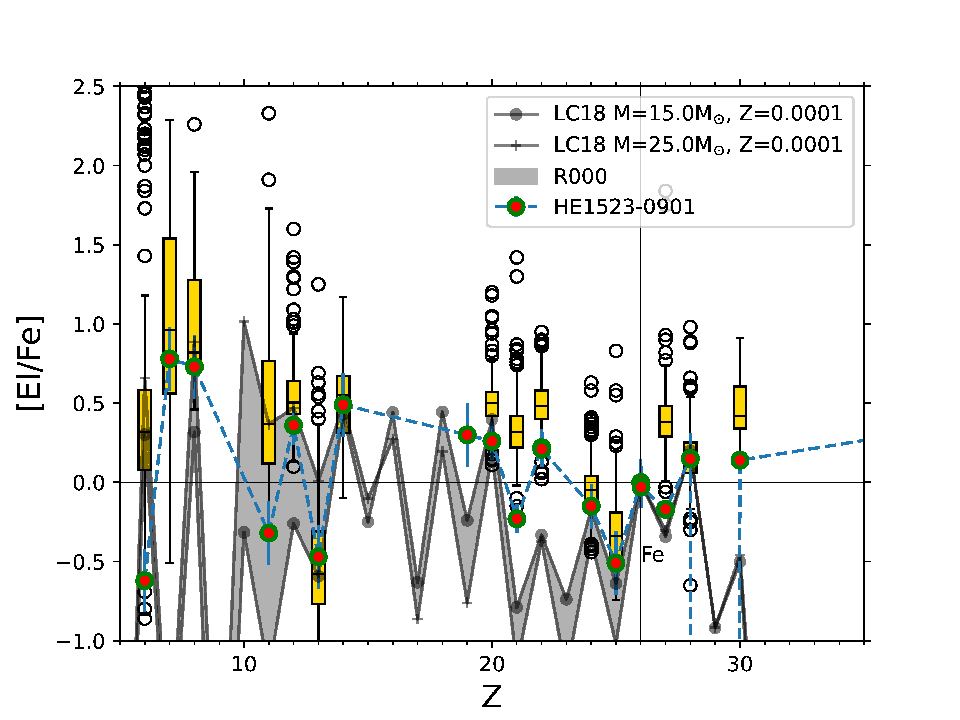}     &\includegraphics[width=8.0cm]{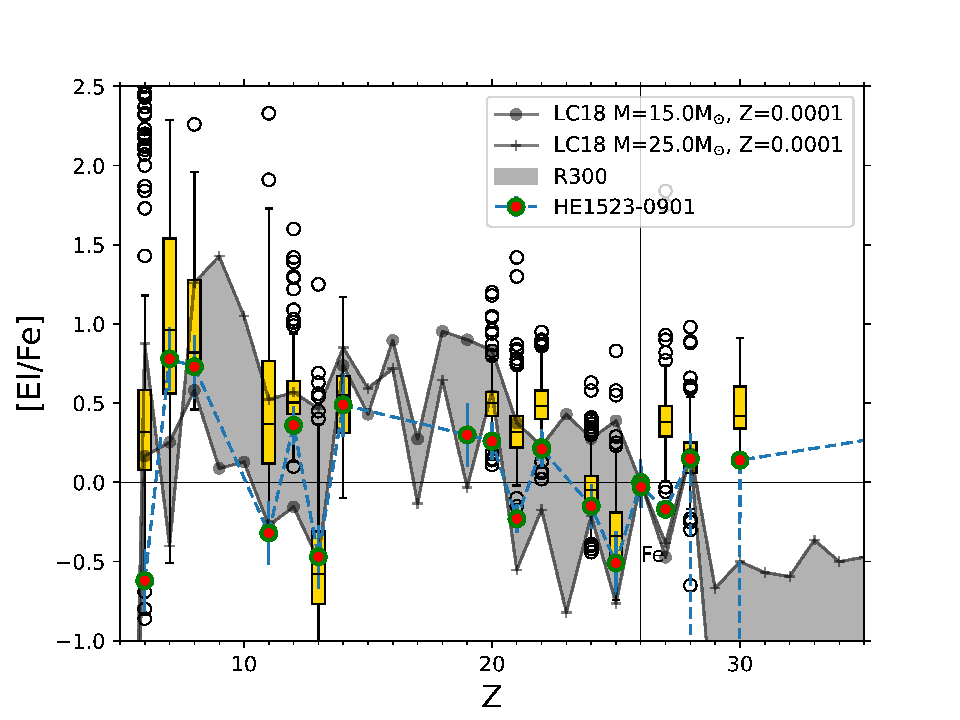}\\
\includegraphics[width=8.0cm]{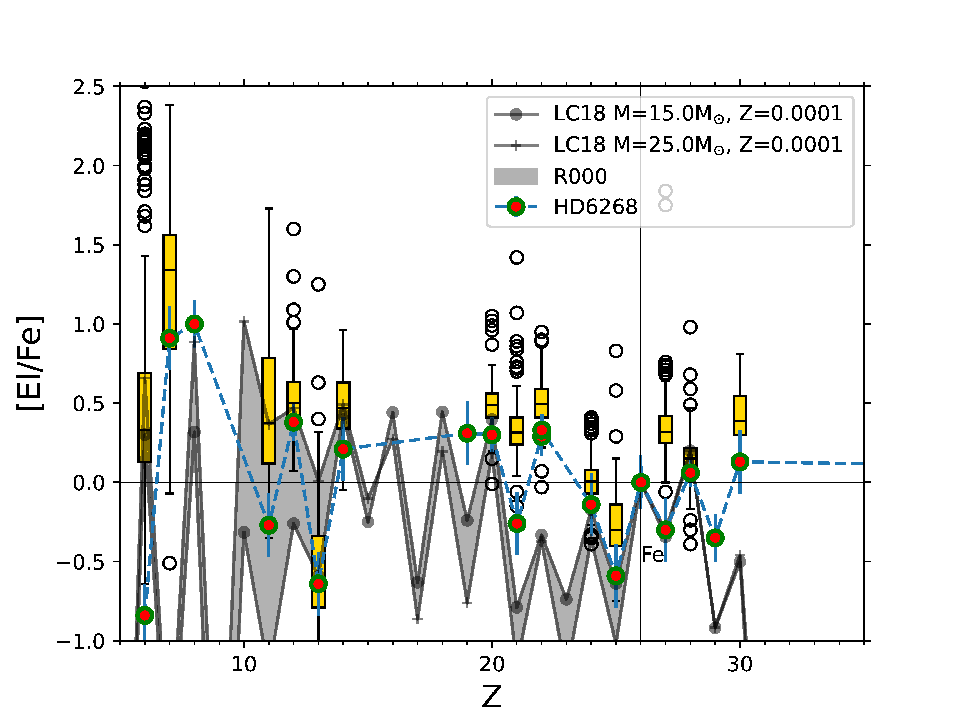}   &\includegraphics[width=8.0cm]{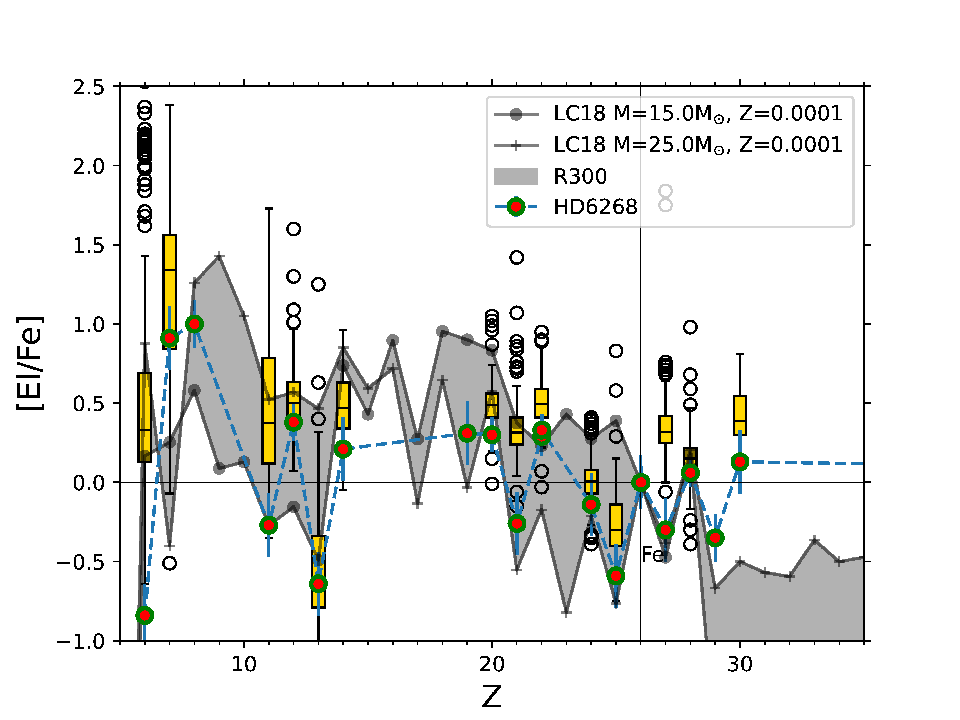}\\
\includegraphics[width=8.0cm]{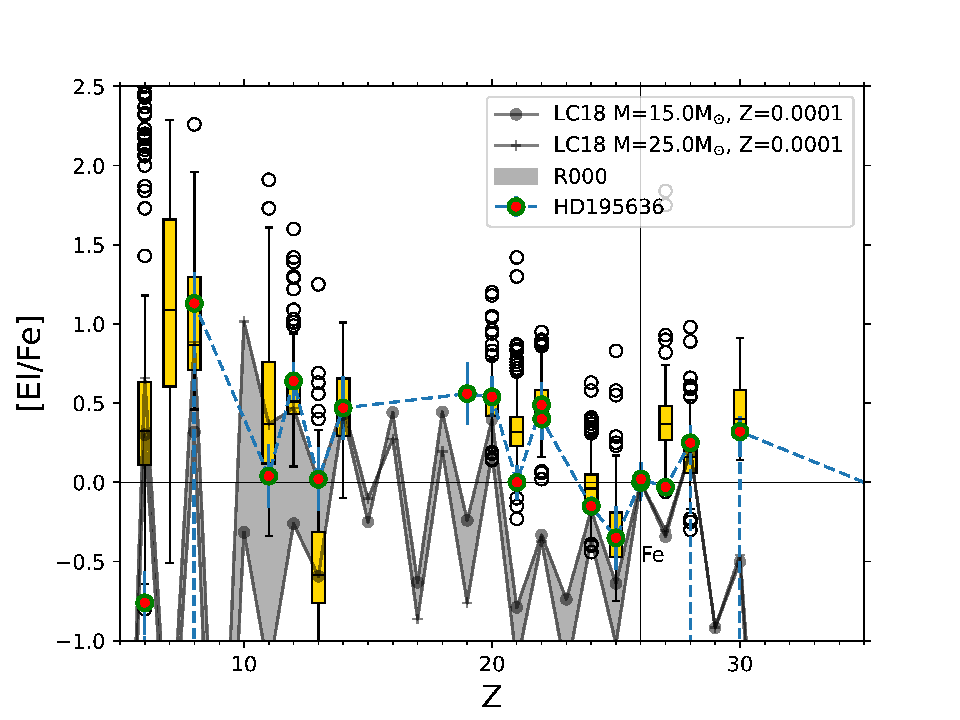} &\includegraphics[width=8.0cm]{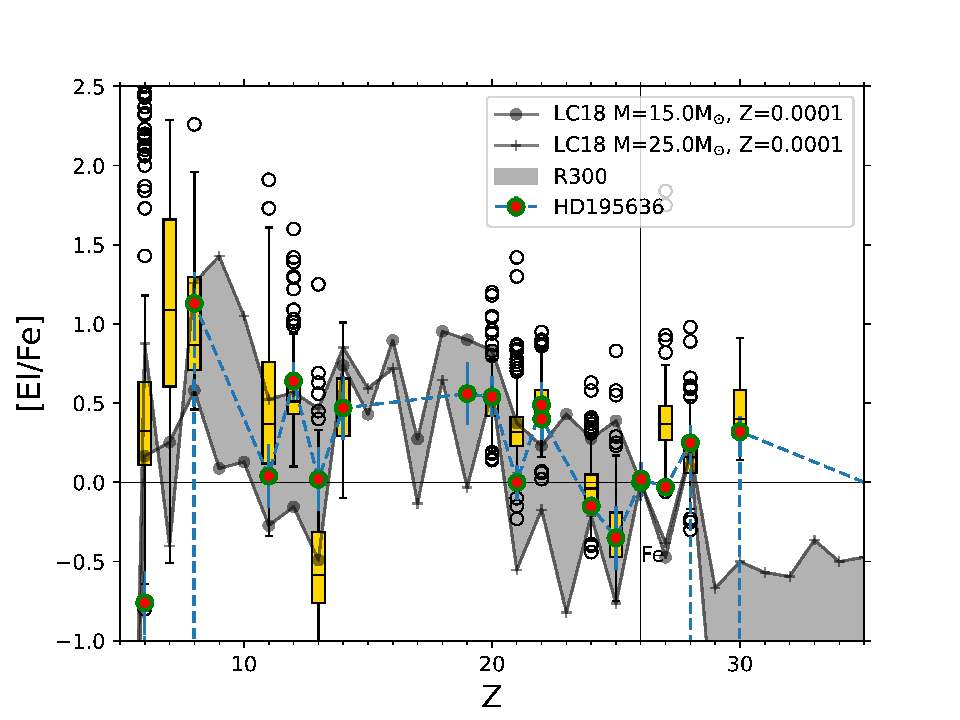}\\
\end{tabular}
\caption{Abundance distribution up to Zn of HE 1523--0901 (upper panels), HD 6268 (central panels) and HD 195636 (lower panels) are compared to other stars with similar [Fe/H] (within 0.5 dex) by \protect\cite{frebel:10} and to stellar model predictions for CCSNe of initial mass 15 M$_{\odot}$ and 25 M$_{\odot}$ and metallicity Z = 0.0001, non rotating (left panels) and with initial rotation equal to 300 Km per second (right panels), by \protect\cite{limongi:18}. }
\label{fig: stellar_archaeology}
\end{figure*}

\textbf{HE 1523--0901, HD 6268 and HD 195636}. In figure \ref{fig: stellar_archaeology} the HE 1523--0901, HD 6268 and HD195636 star abundances are compared with observations by \cite{frebel:10} and with stellar calculations by \cite{limongi:18} at comparable metallicities.
Compared to most of other stars in the reference sample, HE1523-0901 shows a low abundance of Na (Z=11), Ti (Z=22) and Zn (Z=30) with respect to Fe but still compatible with the errors. The low Sc (Z=21) and Co (Z=27) instead classify HE 1523--0901 as an outlier. The [C/Fe] abundance is much lower than for most of other stars, but we have already seen that this is due to the instrinsic HE 1523--0901 nucleosynthesis and it is not relevant here. We may instead expect that other peculiar HE1523-0901 abundances are indicative of the dominant contribution of one or a few CCSNe affecting the pristine gas from where the observed star formed. 

The CCSNe yields of the $\alpha$-elements O and Mg and of the odd element Na are mostly the products of hydrostatic He and C-burning in their massive star progenitors, and therefore we may expect that their production with respect to Fe is increasing with the initial progenitor mass. The heavier $\alpha$-elements Si and Ca are instead resulting from explosive O and Si-burning together with the iron group elements (including Ti, sometimes defined as an $\alpha$-element). Therefore, their production is more affected from the physics properties of the CCSN explosion and of the following propagation of the CCSN shock in the inner ejecta \citep[e.g.][]{thielemann:96, rauscher:02, pignatari:16, sukhbold:16}. In these conditions, typical one-dimensional stellar models with parametrized CCSN explosions historically struggle to reproduce observations. Low stellar predictions for Ti and Sc with respect to Fe are good examples, causing a systematically low [Ti/Fe] and [Sc/Fe] in GCE models compared to Milky Way stars at different metallicities \citep[e.g.][]{timmes:95, goswami:00, prantzos:18, kobayashi:20}. \cite{sieverding:23} recently showed that such a limited production disappears in a three-dimensional CCSN explosion. However, these simulations are computational expensive since the evolution of the neutrino-driven explosion needs to be followed for several seconds after core bounce to get the complete nucleosynthesis. More models will be required to provide a benchmark for observations and guidance to more simple parametrized CCSN stellar sets. PUSH models for CCSNe mimic the multi-dimensional nucleosynthesis of CCSNe still with an approximated spherical approach, but permitting to follow the correct $Y_e$ behavior resulting from neutrino interactions during the collapse and explosion \citep[][]{Perego.Hempel.ea:2015,Thielemann.Diehl.Heger.ea:2018,Curtis.Ebinger.ea:2019, Ebinger.Curtis.ea:2020,Ghosh.Wolfe.Frohlich:2022}. Also for these models the endemic underproduction of Sc compared to Fe seems to be solved, while Ti is still problematic. \cite{ritter:18} also found that the occurrence of C-O shell mergers during the hydrostatic evolution of the massive star progenitor may boost the production of Sc (and other odd elements like K, Z=19). However, C-O shell mergers are convective-reactive events where the predictive power of one-dimensional models is limited, and multi-dimensional hydrodynamics models are required to capture their impact on the nucleosythesis and on the stellar structure \citep[e.g.][]{andrassy:20}. Finally, fast-rotating massive stars may show relevant variations in the production of intermediate-mass elements compared to non-rotating progenitors. 

In Fig. \ref{fig: stellar_archaeology}, we compare the abundance pattern of our stars (including HE 1523--0901 in the upper panels) with non-rotating models for 15 M$_{\odot}$ and 25 M$_{\odot}$ stars and with their analogous rotating models. We choose the models with a high rotation velocity from \cite{limongi:18}, to maximize the impact of rotation on the CCSN yields. As expected, the non-rotating models underproduce Ti and the odd elements Sc and K with respect to Fe. Additionally, they underproduce Co and in particular Zn. The rotating models shown in the figure only marginally overproduce Si and Ca, but they are overall compatible with all the HE 1523--0901 abundances up to Fe. While the models are still underproducing the Ti observed for the reference stellar sample with respect to Fe, the Ti-poor HE 1523--0901 is reproduced. On the other hand, rotation does not provide a solution for the low production of Co and Zn.    

The production of the element Zn is another interesting puzzle for nuclear astrophysics. As we discussed for HE 1523--0901, the yields of typical CCSNe historically failed to reproduce the high abundance of Zn with respect to Fe observed in particular in metal-poor stars in the Milky Way. Hypernovae (HNe, with explosion energies at least ten times larger than typical CCSNe) yields show high Zn abundances with respect to Fe \citep[e.g.][]{nomoto:06, fryer:06, grimmett:21}, due to higher entropies attained in the explosion \citep[e.g.][]{Tsujimoto.Nishimura:2018,farouqi:22}. Therefore, it has been proposed that Zn is a signature of HNe contribution to GCE. \cite{kobayashi:11} and \cite{kobayashi:20}, however, showed that in order to reproduce the Zn observations, at low metallicity 50$\%$ of all CCSNe with progenitor masses larger than 20 M$_{\odot}$ should explode as HNe, which is at least an order of magnitude too high compared to observations \citep[e.g.][]{podsiadlowski:04}.        
A potential solution for the present Zn underproduction may be found as a result of slightly proton-rich ($Y_e>0.5$) explosive Si-burning in the innermost ejecta of CCSNe \citep[e.g.][]{Frohlich.Martinez.ea:2006}, wich can be achieved in several PUSH models also for typical CCSN explosion energies \citep[e.g.][]{Curtis.Ebinger.ea:2019, Ghosh.Wolfe.Frohlich:2022}. 

Compared to other stars in the reference sample, in Fig. \ref{fig: stellar_archaeology} HD 6268 shows a low abundance of Na and Mn (Z=25) with respect to Fe but still compatible with the errors. The low Sc and Co classify HD 6268 as an outlier. Non-rotating CCSN models underproduce the HD 6268 abundances of K, Sc, Ti, Cu (Z=29) and Zn. The rotating-models overproduce Si and Ca, and confirm the issue fitting Cu and Zn. 

Finally, compared to observations HD 195636 show a high abundance of Al (Z=13) and low abundance of Sc and Co with respect to Fe but still compatible with the errors. Non-rotating CCSN models underproduce the HD 195636 abundances of K, Sc, Ti, Co and Zn. The rotating-models overproduce Si, and confirm the issue fitting Co and Zn.

\subsection{Beyond Zn: neutron-capture elements}
\label{sec: beyond_zn}

Neutron-capture elements are 
made in different types of stars, 
during stellar hydrostatic evolution and explosions. In addition, the relative contribution by each process depends on the stellar source and at which time during galactic evolution the different processes are contributing. For the solar abundances, it is an established paradigm that about half of the abundances beyond Fe are made by the $r$-process \citep[][and references therein]{cowan:21}, and half by the $s$- process \citep[][and references therein]{kaeppeler:11}. Nevertheless, observational evidences are supporting the existence of additional processes or stellar sources predicted from theoretical stellar simulations, which may have also contributed to the solar abundances. Among others we mention here neutrino-driven winds ejecta from CCSNe \citep[e.g.][]{Frohlich.Martinez.ea:2006, farouqi:09, roberts:10, arcones:11, wanajo:11, arcones:13}, electron-capture supernovae \citep[e.g.][]{wanajo:11a, jones:19} and the intermediate neutron-capture process from different types of stars \citep[$i$-process, e.g.][]{herwig:11, lugaro:15, roederer:16, jones:16, cote:18, choplin:22}. Based on GCE calculations and stellar archaeology, \cite{travaglio:04} argued about the existence of the Lighter Element Primary Process or LEPP, an additional $s$-process component, to fully reproduce the solar abundance distribution of the lighter heavy elements beyond Fe up to Ba. The $s$-process in fast-rotating massive stars would represents an ideal candidate for such a component \citep[e.g.][]{pignatari:08, Frischknecht.ea:2012, Frischknecht.ea:2016, limongi:18}. However, the LEPP existence or its nature in the solar abundances is still a matter of debate \citep[][]{cristallo:15, bisterzo:17, prantzos:20, vincenzo:21, rizzuti:21}, and a direct connection between the solar composition and anomalous abundances observed in metal-poor stars is uncertain \citep[e.g.][and references therein ]{montes:07, qian:08}. 

For stars younger than the Sun in open clusters, the existence of an anomalous $i$-process contribution causing a Ba enhancement compared to solar was not expected \citep[][]{mishenina:15}, and it is still matter of debate if this may be a (still not identified) problem with Ba observation \citep[e.g.][and references therein]{baratella:21, dorazi:22}. 

While metal-poor stars may be used to trace the chemical enrichment history of the Galaxy, leading eventually to the imprint of the solar abundances \citep[][and references therein]{matteucci:21}, their heavy element abundances will carry the signature of a mixture of neutron capture processes different from the Sun \citep[][]{Molero.Magrini.ea:2023}. For instance, the $s$-process contribution from AGB stars to the GCE of the Milky Way became relevant for Pb at [Fe/H] $\gtrsim$ -1.5 \citep[e.g.][]{travaglio:01, kobayashi:20} and for Ba and La at [Fe/H] $\gtrsim$ -1 \citep[][]{travaglio:99, travaglio:01a}, i.e. at higher metallicities compared to the stars presented in this work (except for HD 121135, but we do not have data for Pb).  
Concerning the $r$-process, observations seem to support the existence of a number of different active sources, with their relative relevance changing over Galactic evolution time \citep[e.g.][]{roederer:10, wehmeyer:15,cote:19,farouqi:22,mishenina:22,vanderswaelmen:23}. In general, for observations of heavy elements at low metallicities the most significant star-to-star abundance scatter is obtained in the Sr-Pd region, where different stellar sources are potentially contributing to the galactic enrichment \citep[e.g.][]{sneden:08, cescutti:13, hansen:14,roederer:22}. On the other hand, when heavy element abundances are considered for single stars and compared with stellar simulations, a major level of complexity is a likely outcome, well beyond the expectations of nucleosynthesis paradigms built when starting from the solar abundances \citep[e.g.][]{arlandini:99, travaglio:04, sneden:08}.   

\begin{figure*}
\begin{tabular}{cc}
\includegraphics[width=8.0cm]{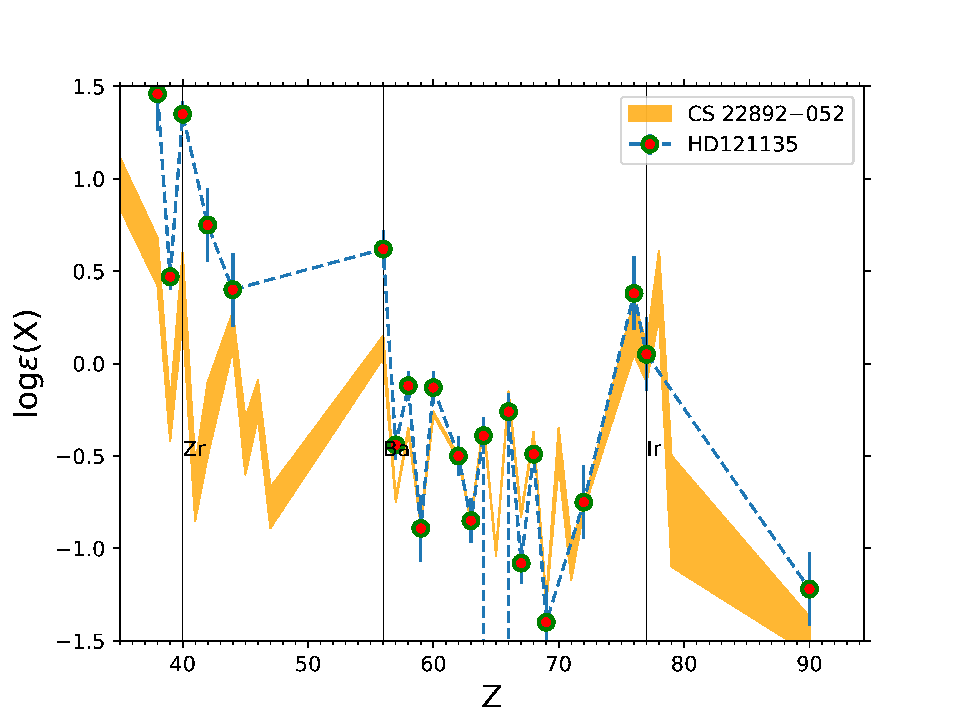}     &\includegraphics[width=8.0cm]{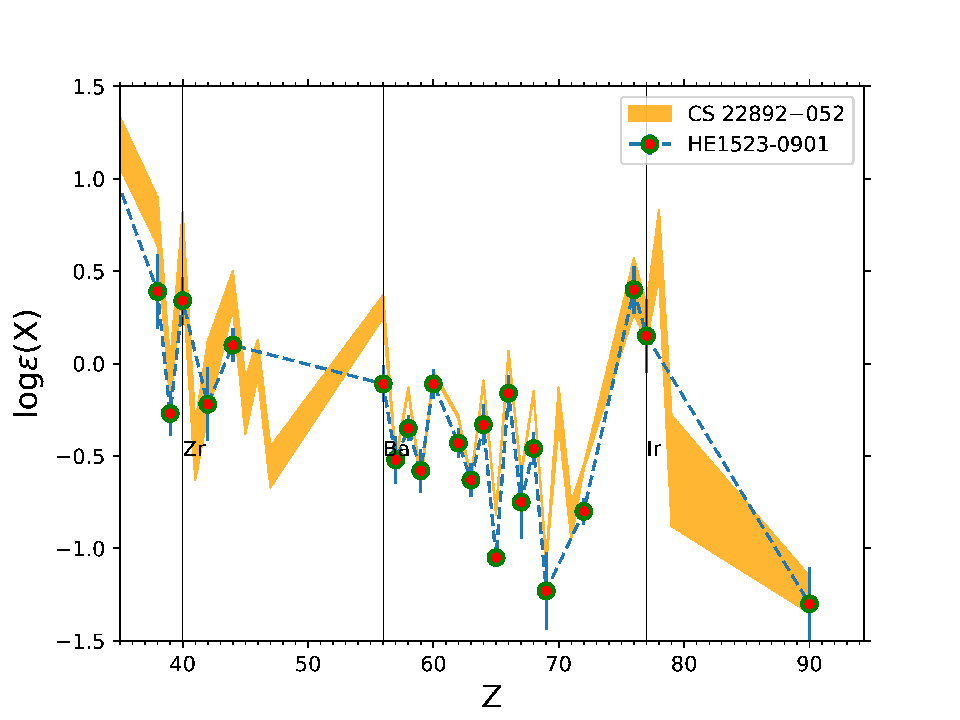}\\
\includegraphics[width=8.0cm]{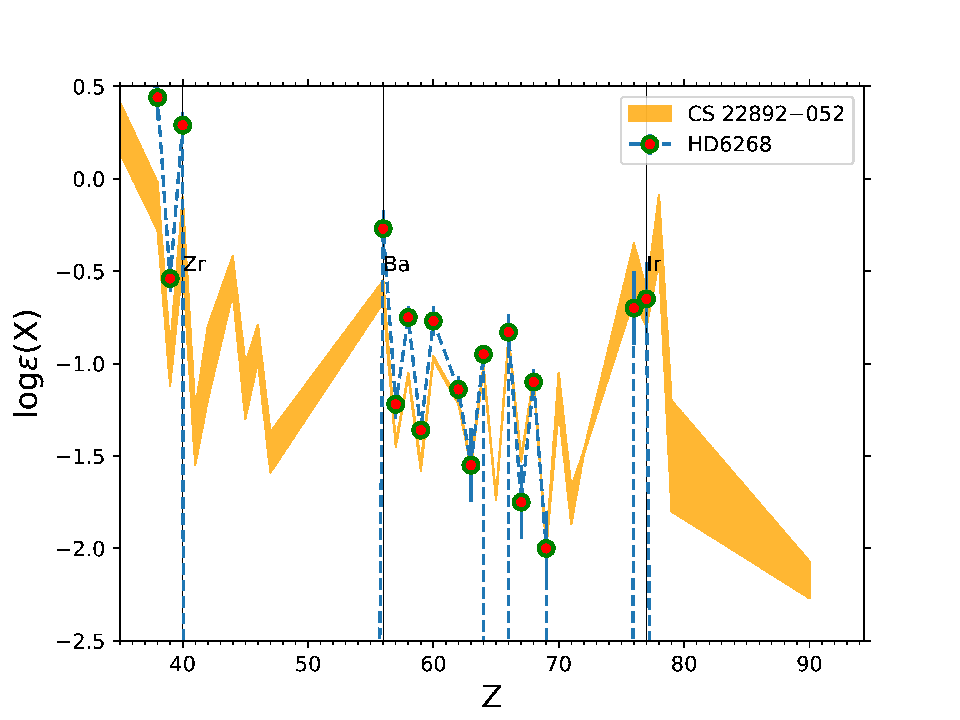}   &\includegraphics[width=8.0cm]{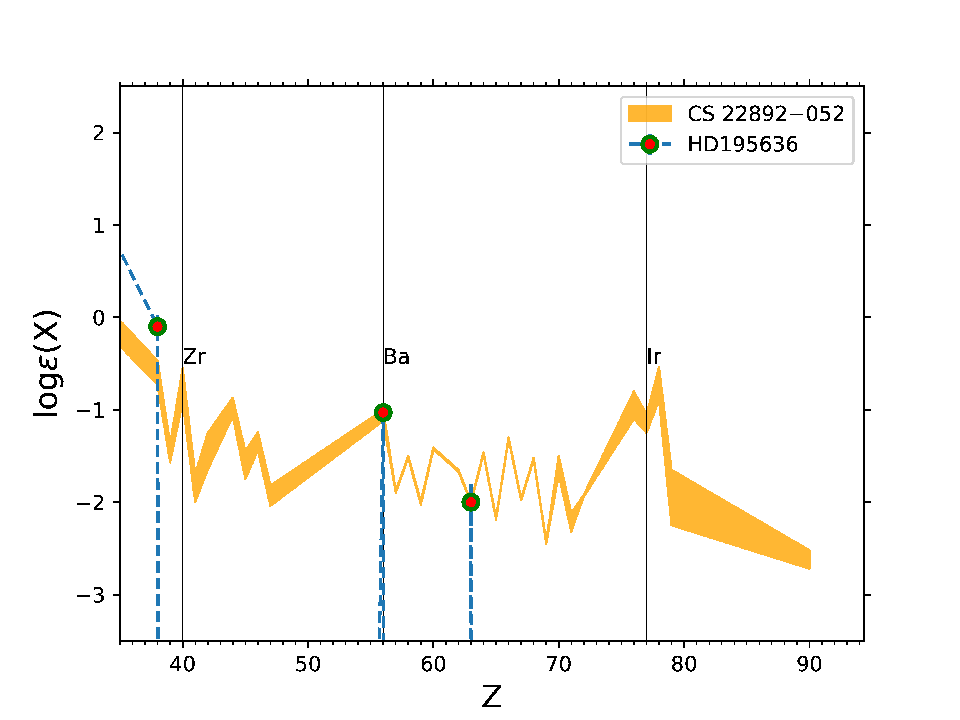}\\
\end{tabular}
\caption{Heavy element abundance distributions of HD1211135, (upper left) HE1523-0901 (upper right), HD6268 (lower left) and HD195636 (lower right) are compared to the abundance pattern of the r-process star CS22892-052 \citep[][]{sneden:03, sneden:09}. }
\label{fig: vs_sneden_star}
\end{figure*}

\textbf{HD 121135} As we have discussed in the first part of this section, for HD 121135 metallicity ([Fe/H] = -1.4) we would not expect any significant contribution from the s-process in AGB stars yet to be relevant (excepting for Pb, which abundance is not known for this star). Therefore, the r-process could be the only nucleosynthesis signature observed. 
However, we have discussed the existence of other nucleosynthesis processes contributing up to the Pd mass region, appearing even earlier than the $r$-process
\citep[e.g.][]{truran:02}. 

In Fig. \ref{fig: vs_sneden_star} (upper-left panel), HD 121135 is compared to the abundance of the classical $r$-process star CS22892-052 \citep[][]{sneden:03, sneden:09}. The two stellar abundances are shown on the same scale, using Eu as a normalization reference. HD 121135 abundances relative to the $r$-process element Eu are much higher than CS22892-052 up to Ce included (Z=58). Nd (Z=60) and Ho (Z=67) are also enhanced even taking into account observational uncertainties.

\textbf{HE 1523--0901} The HE 1523--0901 abundance pattern in Fig.  \ref{fig: vs_sneden_star}, upper right panel, shows in first approximation an $r$-process pattern. Taking into account 1$\sigma$ errors for our measurements and the $r$-process reference star, we observe a lower production for most of the lighter r-process elements with respect to Eu. Ba shows the most significant underproduction, in the order of 0.4 dex. On the other hand, there is no obvious departure from the $r$-process pattern, that would suggest an additional or alternative nucleosynthesis contribution to HE 1523--0901. Also the abundance signature in the actinide region seems to be consistent with the reference star CS22892-052.

\textbf{HD 6268} In Fig. \ref{fig: vs_sneden_star}, lower left panel, HD 6268 show an $r$-process pattern for the observed elements heavier than Sm (Z = 62) included. On the other hand, the elements at the Sr peak and between Ba and Nd (Z = 60) in HD 6268 are clearly overproduced compared to the CS22892-052 $r$-process pattern. The peculiar heavy element abundances of this star show therefore a behavior similar to HD 121135.

\textbf{HD 195636}. If we consider the stars presented in this work, for HD 195636 we have beyond Zn only data available for Sr (Z=38), Ba (Z=56) and Eu (Z=63) (Table \ref{abund}). Within errors, [Sr/Fe] is consistent with solar, [Eu/Fe] is mildly super-solar or solar, and [Ba/Fe] is about a factor of four lower than solar.  Figure \ref{fig: stellar_archaeology} does not show the HD 195636 abundances of those elements, which would be plotted a little lower than the bulk of other stars with similar metallicity in the boxplot. 
In Fig. \ref{fig: vs_sneden_star}, (lower-right panel), HD 195636 shows a similar signature for Ba and Eu compared to CS22892-052, and a mild Sr enrichment. With the limited number of elements available, it is not possible to clearly distinguish additional nucleosynthesis components in this star except for the $r$-process.  

HE 1523--0901 and HD 195636 abundances therefore seem to fit quite well with a classical understanding of nucleosynthesis of the heavier $r$-process elements in the early Galaxy as we introduced earlier. On the other hand, HD 121135 and HD 6268 show boosted abundances departing from the $r$-process CS22892-052 pattern at the Sr peak \citep[which are quite common, e.g.][]{cowan:21,roederer:22} and beyond it, involving several lanthanide elements \citep[which is not typically expected even considering the contribution from the LEPP component, see][]{travaglio:04}. For the metallicities of these stars ([Fe/H] = $-$1.4 and [Fe/H] = $-$2.55 respectively), the $s$-process in fast-rotating massive stars and the $i$-process could affect the local GCE signature of lanthanide elements and possibly explain the departure from the $r$-process signature. 

Concerning fast-rotating massive stars, for instance, the effective capability for the $s$-process to efficiently contribute to the Ba mass region and above is still quite uncertain, with widely different results shown in the literature, depending on the stellar models used and the nuclear reaction rates \citep[e.g.][]{pignatari:08, best:13, limongi:18, frost-schenk:22, lugaro:23}. A more effective impact at the Sr peak is, however, predicted from most of models of fast-rotating massive stars.  

\begin{figure}
\begin{tabular}{c}
\includegraphics[width=7.0cm]{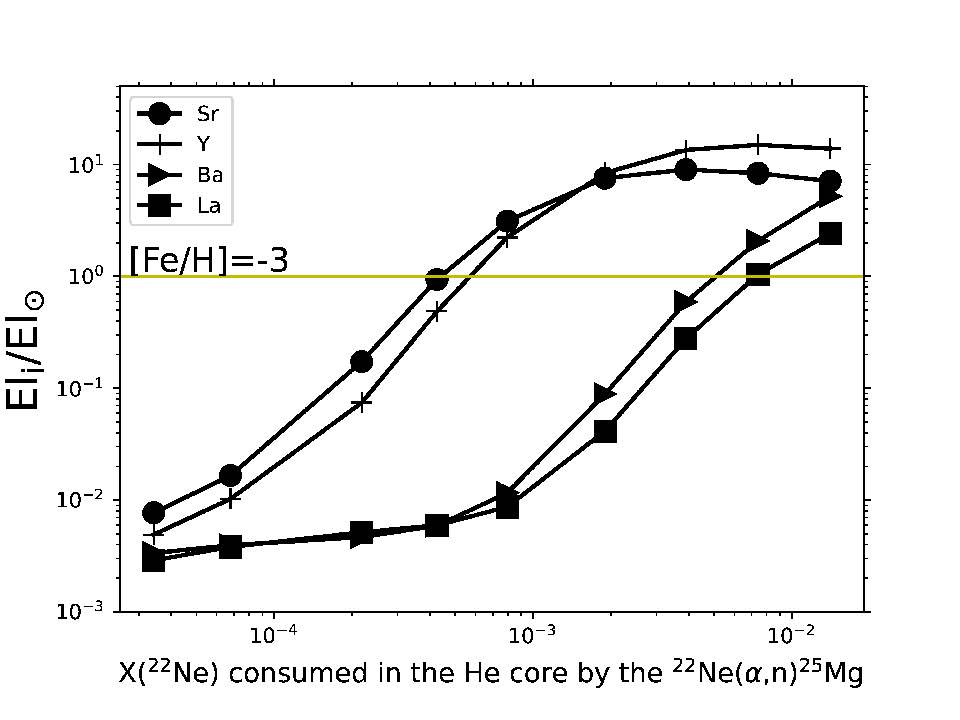}\\
\includegraphics[width=7.0cm]{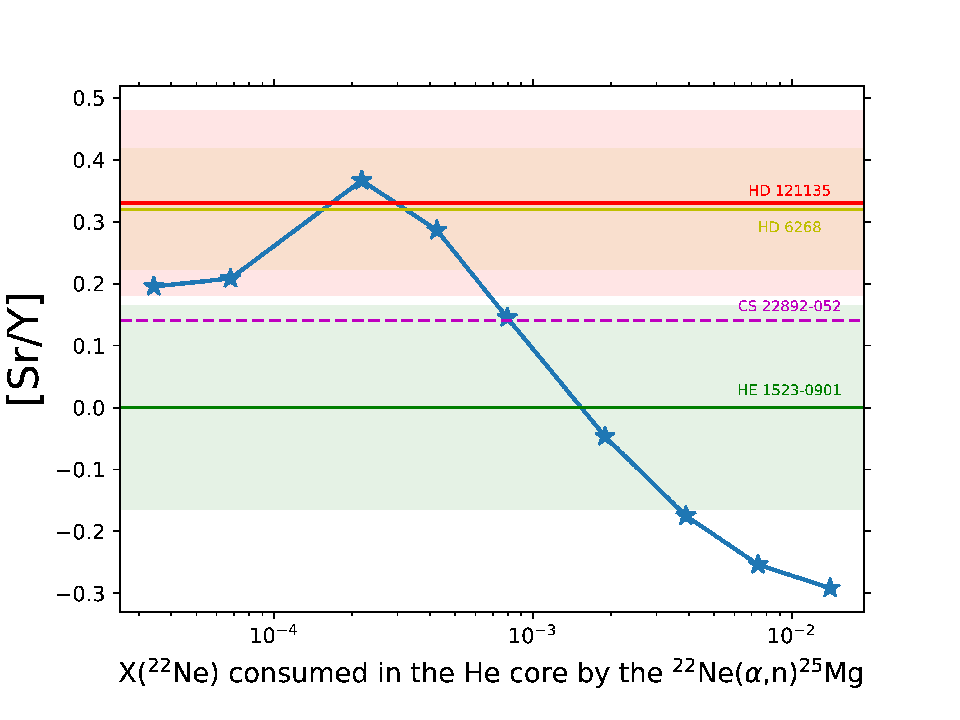}\\
\includegraphics[width=7.0cm]{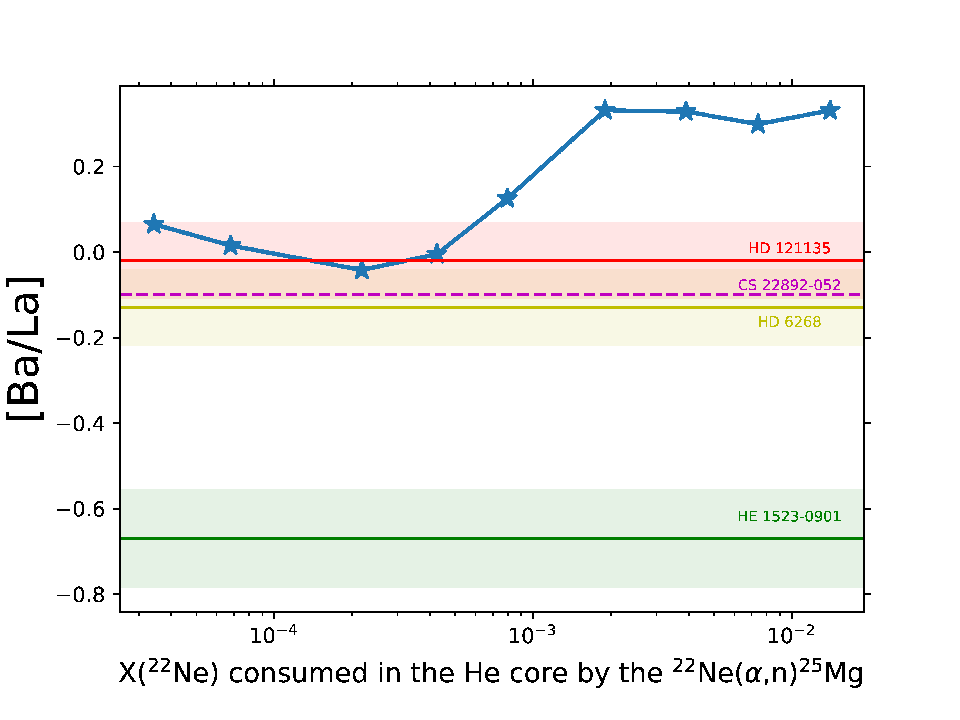}\\
\end{tabular}
\caption{Production of Sr, Y, Ba and La is shown for nucleosynthesis models of fast-rotating massive stars for different amounts of $^{22}$Ne consumed by the $^{22}$Ne($\alpha$,n)$^{25}$Mg reaction rate (top). The trends of [Sr/Y] (middle) and [Ba/La] (bottom) are shown with the stars observed in this work with current error ranges if available, in comparison with the $r$-process star CS22892-052 \citep[][]{sneden:03, sneden:09}. }
\label{fig: vs_fast_rotators_sproc}
\end{figure}

In Fig. \ref{fig: vs_fast_rotators_sproc} we compare the observed [Sr/Y] and [Ba/La] ratios with nucleosynthesis calculations representative of the production in fast-rotating massive stars \citep[][]{pignatari:08, Frischknecht.ea:2016, choplin:18,limongi:18}. The simulations are performed with the aid of the post-processing network code PPN 
\citep[e.g.][]{pignatari:12}, applied to a single-zone trajectory from a complete 25 M$_{\odot}$ star, resulting from calculations with the Geneva stellar evolution code GENEC \citep[][]{hirschi:04}. It is representative of the $s$-process production during both central He-burning and shell C-burning \citep[][]{nishimura:17}. The initial metallicity used for the calculations is [Fe/H] = $-$3.
In order to take into account the possible range of $s$-process production from different fast-rotating massive star models, the same parametric approach as applied, for instance, in \cite{pignatari:13} and \cite{roederer:22} is used here, where the abundance of $^{22}$Ne, consumed by the $^{22}$Ne($\alpha$,n)$^{25}$Mg reaction during core He burning, is used as a main parameter. 

In the upper panel of Figure \ref{fig: vs_fast_rotators_sproc}, the obtained $s$-process abundances for the elements considered are shown. In particular, the calculations show that an efficient production of the Sr $s$-process peak starts when the $^{22}$Ne($\alpha$,n)$^{25}$Mg reaction burns primary $^{22}$Ne in the order of few per mill in mass fraction. When the amount of $^{22}$Ne approaches values in the order of 1 per cent or more, also the Ba-peak elements are produced in comparable amount. In the present framework, the typical range of $s$-process production in fast rotating massive star models is represented \citep[e.g.][]{choplin:18, limongi:18}. In the central and lower panels of Fig. \ref{fig: vs_fast_rotators_sproc}, the HD 121135 and HD 6268 data show similar [Sr/Y] and [Ba/La] ratios within observational uncertainties. These observations would be both compatible with $s$-process abundances obtained for an amount of $^{22}$Ne consumed by the $^{22}$Ne($\alpha$,n)$^{25}$Mg neutron source in the order of a few 10$^{-4}$. However, the upper panel of the same figure shows that no significant production of $s$-process elements should be expected from fast-rotating models from such a low amount of primary $^{22}$Ne generated by internal rotation. On the other hand, for larger amounts of primary $^{22}$Ne consumed to generate neutrons both the predicted [Sr/Y] and [Ba/La] ratios are not compatible with HD 121135 and HD 6268. In this case [Ba/La] is mostly indicative, where an $s$-process production at the Ba peak would yield a ratio in the order of 0.3 dex instead of solar (HD 121135) or slightly sub-solar (HD 6268). Considering the higher metallicity of HD 121135, we cannot exclude that also fast-rotators would have the time to contribute to shape the pristine stellar abundances. However, they do not clearly provide their dominant signature, which is much closer to the $r$-process star CS22892-052 for the [Ba/La] ratio. We have seen that HE 1523--0901 mostly shows an $r$-process pattern, but with some significant variation. In Fig. \ref{fig: vs_fast_rotators_sproc}, we can highlight again this aspect, where the [Sr/Y] is consistent with 
CS22892-052 within uncertainties, but the [Ba/La] is significantly lower. This could suggest that the $r$-process abundances in HE 1523--0901 was produced from a different stellar source that made CS22892-052, and whose signature would not dominate the $r$-process residual in the Solar System \citep[more consistent with CS22892-052, e.g.][]{bisterzo:17}. An open question to be addressed by future theoretical studies is what are the $r$-process nucleosynthesis conditions allowing to yield the lowest Ba/La ratios observed in the early Galaxy, but without greatly affecting the overall $r$-process elemental pattern. 

The $i$-process can also be activated in massive stars \citep[][]{roederer:16, banerjee:18, roederer:22}, and therefore it could have contributed or contaminated the observed abundances of HD 121135 and HD 6268. However, neither of these stars seem to have abundances compatible with an $i$-process signature. While Ba and La are both enhanced compared to CS22892-052 (Figure \ref{fig: vs_sneden_star}), the [Ba/La] ratio is consistent with an $r$-process production (see Figure \ref{fig: vs_fast_rotators_sproc}). The $i$-process instead predicts a [Ba/La] significantly larger than solar \citep[e.g.][]{bertolli:13}. 
Also the abundances at the Sr peak do not seem to be compatible with an $i$-process pollution scenario. 

\begin{figure}
\begin{tabular}{c}
\includegraphics[width=7.0cm]{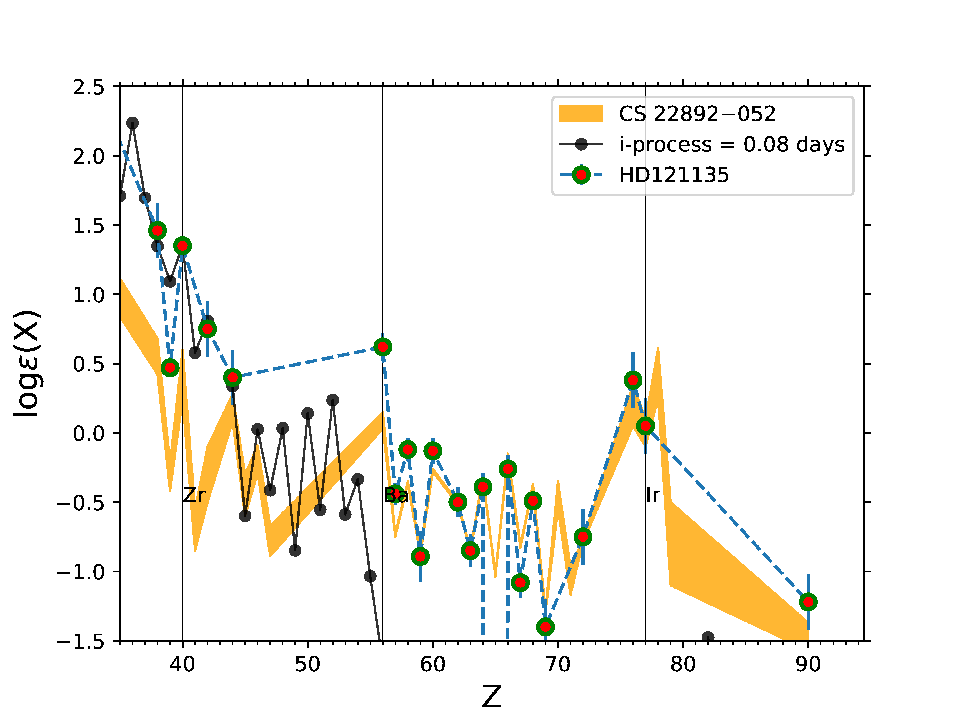}\\
\end{tabular}
\caption{As in Fig. \ref{fig: vs_sneden_star}, the heavy element abundance distributions of HD 121135 are compared to the abundance pattern of the $r$-process star CS22892-052, and to the $i$-process abundances normalized to Zr.}
\label{fig: vs_iproc}
\end{figure}

In Fig. \ref{fig: vs_iproc}, we compare the HD 121135 abundance pattern with the results from $i$-process conditions obtained for a mild neutron exposure feeding mostly the Sr-peak elements. We use here the simplified nucleosynthesis setup also adopted by e.g. \cite{bertolli:13, mishenina:15, roederer:16, roederer:22}. Compared to HD 121135, we find that the $i$-process overproduces Y by about 0.5 dex, with a departure that is significantly larger than the observational uncertainties.

In Appendix \ref{sec: r-process interpretation}, we discuss  the astrophysical interpretation of the abundances beyond iron observed in our stars in greater detail, within the context of stellar archaeology and the current debate about origin of the $r$-process.

\section{Conclusions}
\label{sec: conclusions}

We present a new study of the main properties and abundances for four metal-poor stars: HE 1523-0901 ([Fe/H]=–2.91$\pm$0.15), HD 6268 ([Fe/H]=–2.55$\pm$0.19), HD 121135 ([Fe/H]=–1.63$\pm$0.14), and HD 195636 ([Fe/H]=–2.79$\pm$0.04). Using the stellar spectra obtained with the HRS echelle spectrograph at the Southern African Large Telescope, for each star we derive the abundances of up to 40 elements with errors between 0.10 and 0.20 dex. We adopted the LTE approach for all the elements and NLTE corrections were applied for O, Na, Mg, Al, K, Mn, Cu, Sr, and Ba. For the other elements, we expect the NLTE corrections to be much smaller than current observational errors. HE 1523-0901, HD 6268, and HD 121135 are in the RGB phase, while HD 195636 already is on the HB. We therefore show that the observed C and N abundances cannot be considered as representative of their pristine composition, but instead were modified by their intrinsic nucleosynthesis. On the other hand, the other elements may be used to study the origin and the formation of the four stars.  
The main results are the following:
\begin{itemize}
\item \textbf{HE 1523-0901.} Based on its dynamic properties, we show that HE 1523–0901 is a halo star. 
Compared to other galactic stars in the same metallicity range, HE 1523–0901 has lower Sc and Co abundances, while the abundances of Na, Ti and Zn with respect to Fe are also low but still compatible with the errors. Compared to a set of representative non-rotating CCSN models, HE 1523–0901 shows much higher Ti and the odd elements Sc and K with respect to Fe. Compared to rotating models, the star shows lower abundances of Si and Ca, but it is overall compatible with all abundances up to Fe. In particular, the Ti-poor signature of HE 1523–0901 is reproduced. On the other hand, models with rotation do not provide a good solution for the low production of Co and Zn.
Beyond Zn, HE 1523-0901 shows an abundance signature similar to the $r$-process, although most of the elements are underproduced relatively to Eu compared to the abundandes in the $r$-process reference star CS22892-
052. Ba shows the most significant underproduction compared to the $r$-process pattern of CS22892-052), in the order of 0.4 dex. Therefore, we can classify HE 1523–0901 as an r-II stars ([Eu/Fe]=1.67), but showing anomalous abundance features like the really low Ba/La ratio, which are difficult to explain with classical $r$-process simulations.

\item \textbf{HD 6268.} We find that HD 6268 is a rare prototype of very metal-poor thick disk stars. The low Sc and Co abundances make HD 6268 an outlier compared to other stars with similar metallicity.
The CCSN models from non-rotating progenitors considered in our comparison underproduce K, Sc, Ti, Cu and Zn for HD 6268, while the models from rotating progenitors overproduce
Si and Ca and still under-produce Cu and Zn.
Beyond Zn, HD 6268 shows peculiar features that will definitely require further investigation: we find an r-process pattern beyond Sm, while the elements at the Sr peak and between Ba and Nd are relatively overproduced compared to the heaviest elements. Therefore, 
we could classify HD 6268 as an r-I star based on the Eu abundance, but it also show typical features of a limited-r star if we look at the relative high abundances of Sr, Y and Zr. On the other hand, the marginal actinide boost observed would be more compatible with r-II star features.
\item \textbf{HD 121135.} Its peculiar dynamic features make unclear if this star is a halo, a thick disk or an accreted star. 
Below Zn, for most of the elements, HD 121135 is compatible with other galactic disk stars with the same metallicity. However, the odd elements Al and Sc are significantly lower compared to other stars, which would be more expected for stars with lower metallicities than HD 121135, belonging to the galactic halo.
Beyond Zn, HD 121135 shows features similar to HD 6268.
Compared to the classical $r$-process star CS22892-052, the abundances in our star are enhanced up to Ce (included). Nd and Ho are also enhanced. At this stage it is difficult to identify if there is a significant contribution from some neutron-capture process other than the $r$-process, which would be reasonable considering the relatively higher metallicity of HD 121135 compared to the other stars discussed here. 
In general HD 121135 displays limited-r characteristics, but the Th detection would argue for an enriched r-I contribution instead. 
\item \textbf{HD 195636.} This is clearly a halo star, with the typical high proper motions. 
Compared to other stars, below Zn HD 195636 shows a
high abundance of Al and low abundance of Sc and
Co with respect to Fe, but still compatible with the observational errors. The non-rotating CCSN models underproduce the HD 195636
abundances of K, Sc, Ti, Co and Zn. The rotating-models
overproduce Si, and confirm the issue with Co and Zn.
Beyond Zn, with the limited number of elements available for HD 195636 it is not possible to clearly distinguish
different nucleosynthesis components. Sr, Ba and Eu abundances would be compatible with the r-process. Therefore, HD 195636 would be a borderline case between limited-r and r-I features.
\end{itemize}

In this work, we have provided the abundance pattern measured for a large number of elements in four metal-poor stars, and we have discussed them within the more general context of stellar archaeology. Among different abundance anomalies, we have identified, for instance, clear departures in the relative abundance of Ba and La from the classical solar $r$-process pattern for the r-II star HE 1523-0901. In general, even significant anomalies could not be identified with only a small number of measured elements available. This highlights the crucial importance to measure the chemical composition of a large number of stars, but also to provide stellar abundances from high-resolution spectra for as many elements as possible.      

\section{Acknowledgements}
MP acknowledges significant support to NuGrid from NSF grant PHY-1430152 (JINA Center for the Evolution of the Elements) and STFC (through the University of Hull's Consolidated Grant ST/R000840/1), and access to {\sc viper}, the University of Hull High Performance Computing Facility. MP acknowledges the support from the "Lend\"{u}let-2014" Programme of the Hungarian Academy of Sciences (Hungary).
FKT acknowledges support from the European Research Council (FP7) under ERC Advanced Grant Agreement 321263 FiSH.
 MP acknowledges support from the ERC Consolidator Grant (Hungary) funding scheme (project RADIOSTAR, G.A. n. 724560) and from the National Science Foundation (USA) under grant No. PHY-1430152 (JINA Center for the Evolution of the Elements). This article is based upon work from the ChETEC COST Action (CA16117), supported by COST (European Cooperation in Science and Technology).
We thank the ChETEC-INFRA project funded from the European Union's Horizon 2020 research and innovation programme (grant agreement No 101008324), and the IReNA network supported by NSF AccelNet. MP also thank the UK network BRIDGCE. TM thanks V. Kovtyukh for several comments.  TM also is grateful to the Laboratoire d'Astrophysique de l'Universite de Bordeaux for their kind hospitality. The authors would like to thank Piercarlo Bonifacio as a referee for the useful and detailed comments provided on the manuscript.

\bibliography{pecular}

\begin{appendix}

%\twocolumns

\section{Interpreting the heavy element abundances in terms of $r$-process contributions}
\label{sec: r-process interpretation}

Based on the observations of heavy elements discussed in Section \ref{sec: beyond_zn}, 
we could argue that for all our stars the abundance signature beyond the Sr-peak elements is due to or has been mostly shaped by the $r$-process. We have already mentioned that the $r$-process is not a universal process in nature as it has been assumed classically, and that different stellar sources are most likely responsible for its production in nature \citep[][and references therein]{cowan:21}. In particular, it is well established now that there is not a unique picture in low-metallicity stars to show a solar-type $r$-process pattern \citep[see e.g.][]{beers:05,Cain.ea:2020,Holmbeck.ea:2020,cowan:21,farouqi:22}. At least three different $r$-process patterns have been identified: \\
(a) limited-r stars which show elements as heavy as Eu but no third $r$-process peak elements and essentially no elements with $Z>70$. Its limiting value for [Eu/Fe] has been defined to be less than 0.3, but \cite{farouqi:22} have argued that it could be reduced to [Eu/Fe]$<$0. Additional constraints have been provided for Sr on the light (heavy) element side [Sr/Ba]$>$0.5, [Sr/Eu]$>$0 (in most cases $>$100).\\
(b) r-I stars are clearly $r$-process enriched and show a close to solar $r$-process pattern with
[Eu/Fe]$>$0-0.3 and $<$1 with in all cases [Ba/Eu]$<$0, i.e. the Ba/Eu ratio being smaller than a solar-like $s$-process-dominated composition.\\
(c) r-II stars are highly $r$-process enriched with [Eu/Fe]$>$1 and follow a [Ba/Eu] constraint as in (b). Among the r-II stars there exists an important fraction with a so-called actinide boost and an over-abundance of [Th/Eu] relative to the solar system value. When utilizing the solar system value of Th at the time of its birth, this corresponds to (Th/Eu)$>$0.42 (see Table 6 in \cite{farouqi:22}) which applies to more than 30\% of r-II stars, but also to some r-I stars.\\

If we apply these criteria to the stars discussed in this work, HE 1523--0901 is a clear representative of r-II stars (with [Eu/Fe]=1.67); HD 6268 qualifies as an r-I candidate ([Eu/Fe]=0.47)\footnote{Notice however that when comparing our abundance data to Fig. D1 in \cite{Saraf.ea:2023}, which provides also an extended overview of all existing observations, HD 6268 represents for the light elements like Sr, Y, Zr a borderline case between limited-r and r-I, and for some abundances of heavier elements a borderline case between an r-I and an r-II star. 
\cite{farouqi:22} list the star in their Table 6 as a marginal actinide-boost candidate. This could argue for a superposition of two events, contributing to these features.}; and HD 195636 shows a mild $r$-process enrichment as a borderline case between limited-r and r-I stars ([Eu/Fe]=0.26), if following suggestions by \cite{farouqi:22} to move down the limit between limited-r and r-I stars from [Eu/Fe]=0.3 to 0. For our star at the moment this is not clear, given the observational uncertainties and the limited number of elemental abundances known (for instance, there are no data for the third $r$-process peak elements). 
Finally, with [Eu/Fe]=0.03, HD 121135 is clearly a limited-r candidate, although the Th detection with [Th/Fe]=0.16 argues for an r-I star. 
Overall our four stars with their abundance patterns compare well with the extended overview of existing abundance observations of [El/Fe] shown in Fig. D1 of \cite{Saraf.ea:2023} and the 
limited-r, r-I, and r-II classification. 

The stars studied in this work 
seem to be consistent with a scenario where the solar $r$-process pattern it is not a universal product of $r$-process sites in nature (see also the variations shown in Fig. \ref{elfeat}).
In principle, the source of such a variation could be explained by a unique astrophysical site but with varying conditions from event to event, or alternatively by a number of $r$-process sites contributing to the chemical evolution. Interestingly, we obtained similar conclusions also from Th and Eu observation of more metal-rich MW disk stars \citep[][]{mishenina:22}. 
Among the different stellar sites, recent $r$-process studies include regular CCSNe with a very weak $r$-process \citep[see e.g.][]{Ghosh.Wolfe.Frohlich:2022, Wang.Burrows:2023}, magneto-rotational (MR) SNe with a mostly relatively weak but varying $r$-process pattern \citep[e.g.][]{winteler:12,Moesta.ea:2014,Nishimura.Sawai.ea:2017,moesta20,Reichert.Obergaulinger.ea:2021,Siegel:2022}, collapsars/hypernovae \citep[e.g.][]{siegel:19,Siegel:2022}, as well as neutron star and neutron star-black hole mergers \citep[e.g.][]{Wanajo.Sekiguchi.ea:2014,Just.Bauswein.ea:2015,Wu.Fernandez.Martinez.ea:2016,Siegel:2022,Holmbeck.Sprouse.ea:2023}.
Depending on their nature, these events may be identified as frequent or rare events \citep[][]{cowan:21}. The most frequent stellar explosions are CCSNe and their imprint is seen in GCE as early as [Fe/H]$<$-5. From the observational side, \cite{farouqi:22} showed in their Fig. 26 \citep[extracted from][]{JINAbase:2018} \footnote{https://jinabase.pythonanywhere.com/}
that the first limited-r events are observed around [Fe/H]$\sim$ -4, while r-I and r-II stars start to appear at slightly higher metallicity, with [Fe/H] between -4 and -3.5. This could indicate that limited-r producing events are apparently the most frequent of the three $r$-process sites and probably related to a rare class of supernovae (most probably magneto-rotational (MR) supernovae). 
On the other hand, r-I and r-II pattern producing events would be rarer 
and suggested to be related to compact binary mergers and collapsars. We also should consider the possibility that a fraction of these could also be responsible for actinide boost stars.

\cite{farouqi:22} came to the conclusion that the ratio between MR supernovae and CCSNe is about 1/10 \citep[related to the ratio between magnetars and the total number of neutron stars,][]{Beniamini.Hotokezaka.Horst.ea:2019} and the ratio between compact binary mergers or collapsars and CCSNe is 1/120 or 1/125 (i.e. very similar). \cite{Fukagawa.Prantzos:2023} assumed instead 1/1000 to 1/100 for the ratio of mergers to CCSNe and about 1/4 to 1/5 between collapsars and mergers. 
The reasoning of these different sites is of course still preliminary. Especially whether compact binary mergers and/or collapsars are responsible for r-I and r-II (as well as actinide boost) abundance patterns or possible variations in these two classes of events might bare the chance to produce the full spread of the r-I and r-II stars is still not decided \citep[see e.g.][]{holmbeck19b,farouqi:22}.

Thus, the observed abundance patterns for elements up to Zn (or slightly beyond) are shaped by the nucleosynthesis in (frequent) CCSNe, and show up already in stars with [Fe/H] $\lesssim$ -5.  
For higher metallicities, it is becoming more and more challenging to identify the  
signature of single CCSNe in the observed stars.
The rarer $r$-process events (leading to limited-r, r-I, or r-II abundance patterns), appear observationally only for [Fe/H] $\gtrsim$-4 \citep[][and references therein]{farouqi:22}. For those observations of heavy elements, the imprint of a single (r-process) event is probably still a good assumption. This would imply that a superposition of typical CCSN patterns with a limited-r, r-I or r-II pattern applies for most r-enriched low-metallicity stars. 
Within this same scenario, a stellar abundance pattern resulting from contributions by CCSNe, a rare CCSN event (causing a limited-r pattern), and an r-I or r-II contribution could be also found for some cases \citep[see e.g.][for J1424-2542]{Placco.Almeida.ea:2023}.

If we apply this understanding to our observations, HE 1523--0901 is a clear representative of r-II stars, also showing low abundances of the light $r$-elements, and argues for a single imprint among the heavy elements, but also already a CCSN contribution for the elements up to Zn. HD 6268, on the other hand, displays r-I features, but also higher abundances for Sr, Y, Zr (pointing to a limited-r contribution) and a marginal actinide boost might even argue for r-II features. HD 195636 represents a borderline case between limited-r and r-I features. And HD 121135, displaying clearly limited-r characteristics but also a Th detection arguing for an enriched r-I contribution. Thus, for some of these borderline cases the observed abundance patterns are not necessarily pointing to only one polluting site. In fact, for [Fe/H]$>$-2 already a sufficient amount of (possibly different) $r$-process events took place to lead to an averaged $r$-process pattern which
resembles one dominating $r$-process source \citep[][and references therein]{cote:19}. 

A number of independent investigations by \cite{Hartwig.Ishigaki.ea:2019, Hartwig.Ishigaki.ea:2023} and e.g. \cite{Cescutti.Morossi.ea:2021,Han.Yang.Zhang.ea:2021,Scannapieco.Cescutti.ea:2022,Molero.Magrini.ea:2023,Carrillo.Ness.ea:2023} (but only for lighter elements up to Fe) as well as \cite{siegel:19,Fukagawa.Prantzos:2023,Skinner.Wise:2023,Kolborg.Ramirez-Ruiz.ea:2023} have followed along such lines and discussed contributions from different sources (including rare events) within a galactic chemical evolution framework. In summary, we can conclude that low-metallicity stars with [Fe/H]$<$-2.5 already show, in all cases, a CCSN abundance contribution for the elements lighter than Zn, while for heavier elements a single (or in rare cases very few, possibly different) $r$-process sites have contributed. On the other hand, 
the argument of a dominating $r$-process source shaping the abundances of heavy elements could still be made for the majority of MW disk stars \citep[][and references therein]{cote:19}.

\end{appendix}

\end{document}